\documentclass[11pt]{amsart}

\usepackage[T1]{fontenc}
\usepackage{lmodern}
\usepackage{graphicx}
\usepackage{tikz}
\usepackage{amsthm}
\usepackage{amssymb}
\usepackage{mathrsfs} 
\usepackage{cite}
\usepackage{xcolor}
\usepackage{booktabs}
\usepackage[margin=1in]{geometry}
\usepackage[colorlinks=true]{hyperref}

\theoremstyle{plain}
\newtheorem{theo}{Theorem}[section]
\newtheorem{lemma}[theo]{Lemma}
\newtheorem{proposition}[theo]{Proposition}
\newtheorem{corollary}[theo]{Corollary}

\theoremstyle{definition}
\newtheorem{definition}[theo]{Definition}

\newtheorem{notation}[theo]{Notation}

\theoremstyle{remark}
\newtheorem{remark}[theo]{Remark}

\def\A{{\rm A}} 
\def\B{{\rm B}}

\def\F{{\rm F}}
\def\G{{\rm G}}
\def\K{{\rm K}}
\def\L{{\rm L}}
\def\N{{\rm N}}
\def\P{{\rm P}}
\def\Q{{\rm Q}}
\def\S{{\rm S}}
\def\T{{\rm T}}
\def\U{{\rm U}}
\def\V{{\rm V}}
\def\X{{\rm X}}
\def\Y{{\rm Y}}

\def\GL{{\rm GL}}
\def\Ad{{\rm Ad}}
\def\Id{{\rm Id}}
\def\SO{{\rm SO}}
\def\Sp{{\rm Sp}}

\def\Spin{{\rm Spin}}
\def\Stab{{\rm Stab}}
\def\Span{{\rm Span}}
\def\SU{{\rm SU}}
\def\vol{{\rm vol\,}}

\newcommand{\pd}{\partial}
\newcommand{\mc}{\mathcal}
\newcommand{\mf}{\mathfrak}

\newcommand{\Li}{\mathrm{Li}}
\newcommand{\floor}[1]{\left\lfloor #1 \right\rfloor}

\newcommand{\bs}{\backslash}
\newcommand{\tr}{\mathrm{tr}}
\newcommand{\JacTh}[4]{\vartheta\begin{bmatrix}#1\\#2\end{bmatrix} \left(#3 \,\middle|\, #4\right)}
\newcommand{\abs}[1]{\left\lvert#1\right\rvert}

\usetikzlibrary{patterns}

\begin{document}

\title{Representation theory and massive spectral zeta functions on spherical space forms}

\author{Victoria Martin}
\email{victoria.martin@unf.edu}
\address{Department of Physics and Astronomy, University of North Florida \\ 1 UNF Drive, Jacksonville, FL 32224, USA}

\author{Allan Merino}
\email{allan.merino@unf.edu}
\address{Department of Mathematics and Statistics, University of North Florida \\ 1 UNF Drive, Jacksonville, FL 32224, USA}

\author{Rahul Poddar}
\email{rahul.poddar@unf.edu}
\address{Department of Physics and Astronomy, University of North Florida \\ 1 UNF Drive, Jacksonville, FL 32224, USA}

\author{Simon Roby}
\email{simon.roby@ac-nancy-metz.fr}
\address{Lyc\'ee Fabert \\ 12 Rue Saint-Vincent, 57000 Metz, France}

\begin{abstract}
  
We study the spectral theory of the massive scalar kinetic operator on spherical space forms $\Gamma \bs S^{3}$,  where $\Gamma$ is a finite subgroup of $\SO(4)$ acting freely on $S^{3}$.  
Using the Peter-Weyl theorem and Molien series we compute the spectrum of the Laplace-Beltrami operator on $\Gamma\bs S^3$ directly from representation theory, obtaining the eigenvalues and their multiplicities for all lens spaces $L(p,q)$ and for each ADE-class of finite subgroups of $\SU(2)$, which includes homogeneous lens spaces, prism spaces and quotients by the binary tetrahedral, octahedral and icosahedral groups, the last also known as the Poincar\'e homology sphere.

For homogeneous quotients by subgroups of even order, we construct the meromorphic continuation of the spectral zeta function of the massive kinetic operator and obtain the general zeta-regularized determinant in closed form.
The determinant is expressed as a finite sum of polylogarithms, valid for all real mass, and depends on $\Gamma$ only through its order and a finite set of parameters which we call twist coefficients fixed by the dimensions of the $\Gamma$-invariant subspaces.

As a complementary approach, we express the heat trace of the kinetic operator on spherical space forms in terms of Jacobi theta functions, whose modular properties relate the spectral zeta function to geometric data, the volume and closed geodesic lengths of $\Gamma\bs S^3$. 
We find that the spectral zeta function admits a rapidly convergent expansion in the heavy mass sector in terms of modified Bessel functions, and provides an independent check on the determinant. 

This framework extends to quotients of higher-dimensional manifolds directly, which we outline for future work. 

\end{abstract}


\maketitle
\tableofcontents 
\section{Introduction}

Semiclassical gravity provides a framework for studying quantum fluctuations about fixed, classical saddle geometries. In Euclidean quantum gravity, the gravitational path integral is approximated by a sum over saddle points, each weighted by the exponential of its classical action and by one-loop functional determinants of the kinetic operators for fields propagating on that fixed background \cite{gibbons1977action, Christensen:1979iy}. These determinants give the leading-order quantum correction to each saddle’s contribution. For this reason, it is of great interest to develop a systematic means of computing one-loop determinants on all non-trivial saddle geometries \cite{Maloney:2007ud,Castro:2011xb,Anninos:2020hfj,Bourne:2025azc}.

An important class of nontrivial Euclidean saddles are constant curvature quotient manifolds $\mathcal{M}/\Gamma$, where $\Gamma$ is a discrete subgroup of the group of automorphisms on $\mathcal{M}$. While such manifolds have been studied extensively \cite{Wolf,Giombi:2008vd, Yin:2007gv,Camporesi:1990wm,Dowker:2013qda,David:2009xg,Dijkgraaf:2000fq,bytsenko1996quantum}, the Laplacian eigenvalue multiplicities can become highly nontrivial after quotienting by a discrete group. Here we work towards developing a general framework that determines these multiplicities (and hence the associated one-loop partition functions) directly from the symmetries of the background geometry. While the program that we pose is generic, here we focus on a subset of spherical quotient manifolds $\Gamma\bs S^3$. Our intention is to pave a clear path forward for future work, extending methods developed here to $S^N$ and non-compact manifolds. In addition, for now we restrict our one-loop analysis to the case of a massive scalar field propagating on $\Gamma \bs S^3$, which is the building block for matter loops in dS$_3$ quantum gravity \cite{Castro:2023dxp} and a template for gravitational fluctuation (graviton) kinetic operators. 

Analyzing spectral data on spherical quotients has a rich and well-established history. Mathematicians interested in the topological classification of smooth manifolds introduced Reidemeister torsion \cite{reidemeister1935homotopieringe,franz1935torsion}, which was later recast in terms of functional determinants as Ray-Singer torsion \cite{Ray:1971,Ray1970}. The Laplace-Beltrami eigenvalue multiplicities on spherical quotients have also been computed \cite{IkedaYamamoto1979,Ikeda1980,MR1317006}. In a particularly nice work \cite{Nash:1992sf}, Nash and O'Connor compute Ray-Singer torsion\footnote{The Ray-Singer torsion is built from determinants of the Laplacians $\Delta_q$ on $q$-forms, $\log T=\sum_{q=0}^{\dim\mathcal M}(-1)^{q}q\log\det{}'\Delta_q$ \cite{Nash:1992sf}. The scalar determinants considered in \cite{Nash:1992sf} are defined with a flat bundle and are massless, whereas the determinants here are untwisted and massive.}
for certain lens spaces $L(p,q)$. Determinants of massless and massive fields on spherical quotients have also been studied in \cite{Bauer,Dowker:2004nh, Dowker:2004sm,Dowker:2013qda}. Determinants of spin-2 fields are considered in \cite{Castro:2011xb}.
Recent work involving lens spaces includes \cite{2026arXiv260931757G}.

Our work adds to this program by treating massive scalars and finding closed form expressions.
Shifting the Laplacian by a mass term, $\mathcal K=\Delta+\mf m^2\ell^2$, does not alter the underlying representation-theoretic framework, but requires some additional analytic considerations when computing the spectral zeta function and its regularized determinant. This extension is nevertheless essential for physical applications.
We identify four mass-sectors: the massless limit, the light sector, the critical point, and the heavy sector.
We obtain formulae which are holomorphic in the mass parameter in specific region of the complex plane, and treat the massless limit carefully by removing a logarithmic divergence arising from the zero mode, and as an independent check compute the massless case by itself. 

For each quotient we obtain the multiplicities from the Peter-Weyl theorem and Molien series \cite{stanley1979invariants}, and use them to compute the spectral zeta function and its regularized determinant.
As a complementary approach, we construct the heat trace using the multiplicities and express it in terms of Jacobi theta functions, and, via modular transformations, a form of the heat trace and spectral zeta function are expressed through the volume of $\Gamma\backslash S^3$ and the lengths of closed geodesics.
In the heavy sector this yields rapidly convergent Bessel-function expansions of the zeta function for sufficiently large mass.
All determinants are given as finite sums of polylogarithms, valid for all real masses and for quotients by all even order subgroups of $\SU(2)$.
For $S^3$ itself, we give a derivation of the light-sector determinant, whose analytic continuation from the heavy sector was proposed in \cite{Castro:2023dxp}. 
In addition, since we explicitly calculate the scalar determinants for all even homogeneous lens spaces, we hope this will be a guide for deriving a Wilson spool \cite{Castro:2023dxp} for lens spaces in future work, and can be checked against the spool proposal made in \cite{Haupfear:2025grb}.

The reason that $\Gamma \bs S^3$ is a particularly rich arena to begin with is that the finite subgroups of $\SU(2)$ are completely classified (known as the ADE classification, see \cite{Wolf}). Up to conjugation, they are given by the following families:
\begin{itemize}
\item The cyclic groups $\mathbb{Z}/n\mathbb Z$ of order $n$, for all $n \geq 1$\,,
\item The binary dihedral groups $D_{n}$ of order $4n$, for all $n \geq 2$\,,
\item The binary tetrahedral group $\widetilde{\mathcal{T}}$ of order $24$\,,
\item The binary octahedral group $\widetilde{\mathcal{O}}$ of order $48$\,,
\item The binary icosahedral group $\widetilde{\mathcal{I}}$ of order $120$\,.
\end{itemize}
While the cyclic groups are abelian, all of the others are non-abelian. The dihedral quotients are known as prism spaces \cite{aurich2012well}, and the icosahedral quotients give the Poincar\'e homology sphere \cite{kirby1979eight}. 
Further, the McKay correspondence relates each of these families of finite groups to the affine Dynkin diagram of ADE type: $\mathbb Z/n \mathbb Z \mapsto \hat A_{n-1} \cong \hat{\mf{su}}(n),\, D_n \mapsto \hat D_{n+2} \cong \hat{\mf {so}}(2n+4)$ and the last three mapping to the affine Dynkin diagrams of the exceptional Lie algebras $\hat E_6$, $\hat E_7$ and $\hat E_8$.
For each of these groups, the multiplicities of the eigenvalues of the Laplace-Beltrami operator on the corresponding spherical space form may be determined using the representation-theoretic methods developed in this paper.

Another spherical quotient geometry that has garnered much  attention in theoretical physics is the inhomogeneous lens space $L(p,q)$, $q\neq \pm 1 \bmod p$ and coprime with $p$, which have found applications in the de Sitter Farey tail \cite{Castro:2011xb} and cosmology \cite{Lehoucq:2002wy}. 
Inhomogeneous lens spaces are an example of quotients that are beyond the standard ADE classification, since the quotient group is a subgroup of $\SO(4)$ rather than $\SU(2)$.
We obtain a formula for the multiplicities of the eigenvalues of the Laplace-Beltrami operator for general co-prime $p$ and $q$. 

\subsection{Methods}

In physics, standard techniques for calculating one-loop determinants include the heat kernel method \cite{Vassilevich:2003xt} and the quasinormal mode method \cite{Denef:2009kn}. Our goal is to read off the spectrum of a scalar field on $\Gamma \bs S^3$ from symmetry alone, without ever solving for the heat kernel or relying upon Lorentzian input. We begin by specifying an embedding of the discrete subgroup $\Gamma\subset\SO(4)$, and determining the eigenvalue multiplicities in two ways:
\begin{itemize}
\item \textbf{Peter-Weyl theorem.} Peter-Weyl tells us that the space of square-integrable functions on a compact group $\G$ splits into irreducible representations, each appearing with a multiplicity equal to its dimension. When considering the quotient $\Gamma\bs S^3$, one must keep only those field configurations that remain invariant under the action of $\Gamma$. In general, the multiplicities are reduced, and the problem of finding the spectrum therefore becomes a counting problem: how many states survive the $\Gamma$ projection? For cyclic and binary dihedral groups we count these directly. 
For the subgroups corresponding to the E-class of the ADE classification, i.e. the binary tetrahedral, octahedral and icosahedral groups, we use a generating function known as the Molien series. 
\item \textbf{Molien series.} The Molien series is a generating function for the multiplicities. It is built from an average over the elements of $\Gamma$ which collapses to a compact rational function from which the multiplicities can be easily extracted. Extensions of the Molien series have been studied in the context of high energy theory in \cite{Benvenuti:2006qr} and has been used to compute holographic thermal propagators in a recent work \cite{Bajc:2026wzi}.
This method of calculating multiplicities will be the main tool in extending to higher dimensions.
\end{itemize}

Once the degeneracies are known, we use them to define the spectral zeta function of the massive kinetic operator 
\begin{equation*}
\zeta_{\mc K}(s)=\sum_n \lambda_n^{-s}\,,
\end{equation*}
where $\lambda_n$ denotes the eigenvalues of the shifted Laplacian $\mathcal{K}=\Delta+\mf m^2\ell^2$, and $\mf m$ is the mass of the scalar field and $\ell$ is the associated length scale. We then construct the meromorphic continuation of $\zeta_{\mc K}(s)$ beyond its domain of absolute convergence and evaluate its derivative at the origin. 
The resulting quantity determines the zeta-regularized one-loop determinant \cite{hawking1977zeta}, $\log\det \mathcal{K}=-\zeta'_{\mc K}(0)$, and hence the corresponding one-loop partition function. 
As a complementary approach, we rewrite the heat trace using Jacobi theta functions, which lets us trade the sum over eigenvalues for a sum over closed geodesics, and relates the zeta function directly to the volume and geodesic lengths of $\Gamma\bs S^3$.

We find a general formula for the spectral zeta function, regularized determinant and heat trace of the kinetic operator for all homogeneous spherical space forms where the quotient is generated by a subgroup of even order. 
The formulae for the regularized determinant and heat trace organize themselves into ``central'' terms and a ``twisted sum,'' where the central terms correspond to the elements $\{\pm 1\} \subseteq \Gamma$, and the twisted sum corresponds to the rest of the elements of $\Gamma$. 

This work is organized as follows. 
In the following subsection \ref{sec:ResultSummary}, we share the main results of the paper, a general formula for the determinant of the kinetic operator on $\Gamma\bs S^3$ for $\Gamma$ of even order.
In Section \ref{sec:Review} we set up the problem and review necessary background material, such as spectral zeta functions and representation theory of compact Lie groups. In Section \ref{sec:3sphere}, we warm up by clearly outlining our methodology in the context of $S^3$. 
Here we also define and discuss the mass sectors of the scalar field. 
In Section \ref{sec:GammaS3} we use our techniques to work out the multiplicities for general quotients $\Gamma\bs S^3$. Although these multiplicities have already been calculated elsewhere, it is our methodology that we showcase here and that we aim to adapt to more general scenarios in future work. We focus on even lens spaces as a concrete example. In Section \ref{sec:EvenQuotients} we treat the non-abelian finite subgroups of $\SU(2)$ in general. 
In Section \ref{SectionFutureProjects} we discuss generalizations and future directions. Our conventions and definitions are relegated to Appendix \ref{app:Defs}.

\subsection{Main results}

\label{sec:ResultSummary}

Our main result (Theorem \ref{thm:DetGamma}) is the determinant of the scalar kinetic operator on the homogeneous spherical space form $\Gamma \bs S^3$, where $\Gamma$ is an even order subgroup of $\SU(2)$. 
The kinetic operator is defined 
\begin{equation*}
\mc K_{\Gamma\bs S^3}(\mf m) = \Delta_{\Gamma\bs S^3} + \mf m^2 \ell^2\,,
\end{equation*}
where $\mf m$ is the mass of the scalar field, and we multiply by a length scale $\ell$ to make the operator dimensionless, and the Laplace-Beltrami operator $\Delta$ is non-negative and taken for the unit-radius sphere, (cf.~Equation~\eqref{eq:s3kin}).
Define the mass parameter $\nu = \sqrt{1-\mf m^2\ell^2}$ in the principal branch, and continue away from physical values to $\abs{\Re(\nu)}<1$.
We find that the one-loop effective action $-\log Z$ for a massive free scalar field on $\Gamma\bs S^3$ is determined by
\[
\begin{split}
&2\log Z[\Gamma\bs S^3](\mf m) = -\log\det{}'(\mc K_{\Gamma\bs S^3}(\mf m)) \\
&= \frac{\pi i}{3\abs{\Gamma}}\nu^3 -\frac{4}{\pi^2\abs{\Gamma}}\sum_{q=0}^2 \frac{(-\pi i\nu)^q}{q!} \Li_{3-q}\left(-e^{\pi i\nu}\right) 
- \sum_{r=0}^{\frac12\abs{\Gamma}-1}\beta_r\left[
2 \nu \Li_1\left(e^{\frac{2\pi i\nu}{\abs{\Gamma}}+2\pi i\alpha_r}\right)+\frac{i\abs{\Gamma}}{\pi} \Li_2\left(e^{\frac{2\pi i\nu}{\abs{\Gamma}}+2\pi i\alpha_r}\right)\right].
\end{split}
\]
For massless scalars, $\mf m = 0$ or $\nu = 1$, we have
\begin{eqnarray*}
2\log Z[\Gamma\bs S^3](0) & = & -\log\det{}'(\mc K_{\Gamma\bs S^3}(0)) \\
& = & \frac{\pi i}{12}\abs{\Gamma} - \log\pi - \frac{2\pi i}3 + \frac1{\abs{\Gamma}} \left(\pi i + \log4 - \frac{4}{\pi^2}\zeta_R(3)\right) + \left(1- \frac{2}{\abs{\Gamma}}\right)\log\abs{\Gamma}  \\
&   & - \sum_{r=0}^{\frac12\abs{\Gamma}-2}\beta_r\left(
2\Li_1\left(e^{\frac{2\pi i}{\abs{\Gamma}}+2\pi i\alpha_r}\right)+\frac{i\abs{\Gamma}}{\pi}\Li_2\left(e^{\frac{2\pi i}{\abs{\Gamma}}+2\pi i\alpha_r}\right)\right)\,,
\end{eqnarray*}
where the sphere is of unit radius $\ell=1$. The parameters $\alpha_r$ and $\beta_r$ are provided in Table \ref{tab:ResultSummary}, adapted from Table \ref{tab:EvenGammaDims}.
\begin{table}
\centering
\begin{tabular*}{\textwidth}{@{\extracolsep{\fill}}ccccc@{}}
\toprule
$\Gamma$  & $\abs{\Gamma}$ & $\mathscr R_{\Gamma}$ & $\alpha_r = \dfrac{2r+1}{\abs{\Gamma}}$ & $\beta_r = \dfrac12\dim(\V^{\Gamma}_{2r}) - \alpha_r$ \\
\midrule
$\mathbb Z/2p\mathbb Z$ & $2p$ & --                 & $\dfrac{2r+1}{2p}$ & $\dfrac12-\dfrac{2r+1}{2p}$\\[5mm]
$D_m$                   & $4m$ & --                    & $\dfrac{2r+1}{4m}$ & $\dfrac12\left(\floor{\dfrac rm}+\delta_{r\bmod 2,0}\right)-\dfrac{2r+1}{4m}$\\[5mm]
$\widetilde{\mc T}$     & 24   & $\{0,6,8\}$           & $\dfrac{2r+1}{24}$ & $\dfrac12\left(\floor{\dfrac r6}+\varepsilon_{2r\bmod 12}\right)-\dfrac{2r+1}{24}$ \\[5mm]
$\widetilde{\mc O}$     & 48   & $\{0,8,12,16,18,20\}$ & $\dfrac{2r+1}{48}$ & $\dfrac12\left(\floor{\dfrac r{12}}+\varepsilon_{2r\bmod 24}\right)-\dfrac{2r+1}{48}$\\[5mm]
$\widetilde{\mc I}$     & 120  &  Equation \eqref{eq:Ricosahedral} & $\dfrac{2r+1}{120}$ & $\dfrac12\left(\floor{\dfrac r{30}}+\varepsilon_{2r\bmod 60}\right)-\dfrac{2r+1}{120}$\\\bottomrule
\end{tabular*}\medskip
\caption{The parameters $\alpha_r$ and $\beta_r$ for each of the quotients generated by subgroups of even order.
The notation $\varepsilon_n = 1$ if and only if $n \in \mathscr{R}_{\Gamma}$, $0$ otherwise, and $0\leq r \leq \frac12\abs{\Gamma}-1$.}
\label{tab:ResultSummary}
\end{table}

The $r$ independent terms correspond to the action by the ``central'' elements $\{\pm1\}\subseteq \Gamma$, where $\{\pm 1\}$ is the center of $\SO(4)$, the isometry group of $S^3$.
The sum over $r$ is what we call a ``twisted'' sum since these terms correspond to the rest of the non-central elements of $\Gamma$. 
Indeed, the twisted sum vanishes for $\Gamma\cong \mathbb Z/2\mathbb Z$, and only the central term survives.
So, the central term is always proportional to the determinant on $L(2,1)$ by a factor of $\frac2{\abs{\Gamma}}$ for the massive determinant.

\section{Review} 

\label{sec:Review}

\subsection{Regularized determinants and the one-loop quantum effective action}

Let us begin by motivating the definition of the regularized determinant and the spectral zeta function.
Let $P$ be a non-negative elliptic self-adjoint operator with a (degenerate) discrete spectrum,
\begin{equation}
P \phi_n = \lambda_n \phi_n.
\end{equation}
Then, formally, the determinant of the operator is
\begin{equation*}
\det P = \prod\limits_n \lambda_n,
\end{equation*}
but this is usually either divergent or zero if $P$ has a non-trivial kernel. 
To resolve this, one can resort to various regularization techniques such as zeta function regularization (used to construct the Ray-Singer analytic torsion \cite{Ray:1971}, for example).

\begin{definition}

We define the spectral zeta function,
\begin{equation}
\zeta_P(s) = \sum_{\substack{n\\\lambda_n \neq 0}}\lambda_n^{-s} = \tr' P^{-s}
\end{equation}
which converges for sufficiently large $\Re(s)$, for an appropriate asymptotic growth of the eigenvalues.
The prime on the trace denotes that we omit any zero eigenvalues. 

\end{definition}

The spectral zeta function typically admits a meromorphic continuation to the entire complex plane and is regular at $s=0$.
Now, using $\frac{d}{ds}x^{-s} = -x^{-s} \log x$, we perform the following formal manipulations to motivate the definition of the regularized determinant,
\begin{equation}
\log {\det}' P = \log\prod_{n,\,\lambda_n\neq 0} \lambda_n = \sum_{n,\,\lambda_n\neq 0} \log \lambda_n = -\sum_{n,\,\lambda_n\neq 0} \frac{d}{ds}\lambda_n^{-s}\bigg|_{s=0} = -\zeta_P'(s)\bigg|_{s=0} = -\zeta_P'(0).
\end{equation}
The prime on $\det$ here indicates that we drop any zero eigenvalues.

\begin{definition}

Provided that the analytically continued $\zeta_P$ is regular at $s=0$, we define the zeta-function regularized determinant to be
\begin{equation}
{\det}' P := e^{-\zeta_P'(0)}.
\end{equation}
Note that in order to take the derivative at $s=0$, one needs to find an appropriate analytic continuation and ensure that the spectral zeta function is analytic at $s=0$.

\end{definition}

Now, consider a massive free scalar field $\phi$ defined on a closed $n$-dimensional smooth Riemannian manifold $\mc M$, whose classical dynamics is governed by extremizing the (Euclidean) action functional,
\begin{equation*}
S[\phi;\mf m] = \frac12\int_{\mc M} d^n x \sqrt{\det g} (g^{ij}\pd_i \phi \pd_j \phi + \mf m^2 \phi^2) = \frac12\int_{\mc M} d^n x \sqrt{\det g} (\phi (\Delta_{\mc M}+\mf m^2) \phi)
\end{equation*}
where $g_{ij}dx^i\otimes dx^j$ is the metric tensor, $g^{ij} \pd_i \otimes \pd_j$ is the inverse metric, and $\mf{m}$ is the mass. 
Here the Laplace-Beltrami operator is defined to be a non-negative operator, 
\begin{equation*}
\Delta_{\mc M} \phi = -\frac1{\sqrt{\det g}}\pd_i(\sqrt{\det g}\, g^{ij}\pd_j\phi)\,.
\end{equation*}
Also note that we have performed an integration by parts; the boundary term vanishes since we consider closed $\mc M$.

\begin{definition}

We call the resulting mass-shifted Laplace-Beltrami operator the kinetic operator,
\begin{equation}\label{eq:KineticDef}
\mc K_{\mc M}(\mf m) = \ell^2\Delta_{\mc M}+\mf m^2 \ell^2,
\end{equation}
where we have scaled the operator with the appropriate length scale associated with the manifold in order to make the operator dimensionless.

\end{definition}

The Euclidean path integral quantizing the scalar field is formally given by
\begin{equation*}
Z[\mc M] = \int [\mc D\phi] ~ e^{-S[\phi;\mf m]}\,,
\end{equation*}
which is an integral over the (infinite-dimensional) space of field configurations on $\mc M$, where the fields are appropriately normalized. 
Since $S[\phi;\mf m]$ is quadratic in $\phi$, this integral is formally Gaussian, and evaluating it yields
\begin{equation*}
Z[\mc M] = {\det}'\left(\ell^2\Delta_{\mc M} + \mf m^2\ell^2\right)^{-\frac12} = {\det}'\left(\mc K_{\mc M}(\mf m) \right)^{-\frac12},
\end{equation*}
which is the regularized determinant of the Laplace-Beltrami operator shifted by the mass parameter $\mf m^2$. 
Note that the omission of zero eigenmodes may only be necessary for massless fields.
The path integral measure carries an implicit normalization scale $\ell^2$, which gets inserted into the determinant, rendering the argument dimensionless.

The quantity $W[\mc M] = -\log Z[\mc M]$ is known as the one-loop quantum effective action of the scalar field on $\mc M$.
Computing $Z[\mc M]$ using the spectral zeta function of the kinetic operator, and hence $W[\mc M]$, is the central goal of this paper.
We focus on $\mc M \cong \Gamma\bs S^3$ as instructive examples in this paper. 
To do this we will explicitly construct the spectral zeta function of the kinetic operator whose derivative is proportional to the one-loop quantum effective action,
\begin{equation}
\log Z[\Gamma\bs S^3] = -\frac12\log{\det}'(\mc K_{\Gamma\bs S^3}(\mf m)) = \frac12 \zeta_{\mc K_{\Gamma\bs S^3}(\mf m)}'(0).
\end{equation}

To study the geometric information encoded in the spectral zeta function $\zeta_P(s)$, we can study the corresponding heat operator $e^{-tP}$.
The Mellin transform \eqref{eq:MellinDef} of the heat trace is proportional to the spectral zeta function,
\begin{equation}
\frac1{\Gamma(s)}M[\tr'(e^{-t P})](s) = \frac1{\Gamma(s)}\int_0^\infty \tr'(e^{-t P}) t^{s-1} dt =\frac1{\Gamma(s)}\sum_{\substack{n=0\\ \lambda_n\neq 0}}^\infty\int_0^\infty e^{-t\lambda_n} t^{s-1} dt = \zeta_P(s).
\end{equation}
Indeed the omission of the zero mode allows the Mellin integral to converge as $t\to\infty$ and sufficiently large $\Re(s)$ is required for convergence as $t\to 0$.
Expressing the spectral zeta function as the Mellin transform of the heat trace allows us to easily compute the derivative at $s=0$.
Furthermore, using \eqref{eq:GammaExp}, the derivative at $s=0$ is
\begin{equation}
\label{eq:zetaDer}
\begin{split}
\zeta_{P}'(0)
&= M[\tr'(e^{-tP})](0)\left(\pd_s \frac1{\Gamma(s)} \right)\bigg|_{s=0} + \frac{\big(\pd_s M[\tr'(e^{-tP})](s)\big)|_{s=0}}{\Gamma(0)}\\
&= M[\tr'(e^{-tP})](0) = \int_0^\infty \frac{dt}t \tr'(e^{-tP}),
\end{split}
\end{equation}
which recovers the Schwinger parametrization of $-\log\det' P$ \cite{Schwinger:1951nm}, if $M[\tr'(e^{-tP})](s)$ is regular at $s=0$ after analytic continuation. As written, the integral diverges at $t=0$ without some UV regulator.
Indeed if $M[\tr'(e^{-tP})](0)$ is finite, then $\zeta_P(0) = 0$ since $\frac1{\Gamma(0)} = 0$ (see Equation~\eqref{eq:GammaExp}), which is true for elliptic operators on odd dimensional manifolds, unless one needs to subtract zero modes from the trace, whose contribution may depend on an overall scale.

The heat trace of the Laplace-Beltrami operator can be understood as a worldline path integral of a non-relativistic point particle over closed loops in $\mc M$ \cite{Strassler:1992zr}.
The classical saddle points of this path integral are closed geodesics, with quantum fluctuations around each saddle producing the corresponding fluctuation determinant,
\begin{equation} 
\label{eq:WorldlinePI}
\tr(e^{-t\mc K_{\mc M}(\mf m)}) = \int_{x(0)=x(t)} [\mc D x] e^{-\int_0^t dt' \left(\frac1{4\ell^2} g_{\mu\nu} \dot x^\mu \dot x^\nu + \mf m^2\ell^2\right)}. 
\end{equation}
Typically one solves for the heat kernel, and then one can compute the trace by summing over images of the quotient group, which can be difficult in non-trivial quotients or manifolds.
We will show that the heat trace constructed from representation-theoretic data can be resummed into this geometric perspective without requiring explicit construction of the heat kernel. 
This allows us to express the spectral zeta function in terms of geometric data in addition to the defining spectral data in each of the examples we study.

\subsection{Representation theory of compact Lie groups and algebras}

Let $\G$ be a compact connected Lie group, $\mf g$ be the Lie algebra of $\G$, and $\mf g_{\mathbb C}$ its complexification.
The Haar measure $\mu$ on $\G$ is normalized, $\mu(\G)=1$.
\begin{definition}
A representation is a pair $(\pi, \mathscr H)$ where $\mathscr H$ is a complex Hilbert space and 
\begin{equation}
\pi: \G \to \GL(\mathscr{H})
\end{equation}
is a continuous homomorphism from the group $\G$ to bounded invertible operators on the Hilbert space $\mathscr H$.
\end{definition}

\begin{remark}

Let $\left(\pi, \mathscr{H}\right)$ be a representation of $\G$. We say that 
\begin{itemize}
\item a subspace $\V$ of $\mathscr{H}$ is $\G$-invariant if $\pi(g)(\V) \subseteq \V$ for all $g \in \G$\,,
\item the representation $\pi$ is irreducible if the only closed $\G$-invariant subspaces are $\left\{0\right\}$ and $\mathscr{H}$\,.
\end{itemize}
Similarly, we say that two representations $\left(\pi_{1}, \mathscr{H}_{1}\right)$ and $\left(\pi_{2}, \mathscr{H}_{2}\right)$ are equivalent if there exists a bounded invertible linear map $\T: \mathscr{H}_{1} \to \mathscr{H}_{2}$ commuting with the two representations, i.e. 
\begin{equation*}
\T \circ \pi_{1}(g) = \pi_{2}(g) \circ \T\,, \qquad \left(g \in \G\right)\,.
\end{equation*}
This defines a natural equivalence relation on the set of irreducible representations of $\G$\,.
\end{remark}

\begin{notation}

We denote by $\widehat{\G}$ the set of equivalence classes of irreducible unitary representations of $\G$.

\end{notation}

The representation theory of compact groups is particularly well behaved. Indeed, since $\G$ is compact, every finite-dimensional complex representation is unitarizable and completely reducible, i.e. it decomposes into a direct sum of irreducible representations; see \cite[Chapter IV]{Knapp}. Moreover, the unitary dual of $\G$ is made of finite dimensional representations only as a consequence of the Peter-Weyl theorem, which we shall discuss shortly in Theorem \ref{PeterWeylTheorem}.

\begin{definition} 

\label{def:character}

Let $\left(\pi, \V_{\pi}\right) \in \widehat{\G}$. The character of $\pi$, denoted by $\chi_{\pi}$, is the map given by
\begin{equation}
\chi_{\pi}(g) = \tr(\pi(g))\,, \qquad \left(g \in \G\right)\,.
\end{equation}
  
\end{definition}

\begin{remark}

Characters play a fundamental role in the representation theory of compact groups. Indeed, two finite-dimensional complex representations of $\G$ are equivalent if and only if they have the same character.

\end{remark}

Let $\L^2(\G)$ denote the Hilbert space of square integrable complex-valued functions on $\G$ modulo equality almost everywhere, equipped with the Hermitian positive inner product $\left(\cdot, \cdot\right)$ given by
\begin{equation}
\left(f, h\right) = \displaystyle\int_{\G} f(g) \overline{h(g)} d\mu(g)\,, \qquad \qquad \left(f, h \in \L^2(\G)\right)\,.
\end{equation}
We now introduce a natural unitary representation of $\G$ on the Hilbert space $\L^{2}(\G)$. Understanding the decomposition of $\L^{2}(\G)$ into irreducible representations of $\G$ is one of the central problems of non-commutative harmonic analysis.

\begin{definition}

For every $g\in\G$, we define the operator
\begin{equation*}
\L_{r}(g): \L^{2}(\G) \to \L^{2}(\G)
\end{equation*}
by
\begin{equation}
\left(\L_{r}(g)f\right)(x)=f(g^{-1}x)\,, \qquad \left(f\in\L^{2}(\G), x \in \G\right)\,.
\label{eq:LeftRegularAction}
\end{equation}
The resulting unitary representation $\L_{r}$ of $\G$ on $\L^{2}(\G)$ is called the left-regular representation.

\end{definition}
\noindent The left-regular representation is unitary but not irreducible since $\L^2(\G)$ is infinite dimensional; every irreducible representation of a compact group is finite dimensional.

\noindent One can study the $\G\times\G$ action on $\L^2(\G)$, defined as follows,
\begin{equation}
\left(\left(g, h\right) \cdot f\right)(x) = f(g^{-1}xh), \qquad \left(g, h, x \in \G \,, f \in \L^{2}(\G)\right).
\label{DoubleActionG}
\end{equation}
Since the two factors of $\G \times \G$ commute, the left and right regular actions of $\G$ on $\L^{2}(\G)$ commute as well. Therefore, every $\G$-isotypic component of $\L^{2}(\G)$ with respect to the left-regular representation is invariant under the right-regular action of $\G$.

\begin{notation}

For all $\left(\pi, \V_{\pi}\right) \in \widehat{\G}$, we denote by $\left(\pi^{*}, \V^{*}_{\pi}\right)$ the contragredient representation of $\pi$, where $\V^{*}_{\pi}$ is the dual space of $\V_{\pi}$, given by
\begin{equation}
\left(\pi^{*}(g)\alpha\right)(v) = \alpha\left(\pi(g^{-1})v\right)\,, \qquad \left(g \in \G, v \in \V_{\pi}, \alpha \in \V^{*}_{\pi}\right)\,.
\end{equation}
The representation $\left(\pi^{*}, \V^{*}_{\pi}\right)$ is also an irreducible representation of $\G$, and need not be equivalent to $(\pi,\V_\pi)$. Representations which are equivalent to its contragredient are known as self-dual. 

\end{notation}

\noindent We now go back to the left-regular representation of $\G$. The representation being reducible, it is well-known that $\L^{2}(\G)$ can be decomposed as 
\begin{equation}
\L^{2}(\G) = \mathop{\widehat{\bigoplus}}_{\pi \in \widehat{\G}} m_\pi \V_{\pi}\,,
\label{PeterWeylOne}
\end{equation}
with $m_{\pi} := \dim\left(\mathrm{Hom}_{\G}\left(\V_{\pi}, \L^{2}(\G)\right)\right)$. The next theorem gives us an explicit description of the coefficients $m_{\pi}$\,.

\begin{theo}[Peter-Weyl]   

\label{PeterWeylTheorem}

For all $\pi \in \widehat{\G}$, the coefficients $m_{\pi}$ appearing in Equation \eqref{PeterWeylOne} are non-zero. More precisely, we get
\begin{equation*}
m_{\pi} = \dim\left(\V_{\pi}\right)\,.
\end{equation*}

\end{theo}

\begin{remark}

The Peter-Weyl theorem gives the following decomposition of $\L^{2}(\G)$ as a representation of $\G$,
\begin{equation*}
\L^{2}(\G) = \mathop{\widehat{\bigoplus}}_{\pi\in\widehat{\G}} \dim(\V_{\pi})\V_{\pi}\,.
\end{equation*}
Since the left and right regular actions commute, the right action of $\G$ preserves each isotypic component and acts on its multiplicity space. More precisely, this leads to the following decomposition as a representation of $\G\times\G$,
\begin{equation}
\L^{2}(\G) = \mathop{\widehat{\bigoplus}}_{\pi\in\widehat{\G}} \V_{\pi} \otimes \V_{\pi}^{*}\,,
\label{PeterWeylEquation}
\end{equation}
where $\G$ acts on the first factor through $\pi$ and on the second factor through the contragredient representation $\pi^{*}$.

\end{remark}

We now introduce the notion of a representation of the Lie algebra $\mathfrak{g}$ associated with $\G$. The relationship between representations of $\G$ and those of $\mathfrak{g}$ will be discussed later.

\begin{definition}

A representation of a Lie algebra $\mathfrak{g}$ on a finite-dimensional complex vector space $\V$ is the pair $\left(\Pi, \V\right)$, where $\Pi$ is the linear map $\Pi: \mathfrak{g} \to \mathrm{End}(\V)$ satisfying
\begin{equation*}    
\Pi\left(\left[\X, \Y\right]\right) = \left[\Pi(\X), \Pi(\Y)\right]\,, \qquad \left(\X, \Y \in \mathfrak{g}\right)\,.
\end{equation*}

\end{definition}

Differentiating a unitary irreducible representation $\left(\pi, \V_{\pi}\right) \in \widehat{\G}$ yields a representation $\Pi$ of $\mathfrak{g}$ on the same space, with
\begin{equation*}
\Pi(\X) := d\pi(\X) = \lim\limits_{t \to 0} \frac{\pi(e^{t\X}) - \Id}{t}\,, \qquad \left(\X \in \mathfrak{g}\right)\,. 
\end{equation*}
Since $\G$ is connected, $\pi$ is irreducible if and only if $\Pi$ is also irreducible (see \cite[Chapter~0]{Wolf}). However, not every representation of $\mathfrak{g}$ can be lifted to $\G$. 

\medskip

We give an example here that will be useful in the next section. Let $\G = \SU(2)$ be the subgroup of $\GL_{2}(\mathbb{C})$ given by
\begin{equation*}
\SU(2) := \left\{g \in \GL_{2}(\mathbb{C})\,, gg^{*} = \Id_{2}\,, \det(g) = 1\right\}\,.
\end{equation*}
The Lie algebra $\mathfrak{g}$ of $\G$ is the set of traceless skew-hermitian matrices, and one can see that its complexification is isomorphic to $\mathfrak{sl}_{2}(\mathbb{C})$. Using the classification of finite dimensional representations of $\mathfrak{sl}_{2}(\mathbb{C})$ (see \cite[Theorem~1.66]{Knapp}), we obtain the following theorem.

\begin{theo}

For each $n \geq 0$, there exists a unique, up to equivalence, irreducible representation of $\SU(2)$ of dimension $n+1$\,.

\end{theo}

\begin{remark}

We can even give an explicit realization of the irreducible representations of $\SU(2)$. The action of $\SU(2)$ on $\mathbb{C}^{2}$ can be extended to an action of $\SU(2)$ on $\mathbb{C}\left[z_{1}\,, z_{2}\right]$. We have
\begin{equation*}
\mathbb{C}\left[z_{1}\,, z_{2}\right] = \bigoplus\limits_{n = 0}^{\infty} \mathbb{C}^{n}\left[z_{1}\,, z_{2}\right]\,,
\end{equation*}
where $\mathbb{C}^{n}\left[z_{1}\,, z_{2}\right]$ corresponds to the set of homogeneous polynomials of degree $n$. In particular, we have
\begin{equation}
\mathbb{C}^{n}\left[z_{1}\,, z_{2}\right] = \Span\left\{z^{n-k}_{1}z^{k}_{2}\,, 0 \leq k \leq n\right\}\,.
\label{VectorSpaceVn}
\end{equation}
i.e. $\dim\left(\mathbb{C}^{n}\left[z_{1}\,, z_{2}\right]\right) = n+1$. Using that the action of $\SU(2)$ on $\mathbb{C}\left[z_{1}\,, z_{2}\right]$ preserves the degree, it follows that for all $n \in \mathbb{N}$, $\mathbb{C}^{n}\left[z_{1}\,, z_{2}\right]$ is a representation of $\SU(2)$, denoted by $\pi_{n}$. Moreover, for all $0 \leq k \leq n$ and $z \in \mathbb{C}$ such that $\left|z\right| = 1$, we have
\begin{equation*}
\pi_{n}\begin{pmatrix} z & 0 \\ 0 & z^{-1} \end{pmatrix}\left(z^{n-k}_{1}z^{k}_{2}\right) = z^{\left(n - 2k\right)}z^{n-k}_{1}z^{k}_{2}\,.
\end{equation*}
As explained in \cite{Knapp}, the action of $\SU(2)$ on $\mathbb{C}^{n}\left[z_{1}\,, z_{2}\right]$ is irreducible.

\noindent In particular, it follows from Equation \eqref{PeterWeylEquation} that
\begin{equation*}
\L^{2}(\SU(2)) = \widehat{\bigoplus\limits_{n=0}^{\infty}} \V_{n} \otimes \V^{*}_{n}\,,
\end{equation*}
where $\dim(\V_{n})=n+1$. 

\end{remark}

\medskip

Let $\Ad: \G \to \GL(\mathfrak{g})$ be the adjoint action of $\G$ on its Lie algebra $\mf g$ given by
\begin{equation}
\Ad(g)(\X) = g\X g^{-1}\,, \qquad \left(g \in \G, \X \in \mathfrak{g}\right)\,. 
\end{equation}
Since $\G$ is compact, its Lie algebra $\mathfrak{g}$ admits an $\Ad(\G)$-invariant inner product $\B$. 

\begin{definition}

Let $\left\{\X_{1}, \ldots, \X_{n}\right\}$ be an orthonormal basis of $\mathfrak{g}$ with respect to the inner product $\B$, with $n = \dim(\mathfrak{g})$. The corresponding quadratic Casimir element is defined by
\begin{equation}
\Omega_{\G}:=-\sum\limits_{i=1}^{n}\X_{i}^{2} \in \mathscr{U}(\mathfrak{g}_{\mathbb{C}})\,,
\end{equation}
where $\mathscr{U}(\mathfrak{g}_{\mathbb{C}})$ denotes the universal enveloping algebra of the complexified Lie algebra $\mathfrak{g}_{\mathbb{C}}$, and $\X^{2}_{i}=\X_{i}\X_{i}$ is the product in $\mathscr{U}(\mathfrak{g}_{\mathbb{C}})$. The element $\Omega_{\G}$ is independent of the choice of orthonormal basis and belongs to the center of $\mathscr{U}(\mathfrak{g}_{\mathbb{C}})$.

\end{definition}

The left-regular representation $\L_{r}$ extends from $\mathfrak{g}_{\mathbb{C}}$ to an action of $\mathscr{U}(\mathfrak{g}_{\mathbb{C}})$ by differential operators on the smooth vectors $\mathscr{C}^{\infty}(\G)$ of $\L^2(\G)$.

\begin{definition}

The non-negative Laplace-Beltrami operator is given by
\begin{equation}
\Delta_{\G} = \L_{r}(\Omega_{\G}) =  - \sum\limits_{i=1}^{n} \L_{r}(\X^{2}_{i})\,. 
\end{equation}

\end{definition}

Since the quadratic Casimir is a central element of the enveloping algebra, it follows from Schur's Lemma that for all $\left(\pi, \V_{\pi}\right) \in \widehat{\G}$ there exists a unique scalar $c(\pi) \in \mathbb{C}$ such that
\begin{equation}
\L_{r}(\Omega_{\G})\big|_{\V_{\pi}} = c(\pi) \Id_{\V_{\pi}}\,.
\label{CoefficientsCPi}
\end{equation}
The following theorem gives a formula for the coefficients $c(\pi)$.

\begin{theo}

Let $\pi \in \widehat{\G}$ be an irreducible representation of highest weight $\lambda$. Then the quadratic Casimir element acts on $\V_{\pi}$ by the scalar
\begin{equation}
c(\pi) = \B'(\lambda,\lambda+2\rho)\,,
\end{equation}
where $\rho$ denotes the half-sum of the positive roots of $\mathfrak{g}_{\mathbb{C}}$, and $\B'$ is the inner product
on the dual of a Cartan subalgebra induced by $\B$.

\end{theo}

Recall the corresponding kinetic operator is the Laplace-Beltrami operator scaled and shifted by a mass parameter, whose eigenvalues are therefore
\begin{equation}
\mc K_\G(\mf m)\big|_{\V_\pi} = \left(\Delta_\G + \mf m^2\ell^2\Id\right)\big|_{\V_\pi} = (c(\pi) + \mf m^2 \ell^2)\Id_{\V_\pi}.
\end{equation}
The spectral zeta function of the kinetic operator on $\G$ is therefore
\begin{equation}
\zeta_{\mc K_\G} (s;\mf m) = \sum_{\substack{\pi\in\widehat\G \\ c(\pi)+\mf m^2\ell^2\neq 0}}\frac{(\dim\V_\pi)^2}{(c(\pi)+\mf m^2\ell^2)^s}.
\end{equation}

\section{The 3-sphere} \label{sec:3sphere}

Many of the results in this section have been worked out previously largely using other methods, however this section will serve as a pedagogical blueprint for the following sections involving quotients.

The massless case was considered in \cite{Vardi:1988}, with the explicit closed form in \cite{Choi:1994} which we reproduce. 
The spectral zeta function and heat kernel on many spaces including $S^3$ were considered in \cite{Camporesi:1990wm}, which works out the expansion in terms of (incomplete) Riemann zeta functions using binomial expansions, similar to the procedure we use here. 
The large-mass expansion of the scalar one-loop quantum effective action was first worked out in \cite{Denef:2009kn}.
Numerical evaluation of the spectral zeta function and its derivative was worked out in \cite{Dowker:2014xca} by expressing the spectral zeta function in terms of Barnes-$\zeta$ functions.
The one-loop quantum effective action for the massive scalar expressed as a sum over polylogarithms was worked out using group characters and contour integrals in \cite{Anninos:2020hfj}, and by explicit evaluation of the heat kernel and Wilson spool formalism in \cite{Castro:2023dxp}.


We denote by $S^{3}$ the subset of $\mathbb{R}^{4}$ given by
\begin{equation*}
S^{3} := \left\{x \in \mathbb{R}^{4}\,, \left\|x\right\| = 1\right\},
\end{equation*}
where $\left\|\cdot\right\|$ is the Euclidean norm on $\mathbb{R}^{4}$. Another classical realization of the three-dimensional sphere is obtained from the complex vector space $\mathbb{C}^{2}$. More precisely,
\begin{equation}
S^{3} \cong \left\{\left(z_{1}\,, z_{2}\right) \in \mathbb{C}^{2}\,, \left|z_{1}\right|^{2} + \left|z_{2}\right|^{2} = 1\right\}\,.
\label{DefinitionS3Complex}
\end{equation}
We give two different parameterizations of $S^{3}$.
\begin{itemize}
\item For all $g \in \SO(4)$ and $x \in S^{3}$, we have $\left\|gx\right\| = 1$, i.e. $gx \in S^{3}$. In particular, we get an action of $\SO(4)$ on $S^{3}$
\begin{equation}
\Psi: \SO(4) \times S^{3} \ni \left(g\,, x\right) \mapsto gx \in S^{3}\,.
\label{ActionSO4S3}
\end{equation}
This action is transitive, i.e. for all $x\,, y \in S^{3}$, there exists $g \in \SO(4)$ such that $y = gx$. Let $e_{4} = \begin{pmatrix} 0 & 0 & 0 & 1 \end{pmatrix}^{t}$. Then the stabilizer of $e_{4}$ with respect to the action given in Equation \eqref{ActionSO4S3} is
\begin{equation*}
\Stab_{\SO(4)}(e_{4}) := \left\{g \in \SO(4)\,, ge_{4} = e_{4}\right\} = 
\left\{\begin{pmatrix} \A & 0 \\ 0 & 1 \end{pmatrix}\,, \A \in \SO(3)\right\} \cong \SO(3)\,.
\end{equation*}
In particular, using the transitivity of \eqref{ActionSO4S3}, it follows from the orbit-stabilizer theorem that
\begin{equation*}
S^{3} \cong \SO(4) / \SO(3)\,.
\end{equation*}
\item The map
\begin{equation}
\SU(2) \ni \begin{pmatrix} \alpha & \beta \\ -\overline{\beta} & \overline{\alpha} \end{pmatrix} \longmapsto \left(\alpha\,, \beta\right) \in S^{3}
\label{S3asSU2}
\end{equation}
is a diffeomorphism. In particular,
\begin{equation*}
\SU(2) \cong S^{3}\,.
\end{equation*}
\end{itemize}

\begin{remark}

\begin{enumerate}
\item The groups $\SU(2)$ and $\SO(4)$ are closely related. Indeed, let $\Spin(4)$ denote the connected double cover of $\SO(4)$. As explained in \cite{GOODMANWALLACH}, we have
\begin{equation*}
\Spin(4)\cong\SU(2)\times\SU(2)\,.
\end{equation*}
Therefore
\begin{equation*}
\SO(4) \cong \frac{\SU(2)\times\SU(2)}{\left\{(\Id_{2},\Id_{2})\,,(-\Id_{2},-\Id_{2})\right\}}\,.
\end{equation*}
Moreover, using the identification $S^{3} \cong \SU(2)$ given in Equation \eqref{S3asSU2}, the natural action of $\SU(2) \times\SU(2)$ on $S^{3}$ is given by left and right multiplication,
\begin{equation}
\left(g_{1}, g_{2}\right) \cdot x = g_{1} x g_{2}^{-1}\,, \qquad \left(g_{1}, g_{2}, x \in \SU(2)\right)\,.
\label{ActionSU2SU2a}
\end{equation}
The kernel of this action is
\begin{equation*}
\left\{\left(\Id_{2},\Id_{2}\right)\,, \left(-\Id_{2},-\Id_{2}\right)\right\}\,,
\end{equation*}
and hence the action descends to the natural action of $\SO(4)$ on $S^{3}$.
\item Using Equation \eqref{S3asSU2}, it follows from Peter-Weyl that
\begin{equation*}
\L^{2}(S^{3}) \cong \L^{2}(\SU(2)) = \widehat{\bigoplus}_{n=0}^{\infty} \V_{n}\otimes\V_{n}^{*}\,,
\end{equation*}
where $\V_{n}$ denotes the unique (up to isomorphisms) irreducible representation of $\SU(2)$ of dimension $n+1$.
The $\SU(2)\times\SU(2)$ action \eqref{DoubleActionG} on $\L^2(\SU(2))$ used in Peter-Weyl is identical to the action of $\SU(2)$ on $S^3$ Equation~\eqref{ActionSU2SU2a}.
\end{enumerate}

\label{RemarkS3}

\end{remark}

\subsection{The massive spectral zeta function on \texorpdfstring{$S^3$}{S3}}

Let us now construct the spectral zeta function of the kinetic operator on $S^3$ of radius $\ell$,
\begin{equation}
\label{eq:s3kin}
\mc K_{S^3}(\mf m) = \Delta_{S^3}+\mf m^2 \ell^2 \Id
\end{equation}
from the representation theoretic data established above, where $\mf m$ is the mass of the scalar field on $S^3$.
We have used $\ell^2\Delta_{S^3_{\ell}} = \Delta_{S^3}$, where $S^3_{\ell}$ is the 3-sphere of radius $\ell$.

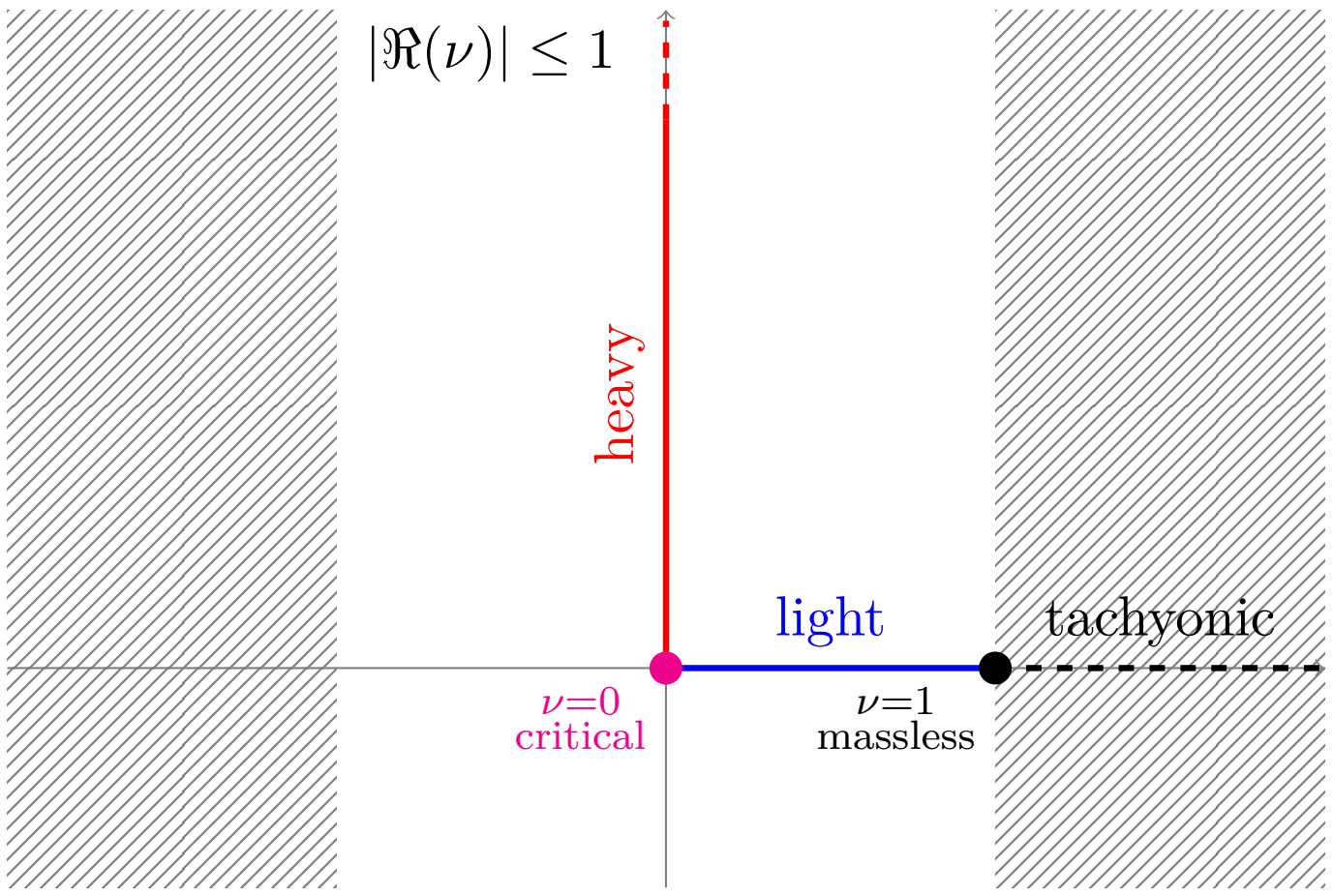
\begin{figure}
\centering
\begin{tikzpicture}[scale=1.5]
\draw[gray,->] (-3,0) -- (3,0);
\draw[gray,->] (0,-1) -- (0,3);
\draw[blue,very thick] (1.5,0) -- (0,0) node [midway, anchor=south] {light};
\draw[red,very thick] (0,0) -- (0,2.5) node [midway, anchor=south, rotate=90] {heavy};
\draw[red,very thick, dashed] (0,2.5) -- (0,2.95);
\fill[pattern={north east lines},pattern color = gray] (1.5,-1) rectangle (3,3);
\fill[pattern={north east lines},pattern color = gray] (-1.5,-1) rectangle (-3,3);
\draw[very thick,dashed] (1.5,0) -- (3,0) node[midway, anchor=south] {\color{black} tachyonic};
\filldraw[black] (1.5,0) circle (2pt) node [anchor = north east] {$\substack{\nu = 1\\\text{massless}}$};
\filldraw[magenta] (0,0) circle (2pt) node [anchor = north east] {$\substack{\nu = 0\\\text{critical}}$};
\node at (-0.8,2.8) {$\abs{\Re(\nu)}\leq 1$};
\end{tikzpicture}
\caption{The light, heavy, and tachyonic sectors in the complex $\nu$-plane for the principal branch of $\nu=\sqrt{1-\mf m^2\ell^2}$. The analysis is restricted to the strip $\abs{\Re(\nu)}\leq 1$; the hatched regions lie outside this strip.}
\label{fig:mass-parameter-diag}
\end{figure}

It will be convenient to define a reparametrized mass parameter $\nu$ and introduce some terminology of the different regimes. 
\begin{definition}\label{def:mass}
The mass parameter $\nu \in \mathbb C$ in terms of the mass of the scalar field is the principal branch of
\begin{equation}
\nu := \sqrt{1 - \mf m^2\ell^2}.
\end{equation}
The four different regimes of the mass parameter which we are interested in are (see figure \ref{fig:mass-parameter-diag}),
\begin{enumerate}
\item \emph{Massless} limit: $\mf m = 0$, $\nu = 1$,
\item \emph{Light} sector: $0<\mf m < \frac1\ell$, $0<\nu<1$,
\item \emph{Critical} point: $\mf m = \frac1\ell$, $\nu=0$,
\item \emph{Heavy} sector: $\mf m > \frac1\ell$, $\nu_H > 0$, where $\nu_H = -i\nu$ and $\nu^2 < 0$.
\end{enumerate}
Our expressions are even and holomorphic in $\nu$ in the open strip $-1 < \Re(\nu) < 1$, analytically continuing away from the physical values of $\nu$.
The massless point at $\nu=1$ lies on the boundary of the strip and needs to be handled separately. 
\end{definition}

\begin{remark}

A conformally coupled massless scalar (cf.~\cite[Equation~(34)]{Klebanov:2011gs}) on $S^3$ is equivalent to a massive scalar with $\mf m^2 \ell^2 = \frac34$ i.e. $\nu=\frac12$, within the light sector. 
    
For completeness we should also define the tachyonic sector, where $\mf m \in i \mathbb R_{>0}$, i.e. $\abs{\Re(\nu)}>1,\,\Im(\nu) = 0$. 
The kinetic operator is no longer non-negative, and requires the removal of a zero mode in the spectral zeta function if $\nu \in \mathbb Z_{>0}$.
Studying the spectrum in this sector is out of scope of this paper, however it will prove useful in future work where we extend our analysis to Lorentzian manifolds.

\end{remark}

The eigenvalues and multiplicities of the kinetic operator are inherited from that of the Laplace-Beltrami operator, where the inner product $\B$ is normalized such that $c(\pi_n) = n(n+2)$,
\begin{equation}
c(\pi_n) + \mf m^2\ell^2 = n(n+2) + \mf m^2\ell^2 = (n+1)^2 -\nu^2.
\end{equation}
The spectral zeta function for the kinetic operator is therefore
\begin{equation} 
\label{eq:s3zeta1}
\zeta_{\mc K_{S^3}} (s;\nu) = \sum_{n=0}^\infty \frac{(n+1)^2}{((n+1)^2-\nu^2)^s} = \sum_{k=1}^\infty \frac{k^2}{(k^2-\nu^2)^s},
\end{equation}
where $k=n+1$, and the sum converges for $\Re(s)>\frac32$.
Note that for the massless case at $\nu=1$, we should start the sum from $n=1$ or $k=2$.

\begin{lemma} 

\label{lem:s3zeta}

Let $\mc K_{S^3}(\mf m) = \Delta_{S^3}+\mf m^2 \ell^2$, with mass parameter $\nu=\sqrt{1-\mf m^2\ell^2}$, with $\abs{\Re(\nu)}\leq 1$ (cf.~\ref{def:mass}). For every $N\in\mathbb Z_{\geq 1}$ such that $|\nu|<N$ then the spectral zeta function of the kinetic operator on $S^3$ admits a meromorphic extension to the complex plane in terms of a sum over Hurwitz zeta function \eqref{eq:HurwitzDef}, 
\begin{equation} 
\label{eq:s3zetaH}
\zeta_{\mc K_{S^3}}(s;\nu)=\sum_{k=1}^{N-1}\frac{k^2}{(k^2-\nu^2)^s}+\sum_{j=0}^\infty\frac{(s)_j}{j!}\,\nu^{2j}\,\zeta_H(2s+2j-2,\,N)\,,
\end{equation}
with at most simple poles at $s=\frac32-j$ with $j\in\mathbb Z_{\geq 0}$, and $s=0$ is regular.
Note that the pole structure is independent of $N$. 

\noindent For $\nu=1$, note that the sum must start at $k=2$, and $N\in\mathbb Z_{\geq 2}$.

\end{lemma}

This lemma extends Camporesi's method \cite{Camporesi:1990wm} of using binomial expansions to express the spectral zeta function of Laplacians on $S^{n}$ as a sum over incomplete Riemann zeta functions. 
We adapt this method for the spectral zeta functions for kinetic operators on $\Gamma\bs S^3$, where this is the trivial case $\Gamma \cong \Id$. 
\begin{proof}

The Dirichlet series defining the massive spectral zeta function on $S^3$ converges for $\Re(s)>\frac32$, so all our manipulations proceeding will be in this range unless stated otherwise.
Furthermore we work in the non-tachyonic sector $\abs{\Re(\nu)}< 1$ where the kinetic operator has no negative or zero eigenvalues, so no modes must be omitted, except in the case of $\nu=1$, where all sums must start from $k=2$ instead to omit the zero mode.
    
First we choose $N\in\mathbb Z_{>0}$ such that $N>|\nu|$, and split the infinite sum,
\begin{equation*}
\zeta_{\mc K_{S^3}}(s;\nu) = \sum_{k=1}^{N-1}\frac{k^2}{(k^2-\nu^2)^s} + \sum_{k=N}^\infty\frac{k^2}{(k^2-\nu^2)^s}.
\end{equation*}
We use the generalized binomial expansion for $k\geq N$, 
\begin{equation*}
(k^2-\nu^2)^{-s} = k^{-2s}\left(1-\frac{\nu^2}{k^2}\right)^{-s} = \sum_{j=0}^\infty \frac{(s)_j}{j!} \nu^{2j} k^{-2s-2j},
\end{equation*}
where $(s)_j$ is the Pochhammer symbol \eqref{def:Pochhammer}.
Rewriting the infinite sum using the binomial sum, we have
\begin{equation*}
\zeta_{\mc K_{S^3}}(s;\nu) = \sum_{k=1}^{N-1}\frac{k^2}{(k^2-\nu^2)^s}+\sum_{k=N}^\infty k^{2-2s}\sum_{j=0}^\infty \frac{(s)_j}{j!}\nu^{2j} k^{-2j}.
\end{equation*}
To swap the sum over $k$ and $j$, note that 
\begin{equation*}
\frac{(s)_j}{j!} = \frac{\Gamma(s+j)}{\Gamma(s)\Gamma(j+1)},
\end{equation*}
then by Stirling's formula, on every compact subset of $\mathbb C$, there exists constants $C,M>0$ such that
\begin{equation*}
\left|\frac{(s)_j}{j!}\right| \leq C (1+j)^M. 
\end{equation*}
Therefore, for $k\geq N$ we have,
\begin{equation*}
\sum_{j=0}^\infty \left|\frac{(s)_j}{j!}\right| \left(\frac{|\nu|}{k}\right)^{2j} \leq \sum_{j=0}^\infty C (1+j)^M \left(\frac{|\nu|}{N}\right)^{2j}
\end{equation*}
which is a $k$ independent bound, ensuring the sum converges since $|\nu|/N <1$. 
Thus we can interchange the sums for $\Re(s)>\frac32$, to express the sum over $k$ as a Hurwitz zeta function, defined in \eqref{eq:HurwitzDef},
\begin{eqnarray*}
\zeta_{\mc K_{S^3}}(s;\nu)  & = & \sum_{k=1}^{N-1}\frac{k^2}{(k^2-\nu^2)^s} + \sum_{j=0}^\infty \sum_{k=N}^{\infty} \frac{(s)_j}{j!}\nu^{2j} k^{-2s-2j+2} \\
& = & \sum_{k=1}^{N-1}\frac{k^2}{(k^2-\nu^2)^s} + \sum_{j=0}^\infty \frac{(s)_j}{j!}\nu^{2j}\zeta_H(2s+2j-2,N) 
\end{eqnarray*}
The first term is finite and therefore entire in $s$ since $\Re(k^2-\nu^2)>0$ (in the massless limit $\nu=1$ we exclude $k=1$, so the inequality holds).
To show that the sum over Hurwitz zetas converges, note that for sufficiently large $j$, 
\begin{equation*}
|\zeta_H(2s+2j-2,N)| = \left|\sum_{k=N}^\infty k^{-2s-2j+2} \right| \leq C_1 N^{-2j},
\end{equation*}
for some constant $C_1$.  Therefore the sum over $j$ converges normally on compact subsets of $\mathbb C$ away from the poles of $\zeta_H(2s+2j-2,N)$.  Therefore 
\begin{equation*}
\sum_{k=1}^{N-1}\frac{k^2}{(k^2-\nu^2)^s} + \sum_{j=0}^\infty \frac{(s)_j}{j!}\nu^{2j}\zeta_H(2s+2j-2,N)
\end{equation*}
defines a meromorphic function on $\mathbb C$ with possible poles at $s=\frac32-j$, with $j\in\mathbb Z_{\geq 0}$.
Since this agrees with $\zeta_{\mc K_{S^3}}(s;\nu)$ for $\Re(s)>\frac32$, the identity theorem implies that it is indeed the meromorphic continuation of $\zeta_{\mc K_{S^3}}(s;\nu)$. 

\end{proof}

\begin{corollary} 
\label{cor:s3zeta1}
    
For the massless sector, $\mf m  = 0$, $\nu = 1$, the $n=0$ term (zero mode) in the sum must be omitted. Choosing $N=2$ we obtain
\begin{equation}        
\zeta_{\mc K_{S^3}}(s;1) = \sum_{j=0}^\infty \frac{(s)_j}{j!} (\zeta_R(2s+2j-2)-1).
\end{equation}
In the light sector $0<\mf m<\frac1\ell$ so $0<\nu<1$, hence we can choose $N=1$.
The spectral zeta function reduces to 
\begin{equation}
\zeta_{\mc K_{S^3}}(s;\nu) = \sum_{j=0}^\infty \frac{(s)_j}{j!} \nu^{2j}\zeta_R(2s+2j-2).
\end{equation}
At the critical point $\mf m = \frac1\ell$, or $\nu=0$ the sum collapses to just the first term, so the spectral zeta function is
\begin{equation}
\zeta_{\mc K_{S^3}} (s;0) = \zeta_R(2s-2).
\end{equation}  
In the heavy sector, we reparametrize $\nu_H^2 = -\nu^2 = \mf m^2\ell^2 - 1$, and choose $N>\nu_H$.
The spectral zeta function now reads
\begin{equation}
\label{eq:zetaS3heavy}
\zeta_{\mc K_{S^3}}(s;i\nu_H)= \sum_{k=1}^{N-1}\frac{k^2}{(k^2+\nu_H^2)^s}+ \sum_{j=0}^\infty \frac{(s)_j}{j!}(-1)^j\nu_H^{2j} \,\zeta_H(2s+2j-2,N).
\end{equation}

\end{corollary}

\begin{corollary}
\label{cor:s3det1}

The regularized determinant of the kinetic operator on $S^3$ for $\abs{\Re(\nu)}<1$,
\begin{equation}
\label{eq:S3det}
\begin{split}
-\log{\det}'(\mc K_{S^3}(\mf m)) 
&= \pd_s\zeta_{\mc K_{S^3}}(s;\nu)\big|_{s=0}\\
&= \frac{i\pi}3 \nu^3 -\frac{1}{2\pi^2}\Li_3\left(e^{2\pi i\nu}\right) +\frac{i\nu}{\pi}\Li_2\left(e^{2\pi i \nu}\right)+\nu^2\Li_1\left(e^{2\pi i \nu}\right)\\
&= \frac{i\pi}3 \nu^3 -\frac1{2\pi^2}\sum_{q=0}^2 \frac{(-2\pi i \nu)^q}{q!} \Li_{3-q}(e^{2\pi i\nu}) . 
\end{split}
\end{equation}
In the massless limit, the regularized determinant of the kinetic operator on $S^3$ is,
\begin{equation}
-\log{\det}'(\mc K_{S^3}(0)) = \zeta_{\mc K_{S^3}}'(0;1) = -\frac{1}{2\pi^2}\zeta_R(3)  -\log\pi.
\end{equation}
The regularized determinant at the critical point $\nu=0$ is
\begin{equation}
-\log{\det}'(\mc K_{S^3}(\tfrac1\ell)) = \zeta_{\mc K_{S^3}}'(0;0) = -\frac{1}{2\pi^2}\zeta_R(3). 
\end{equation}

\end{corollary}

The massless expression was first derived in \cite{Vardi:1988}, and was expressed in the presented form in \cite{Choi:1994}.
Evaluating \eqref{eq:S3det} at $\nu=\frac12$ for the conformally coupled scalar reproduces the free energy result of \cite{Klebanov:2011gs}\footnote{The relative factor of $-\frac12$ between equation (41) in \cite{Klebanov:2011gs} and our \eqref{eq:S3det} evaluated at $\nu=\frac12$ arises since free energy $F = -\log Z = -\frac12\zeta_{\mc K}'(0)$.}. 
The light sector result was motivated in \cite{Castro:2023dxp}, where the heavy sector version is computed using the heat kernel and Wilson spool, and then analytic continuation is proposed to the light sector, which we rigorously obtain here. 
The expression for the regularized determinant in the heavy sector was first derived in \cite{Anninos:2020hfj} using contour integrals over group characters, and by computing the heat kernel and Wilson spool in \cite{Castro:2023dxp}.
Our Theorem \ref{thm:DetGamma} generalizes this result to all homogeneous spherical quotients by subgroups of even order.

\begin{proof}

Let us start with deriving the massless determinant.  Recall the spectral zeta function from Corollary \ref{cor:s3zeta1},
\begin{equation*}
\zeta_{\mc K_{S^{3}}}(s;1) = \sum_{j=0}^\infty\frac{(s)_j}{j!}\left(\zeta_R (2s+2j-2)-1\right),
\end{equation*}
which is regular at $s=0$. To take the derivative at $s=0$, first separate the $j=0$ term,
\begin{equation*}
\zeta_{\mc K_{S^{3}}}(s;1) = \zeta_R(2s-2)-1 + \sum_{j=1}^\infty \frac{(s)_j}{j!}\left(\zeta_R(2s+2j-2)-1\right).
\end{equation*}
Using $\frac{d}{ds}\left(\frac{(s)_j}{j!}\right)\bigg|_{s=0} = \frac1j$, $(0)_j = 0$ for $j\geq 1,$ $j\in \mathbb{N}$, and $\zeta_R(0) = -\frac12$, we have
\begin{equation*}
\zeta'_{\mc K_{S^3}}(0;1)  = 2\zeta'_{R}(-2) + \sum_{j=1}^\infty \frac{\zeta_R(2j-2)-1}{j} = 2\zeta'_{R}(-2) -\frac32 + \sum_{l=1}^\infty \frac{\zeta_R(2l)-1}{l+1}.
\end{equation*}
We use the result (see \cite[Equation~545, Section~3.4]{Choi:2012}) $\sum_{l=1}^\infty \frac{\zeta_R(2l)-1}{l+1} = \frac32-\log\pi$, which can be derived using the Taylor expansion of $\log(1-x)$, Stirling's approximation and telescoping partial sums. 
To evaluate $\zeta_R'(-2)$, one can differentiate the functional equation satisfied by the Riemann zeta function \eqref{eq:RiemannZetaFunct} evaluated at $s=-2$ to obtain
\begin{equation*}
\zeta_R'(-2) = -\frac{1}{4\pi^2}\zeta_R(3),
\end{equation*}
where $\zeta_R(3)$ is commonly called Ap\'ery's constant.
Therefore, we arrive at the desired expression for the massless case, 
\begin{equation*}
\zeta'_{\mc K_{S^3}}(0;1) = -\frac1{2\pi^2}\zeta_R(3)-\log\pi.
\end{equation*}

The regularized determinant at the critical point is immediate with the identity $\zeta_R'(-2) = -\frac1{4\pi^2}\zeta_R(3)$.
Recall $\zeta_{\mc K_{S^3}}(s;0) = \zeta_R(2s-2)$ from Corollary \ref{cor:s3zeta1}, so
\begin{equation*}
\zeta_{\mc K_{S^3}}'(0;0) = 2\zeta_R'(-2) = -\frac1{2\pi^2}\zeta_R(3). 
\end{equation*}
To derive the light sector result, we start the same way as in the massless case, except instead of using Stirling's approximation and telescoping partial sums, we shall utilize a Fourier expansion of the sum over Riemann zeta functions.
    
Recall the spectral zeta function from Corollary \ref{cor:s3zeta1},
\begin{equation*}
\zeta_{\mc K_{S^3}}(s;\nu) = \sum_{j=0}^\infty \frac{(s)_j}{j!} \nu^{2j}\zeta_R(2s+2j-2)\,, \qquad \left(0 < \nu < 1\right)\,.
\end{equation*}
First we separate the $j=0$ term,
\begin{equation*}
\zeta_{\mc K_{S^3}}(s;\nu) = \zeta_R(2s-2) + \sum_{j=1}^\infty\frac{(s)_j}{j!}\nu^{2j}\zeta_R(2s+2j-2). 
\end{equation*}
Again, using $\frac{d}{ds}\left(\frac{(s)_j}{j!}\right)\bigg|_{s=0} = \frac1j$, $(0)_j = 0$ for $j\geq 1,$ $j\in \mathbb{N}$, and $\zeta_R(0) = -\frac12$, we have,
\begin{eqnarray*}
\zeta_{\mc K_{S^3}}'(s;\nu)|_{s=0}  & = & 2\zeta_R'(-2) + \sum_{j=1}^\infty \frac{\nu^{2j}}j \zeta_R(2j-2) = 2\zeta_R'(-2) - \frac{\nu^2}{2} + \sum_{l=1}^\infty \zeta_R(2l)\int_0^\nu 2t^{2l+1} dt\\   & = & 2\zeta_R'(-2) - \frac{\nu^2}{2} + 2\int_0^\nu t \sum_{l=1}^\infty \zeta_R(2l)t^{2l} dt. 
\end{eqnarray*}
The sum and integral swap is justified once again by the summands being non-negative for all $t\in[0,\nu]$. 
    
From the Euler product sine formula $\sin(\pi z) = \pi z \prod\limits_{k=1}^\infty \left(1-\frac{z^2}{k^2}\right)$, one can show
\begin{equation*}
2\sum_{k=1}^\infty z^{2k}\zeta_R(2k) = 1-\pi z\,\cot\pi z.
\end{equation*}
Using the above identity and $\zeta_R'(-2) = -\frac1{4\pi^2}\zeta_R(3)$ again,  we have,
\begin{eqnarray*}
\zeta_{\mc K_{S^3}}'(s;\nu)|_{s=0}  & = & 2\zeta_R'(-2) - \frac{\nu^2}{2} + 2\int_0^\nu t \sum_{l=1}^\infty \zeta_R(2l)t^{2l} dt \\
& = & -\frac1{2\pi^2}\zeta_R(3) - \frac{\nu^2}{2} + \frac{\nu^2}{2} -\pi\int_0^\nu t^2\,\cot\,\pi t\,dt.
\end{eqnarray*}
To evaluate this integral note the Fourier expansion of cotangent,
\begin{equation*}
\cot\pi z = i\frac{e^{i\pi z}+e^{-i\pi z}}{e^{i\pi z}-e^{-i\pi z}} = -i\left(1+\frac{2e^{2\pi iz}}{1-e^{2\pi iz}}\right) = -i\left(1+2\sum_{n=1}^\infty e^{2\pi inz}\right), \quad \Im(z)>0,
\end{equation*}
where $z$ needs to have some imaginary component for the expansion to converge. 
To achieve this, we shift the integration contour of $t$ by $i\epsilon$, $\epsilon\in\mathbb R_+$, which is justified since shifting the contour did not cross any poles of the integrand. 
This shift also allows the following swap between sum and integral.
\begin{eqnarray*} 
\zeta_{\mc K_{S^3}}'(s;\nu)|_{s=0} & = & -\frac1{2\pi^2}\zeta_R(3) +i\pi\lim_{\epsilon\to 0^+}\int_{0+i\epsilon}^{\nu+i\epsilon} t^2 \left(1+2\sum_{n=1}^\infty e^{2\pi int}\right)dt \\
& = & -\frac1{2\pi^2}\zeta_R(3) +\frac{i\pi}3 \nu^3 + \sum_{n=1}^\infty \left(-\frac{e^{2\pi i\nu n}}{2\pi^2 n^3}+\frac{1}{2\pi^2 n^3}+\frac{i \nu  e^{2\pi i\nu n}}{\pi n^2}+\frac{\nu^2 e^{2\pi i\nu  n}}{n}\right)\\
& = & \frac{\pi i}3 \nu^3 -\frac{1}{2\pi^2}\Li_3\left(e^{2\pi i\nu}\right) +\frac{i\nu}{\pi}\Li_2\left(e^{2\pi i \nu}\right)+\nu^2\Li_1\left(e^{2\pi i \nu}\right),
\end{eqnarray*}
where the $\zeta_R(3)$ term is canceled by the $\nu$ independent sum over $n^{-3}$. We have also introduced the polylogarithm \eqref{eq:PolylogDef}.

For the heavy sector, using $\nu_H=-i\nu>0$, and an integer $N>\nu_H$, we differentiate the meromorphic continuation \eqref{eq:zetaS3heavy} at $s=0$ where it is regular.
\begin{equation*}
\pd_s\zeta_{\mc K_{S^3}}(s;i\nu_H)\big|_{s=0}  = -\sum_{k=1}^{N-1}k^2 \log(k^2+\nu_H^2) + 2\zeta_H'(-2,N)+\sum_{j=1}^\infty \frac{(-1)^j\nu_H^{2j}}{j}\zeta_H(2j-2,N).
\end{equation*}
We now show that the dependence on $N$ disappears. 
For the $j=1$ term we use $\zeta_H(0,N) = \frac12-N = -\frac12-\sum_{k=1}^{N-1}1$.
For the $j\geq2$ terms, the series is absolutely convergent, so we may swap the two sums. Thus
\begin{equation*}
\sum_{j=2}^\infty \frac{(-1)^j\nu_H^{2j}}{j}\zeta_H(2j-2,N) = \sum_{k=N}^\infty k^2 \sum_{j=2}^\infty \frac{(-1)^j}{j}\left(\frac{\nu_H^2}{k^2}\right)^j = \sum_{k=N}^\infty \left[\nu_H^2-k^2\log\left(1+\frac{\nu_H^2}{k^2}\right)\right].
\end{equation*}
To tackle the $j=0$ term, we use
\begin{equation*}
\zeta_H(s,N) = \zeta_R(s) - \sum_{k=1}^{N-1}k^{-s},\implies \zeta_H'(-2,N) = \zeta_R'(-2) + \sum_{k=1}^{N-1}k^2\log(k). 
\end{equation*}
Combining all of the above identities, the finite sum combines with the sum beginning at $k=N$, and we obtain
\begin{equation*}
\pd_s\zeta_{\mc K_{S^3}}(s;i\nu_H)\big|_{s=0} = 2\zeta_R'(-2) + \frac{\nu_H^2}2 + \sum_{k=1}^\infty \left[\nu_H^2-k^2\log\left(1+\frac{\nu_H^2}{k^2}\right)\right],
\end{equation*}
which proves the derivative is explicitly independent of the auxiliary integer $N$. Note that 
\begin{equation*}
\nu_H^2-k^2\log\left(1+\frac{\nu_H^2}{k^2}\right) = \frac{\nu_H^4}{2k^2}+ O(k^{-4}),
\end{equation*}
so that the series is absolutely convergent. 

To evaluate the sum, we will use
\begin{equation*}
\sum_{k=1}^\infty \frac1{k^2+z^2} = \frac{\pi}{2z}\coth(\pi z)-\frac1{2z^2},\quad \coth(\pi z) = 1+\frac{2}{e^{2\pi z}-1}= 1+2\sum_{w=1}^\infty e^{-2\pi w z}, \qquad \left(\Re(z) > 0\right).
\end{equation*}
Note that $\pd_s\zeta_{\mc K_{S^3}}(s;0)\big|_{s=0} = 2\zeta_R'(-2) = -\frac1{2\pi^2}\zeta_R(3)$ and 
\begin{equation*}
\pd_{\nu_H}\pd_s \zeta_{\mc K_{S^3}}(s;i\nu_H)\big|_{s=0} = \nu_H+2\nu_H^3\sum_{k=1}^\infty \frac1{k^2+\nu_H^2} = \pi \nu_H^2 \coth(\pi \nu_H).
\end{equation*}
Integrating the above expression using the Fourier expansion of $\coth$, direct computation yields,
\begin{eqnarray*}
\pd_s\zeta_{\mc K_{S^3}}(s;i\nu_H)\big|_{s=0} & = & -\frac{1}{2\pi^2}\zeta_R(3) + \pi \int_0^{\nu_H} t^2 \coth(\pi t) dt \\
& = & \frac\pi3 \nu_H^3 -\nu_H^2\Li_1(e^{-2\pi\nu_H})-\frac{\nu_H}{\pi}\Li_2(e^{-2\pi\nu_H}) - \frac1{2\pi^2}\Li_3(e^{-2\pi\nu_H}) \\
& = & \frac{i\pi}3 \nu^3 -\frac{1}{2\pi^2}\Li_3\left(e^{2\pi i\nu}\right) +\frac{i\nu}{\pi}\Li_2\left(e^{2\pi i \nu}\right)+\nu^2\Li_1\left(e^{2\pi i \nu}\right),
\end{eqnarray*}
where we have used $\nu_H = -i\nu$  in the final step to show that the expression is identical to the light sector, thus completing the proof.  
\end{proof}

\begin{remark}

One can recover the massless result from the light sector result by taking the limit $\nu\to1^-$ and removing the resulting logarithmic divergence which arises due to the presence of the zero mode. 
Using $\Li_s(1) = \zeta_R(s),~s\neq 1$, we have
\begin{equation*}
\lim_{\nu\to 1^-} \zeta_{\mc K_{S^3}}'(s;\nu)|_{s=0}  = 
\frac{i\pi}3 -\frac{1}{2\pi^2}\zeta_R(3) +\frac{i}{\pi}\zeta_R(2)-\lim_{\nu\to1^-}\log(1-e^{2\pi i\nu}),
\end{equation*}
where we have a logarithmic divergence in the last term. 
To remove this divergence, we need to see what the zero mode contributes to the final answer. 
Recall the definition of the spectral zeta function \eqref{eq:s3zeta1}, we have
\[\begin{split}
\zeta_{\mc K_{S^3}}(s;\nu) &= \sum_{k=1}^\infty \frac{k^2}{(k^2-\nu^2)^s} = (1-\nu^2)^{-s} + \sum_{k=2}^\infty \frac{k^2}{(k^2-\nu^2)^s}, \\
\implies \pd_s \zeta_{\mc K_{S^3}}(s;\nu)|_{s=0} &= -\log(1-\nu^2) +\pd_s\left(\sum_{k=2}^\infty \frac{k^2}{(k^2-\nu^2)^s}\right)\bigg|_{s=0}.
\end{split}
\]
So the correct $\nu\to1^-$ limit requires removal of $-\log(1-\nu^2)$ by hand from the derivative of the spectral zeta function in the light sector. 
\begin{eqnarray*}
\lim_{\nu\to1^-} \zeta_{\mc K_{S^3}}'(s;\nu)|_{s=0}  & = & \frac{i\pi}3 -\frac{1}{2\pi^2}\zeta_R(3) +\frac{i}{\pi}\zeta_R(2)+\lim_{\nu\to1}(-\log(1-e^{2\pi i\nu})+\log(1-\nu^2)) \\
& = & \frac{i\pi}3 -\frac{1}{2\pi^2}\zeta_R(3) +\frac{i\pi}{6} +\lim_{\nu\to1}\log\frac{1-\nu^2}{1-e^{2\pi i\nu}} = \frac{i\pi}2 -\frac{1}{2\pi^2}\zeta_R(3)  -\log i\pi\\
& = & -\frac{1}{2\pi^2}\zeta_R(3)  -\log\pi\,,
\end{eqnarray*}
which recovers the massless result in the principal branch, and provides a consistency check on the light sector result as all the imaginary phases coming from each polylogarithm correctly cancel out.
    
The critical point result can also be obtained by taking the $\nu\to 0^+$ limit of the light sector result.
The logarithmic divergence of $\Li_1$ at $\nu\to 0^+$ is canceled by the prefactor $\nu^2$, rendering that term zero. 
The rest of the non-zero terms then yield the critical result.

\end{remark}

\subsection{The heat trace on \texorpdfstring{$S^3$}{S3}}

Let us define the heat operator and its trace from the representation-theoretic spectral data obtained above.

\begin{definition}

The heat operator associated with the kinetic operator $\mc K_{S^3}$ is the exponential $e^{-t \mc K_{S^3}(\mf m)}$, where $t>0$.
The corresponding heat trace is the trace of the heat operator over $\L^2(S^3)$,
\begin{equation}
H_{\mc K_{S^3}}(t;\nu) := \tr'(e^{-t\mc K_{S^3}(\mf m)})\,, \qquad \left(t > 0\right)\,,
\end{equation}
where we exclude any zero modes from the sum as indicated by the prime on the trace. 

\end{definition}

\begin{lemma} 
\label{lem:s3heat}

The heat trace associated with the massive kinetic operator on $S^3$ can be expressed in terms of (a) spectral data, (b) the classical Jacobi theta function $\vartheta_3$ (see \ref{def:JacobiThetaDef}), and (c) a resummed expression highlighting small $t$ asymptotics,
\begin{subequations}
\begin{align}
H_{\mc K_{S^3}}(t;\nu) 
&= \sum_{k=1}^\infty k^2 e^{-t(k^2-\nu^2)} - \delta_{\nu^2,1}\\
&= -\frac12 e^{t\nu^2}\pd_t\vartheta_3\left(\frac{it}{\pi}\right) - \delta_{\nu^2,1} \\
&= \frac{\sqrt \pi}4 t^{-\frac32} e^{t\nu^2} -\delta_{\nu^2,1} + \frac{\sqrt \pi}2  t^{-\frac32} \sum_{w=1}^\infty \left(1-\frac{2\pi^2 w^2}{t}\right)e^{t\nu^2-\frac{w^2\pi^2}t}.
\end{align}
\end{subequations}

\end{lemma}

Our expressions in the massless limit agree with \cite[Proposition~9.4]{Avramidi:2023} which computes the massless heat kernel and integrates it over the sphere coincident limit to find the trace.

\begin{proof}

Recall that the eigenvalues of the kinetic operator are $(n+1)^2-\nu^2$, where $n\in\mathbb Z_{\geq 0}$, with corresponding multiplicity $(n+1)^2$. 
Therefore the heat trace is, for $t>0$,
\begin{equation*}
H_{\mc K_{S^3}}(t;\nu) = \tr'(e^{-t\mc K_{S^3}(\mf m)}) = \sum_{n=0}^\infty (n+1)^2 e^{-t((n+1)^2-\nu^2)} = \sum_{k=1}^\infty k^2 e^{-t(k^2-\nu^2)}\,.
\end{equation*}
 To prove that the sum converges, note that $k^2-\nu^2 > 0$, so each term is exponentially suppressed and the sum converges for all $t>0$.
The massless case requires us to remove the zero mode $n=0$ or $k=1$. 
The contribution of the zero mode to the heat trace in the massless limit is simply $1$, so we subtract it when $\nu = 1$ by introducing a Kronecker delta,
\begin{equation*}
H_{\mc K_{S^3}}(t;\nu) = \sum_{k=1}^\infty k^2 e^{-t(k^2-\nu^2)} - \delta_{\nu^2,1},
\end{equation*}
which recovers the first form of the heat trace in the lemma.
    
To express this as a derivative of the Jacobi theta function, from the definition (\ref{def:JacobiThetaDef}),
\begin{equation*}
\vartheta_3(\tau)  = \sum_{n\in\mathbb Z} e^{\pi i n^2 \tau} = 1+2\sum_{n=1}^\infty e^{\pi i n^2 \tau}\,, \qquad \pd_\tau \vartheta_3(\tau)  = 2\pi i \sum_{n=1}^\infty n^2 e^{\pi i n^2 \tau},
\end{equation*}
where we are allowed to take term-wise derivatives since the series converges normally in compact subsets of the upper half plane. 
So for $\tau = \frac{it}\pi$ we have $\pd_\tau = -\pi i\,\pd_t$ and
\begin{equation*}
H_{\mc K_{S^3}}(t;\nu) = -\frac12 e^{t\nu^2}\pd_t \vartheta_3\left(\frac{it}{\pi}\right) - \delta_{\nu^2,1}\,,
\end{equation*}
which recovers the second form of the heat trace in the lemma. 

To obtain the third form of the heat trace, one can use the modular $\mc S$ transformation on the Jacobi theta function, $\vartheta_3(-1/\tau) = \sqrt{-i\tau}\vartheta_3(\tau)$, (see \ref{def:JacobiThetaDef}) which can be shown via Poisson summation. 
We have
\begin{eqnarray*}
H_{\mc K_{S^3}}(t;\nu) & = & -\frac12 e^{t\nu^2}\pd_t \left(\sqrt{\frac\pi t} \vartheta_3\left(\frac{\pi i}{t}\right)\right) -\delta_{\nu^2,1} \\
& = & -\frac12 e^{t\nu^2}\left(-\frac{\sqrt \pi}2 t^{-\frac32}\vartheta_3\left(\frac{\pi i}{t}\right)+\sqrt{\frac\pi t}\,\pd_t\vartheta_3\left(\frac{\pi i}{t}\right)\right) -\delta_{\nu^2,1} \\
& = &\frac{\sqrt \pi}4 e^{t\nu^2}\sum_{w\in\mathbb Z} \left(1-\frac{2\pi^2 w^2}{t}\right)t^{-\frac32} e^{-\frac{\pi^2 w^2}t} -\delta_{\nu^2,1} \\
& = & \frac{\sqrt \pi}4 t^{-\frac32} e^{t\nu^2} -\delta_{\nu^2,1} + \frac{\sqrt \pi}2  t^{-\frac32} \sum_{w=1}^\infty \left(1-\frac{2\pi^2 w^2}{t}\right)e^{t\nu^2-\frac{\pi^2 w^2}t},
\end{eqnarray*}
which recovers the third form of the heat trace and completes the proof. 

\end{proof}

\begin{remark}
    
The resummed expression of the heat trace contains a lot of information when understood geometrically. 
Recall from section \ref{sec:Review} that the heat trace can be understood as a worldline path integral \eqref{eq:WorldlinePI}. 
Perturbatively expanding the worldline path integral around the classical saddle corresponding to the zero-length geodesic, we recover the so-called integrated Seeley-DeWitt coefficients \cite{DeWitt:1964mxt,Seeley:1967ea,Gilkey:1975iq}. 

The integrated Seeley-DeWitt coefficients are the coefficients of the expansion in the $t\to 0^+$ limit of the heat trace of a Laplace-type operator on a $d$-dimensional manifold without boundary,
\begin{equation*}
H_{\mc K_{\mc M}}(t;\nu) \sim \frac1{(4\pi t)^{d/2}} \sum_{r=0}^\infty a_r(\mc K_{\mc M}) t^r\,,
\end{equation*}
where the coefficients are $a_r(\mc K_{\mc M})$. 
They encode the curvature invariants of the manifold, the fields and their covariant derivatives, integrated over $\mc M$.

In our case, the $w\neq 0$ terms are suppressed in this limit due to the $e^{-\frac{\pi^2 w^2}{t}}$ term, and do not contribute to the local Seeley-DeWitt expansion. 
Expanding the $w=0$ term, we find
\begin{equation*}
\frac{\sqrt\pi}{4} t^{-\frac32} e^{t\nu^2} = \frac{2\pi^2}{(4\pi t)^{\frac32}}\sum_{r=0}^\infty \frac{\nu^{2r}}{r!} t^r \implies a_r(\mc K_{S^3}(\mf m)) = \frac{\vol{S^3}}{r!} (1-\mf m^2 \ell^2)^r
\end{equation*}
where we recognize the factor of $\vol S^3 = 2\pi^2$ as the volume of the unit 3-sphere (unit since we have rescaled the kinetic operator and therefore the heat parameter $t$ to be dimensionless). 
Our results match the $d=3$ results in \cite[Equation~94]{Kluth:2019vkg}, where to compare we use the Ricci scalar of $S^3$ is $R=6$, and the endomorphism $E=-\mf m^2\ell^2$, so $E+R/6 = \nu^2$. 

The $w\neq 0$ terms are the contribution to the heat trace from non-constant saddles of the worldline path integral, which are non-perturbative in small $t$ and are geodesics, i.e. great circles which wrap around the sphere $w$ times. 
This parameter is not a winding number since $\pi_1(S^3) = 0$, however it does count the number of times the geodesic wraps around the 3-sphere, and therefore multiplies the length of the geodesic. 

Recall that the worldline action \eqref{eq:WorldlinePI} has a kinetic term with a $\frac14$ prefactor, so the kinetic term contributes $\frac{L_w^2}{4t}$ to the on-shell action.
The length of a geodesic is $L_w = 2\pi\abs{w}$, so the kinetic term reads $\frac{\pi^2 w^2}{t}$, which is indeed the exponential in the non-perturbative term in heat trace. 
Expressing the heat trace in terms of this geometric data, we have
\begin{equation*}
H_{\mc K_{S^3}}(t;\nu) = \frac{\vol S^3}{(4\pi t)^{\frac32}} e^{t\nu^2}\left[ 1+ 2\sum_{w=1}^\infty \left(1-\frac{L_w^2}{2t}\right) e^{-\frac{L_w^2}{4t}}\right] - \delta_{\nu^2,1}\,. 
\end{equation*}
We will see that this general geometric structure carries forward in the subsequent examples in $\Gamma\bs S^3$.

\end{remark}

\subsubsection{A geometric expansion for the spectral zeta function}

The heat trace allowed us to use Poisson summation to relate spectral and geometric data. 
Since the spectral zeta function is related to the Mellin transform of the heat trace, 
\begin{equation}
M[H_{\mc K_{S^3}}(t;\nu)](s) = \Gamma(s)\zeta_{\mc K_{S^3}}(s;\nu),
\end{equation}
we can use the geometric version of the heat trace to work out the geometric version of the spectral zeta function. 

\begin{proposition} 
\label{prop:s3zeta2}

Let $\mc K_{S^3}(\mf m) = \Delta_{S^3}+\mf m^2\ell^2$, with $\nu^2 = 1 -\mf m^2\ell^2$ be the massive kinetic operator on $S^3$. 
The corresponding spectral zeta function admits an expansion in terms of modified Bessel functions of the second kind in the heavy sector,
\begin{equation}
\zeta_{\mc K_{S^3}}(s;i\nu_H) = \frac{\sqrt\pi}4 \frac{\Gamma(s-\frac32)}{\Gamma(s)}\nu_H^{3-2s} + \frac{\pi^{s-1}\nu_H^{\frac32-s}}{\Gamma(s)} \sum_{w=1}^\infty w^{s-\frac32}\left(K_{\frac32-s}(2w\pi\nu_H) -2w\pi\nu_H K_{\frac52-s}(2w\pi\nu_H) \right)\,.
\end{equation}
The sum over $w$ is entire in $s$, with poles only arising from the first term, with simple poles at $s=\frac32-j$, $j\in\mathbb Z_{\geq 0}$ with residues
\begin{equation*}
\mathrm{Res}_{s=\frac32-j}(\zeta_{\mc K_{S^3}}(s;i\nu_H)) = \frac{\sqrt\pi}{4}\frac{(-1)^j\nu_H^{2j}}{j!\,\Gamma(\frac32-j)}\,.
\end{equation*}

\end{proposition}

\begin{proof}

This derivation follows the strategy of combining the Mellin transform with Poisson resummation to obtain a rapidly convergent Bessel-K expansion, as in the evaluation of lattice sums by van der Hoff and Benson \cite{vanderHoff:1953}, used later in zeta function regularization by Elizalde and Romeo \cite{Elizalde:1989mr} and other physical contexts by Elizalde later \cite{Elizalde:1994sr}. 
Terras obtains similar Bessel-K expansions as a generalization of the Chowla-Selberg formula \cite{Terras:1973}, albeit by a slightly different route using a Hankel transform. 
Here we sketch our proof of the Bessel expansion of the spectral zeta function of the kinetic operator on $S^3$  in the heavy sector. 

Let us first establish the relationship between the heat trace and spectral zeta function using the spectral form of both constructed from the spectral data. 
Recall 
\begin{equation*}
H_{\mc K_{S^3}}(t;\nu) = \sum_{k=1}^\infty k^2 e^{-t(k^2-\nu^2)}\,, \qquad \zeta_{\mc K_{S^3}}(s;\nu) = \sum_{k=1}^\infty \frac{k^2}{(k^2-\nu^2)^s}\,.
\end{equation*}
The Mellin transform of the heat trace is
\begin{equation*}
M[H_{\mc K_{S^3}}(t;\nu)](s)  = \int_0^\infty t^{s-1} \sum_{k=1}^\infty k^2 e^{-t(k^2-\nu^2)} dt  = \sum_{k=1}^\infty k^2 \int_0^\infty t^{s-1} e^{-t(k^2-\nu^2)}dt\,.
\end{equation*}
We have shown in Lemma \ref{lem:s3heat} that the sum is absolutely convergent, so for sufficiently large $\Re(s)$ and $\abs{\Re(\nu)}<1$, the sum and integral may be swapped. 
Using the Gamma integral \eqref{eq:GammaDef}, we find 
\begin{equation*}
M[H_{\mc K_{S^3}}(t;\nu)](s) = \Gamma(s) \sum_{k=1}^\infty \frac{k^2}{(k^2-\nu^2)^s} = \Gamma(s)\zeta_{\mc K_{S^3}}(s;\nu)\,. 
\end{equation*}
Lemma \ref{lem:s3heat} also presents a resummed expression for the heat trace which may be Mellin transformed as well.
We will now work in the heavy sector, $\nu_H = -i\nu \in \mathbb R > 0$, to ensure convergence of integrals.
We also drop the zero-mode subtraction since it is unnecessary in this sector. 
Using the Bessel-K integral \eqref{eq:BesselKDef} and the Gamma integral \eqref{eq:GammaDef}, we have
\begin{eqnarray*}
& & \zeta_{\mc K_{S^3}} (s;i\nu_H) = \frac1{\Gamma(s)} M[H_{\mc K_{S^3}}(t;i\nu_H)](s) \\
& = & \frac1{\Gamma(s)} \int_0^\infty t^{s-1} \left(\frac{\sqrt\pi}4 t^{-\frac32} e^{-t\nu_H^2} + \frac{\sqrt\pi}2  t^{-\frac32} \sum_{w=1}^\infty \left(1-\frac{2\pi^2 w^2}{t}\right)e^{-t\nu_H^2-\frac{\pi^2 w^2}t}\right)dt \\
& = & \frac{\sqrt\pi}4 \frac{\Gamma(s-\frac32)}{\Gamma(s)}\nu_H^{3-2s} + \frac{\pi^{s-1}\nu_H^{\frac32-s}}{\Gamma(s)} \sum_{w=1}^\infty w^{s-\frac32}\left(K_{\frac32-s}(2w\pi\nu_H) -2w\pi\nu_H K_{\frac52-s}(2w\pi\nu_H) \right),
\end{eqnarray*}
where we again use the dominated convergence of the sum proved for Lemma \ref{lem:s3heat} to swap the sum and integral. 

\end{proof}

\begin{remark}
    
It may be useful or interesting to express the Bessel-expanded spectral zeta function in terms of geodesic lengths $L_w = 2\pi|w|$, just as we did for the heat trace after resumming, in the massive sector
\begin{equation*}
\zeta_{\mc K_{S^3}}(s;i\nu_H) =  \frac{\vol S^3}{\pi^{\frac32}\Gamma(s)}\left[\frac18\Gamma\left(s-\frac32\right)\nu_H^{3-2s}+ \frac{\nu_H^{\frac32-s}}{2^{s-\frac12}} \sum_{w=1}^\infty L_w^{s-\frac32}\left(K_{s-\frac32}(L_w\nu_H) -L_w\nu_H K_{s-\frac52}(L_w\nu_H)\right) \right]
\end{equation*}
The Mellin transform of the heat trace $M[H_{\mc K_{S^3}}(t;i\nu_H)](s)$ is regular at $s=0$, then $\zeta_{\mc K_{S^3}}(0;i\nu_H) = 0$, since $\frac1{\Gamma(0)} = 0$ \eqref{eq:GammaExp}.

\end{remark}

\begin{remark}

The regularized determinant of the kinetic operator on $S^3$ in the heavy sector can easily be recovered from the Bessel expansion.
    
Since $M[H_{\mc K_{S^3}}(t;i\nu_H)](s)$ is regular at $s=0$, then the formal manipulation in \eqref{eq:zetaDer} goes through. 
Repeating it here, using \eqref{eq:GammaExp}, we have,
\begin{equation*}
\pd_s \zeta_{\mc K_{S^3}}(s;i\nu_H) \big|_{s=0} = \pd_s\left(\frac1{\Gamma(s)} M[H_{\mc K_{S^3}}(t;i\nu_H)](s)\right)\big|_{s=0} = M[H_{\mc K_{S^3}}(t;i\nu_H)](0).
\end{equation*}
From Proposition \ref{prop:s3zeta2}, we can work out the expression at $s=0$, where the Bessel-K functions become elementary \eqref{eq:BesselKCases}.
We have,
\begin{eqnarray*}
\zeta_{\mc K_{S^3}}'(0;i\nu_H) & = & M[H_{\mc K_{S^3}}(t;i\nu_H)](0) \\ 
& = & \frac{\sqrt\pi \Gamma(-\frac32)}{4} \nu_H^3 + \frac{\nu_H^{\frac32}}{\pi}\sum_{w=1}^\infty w^{-\frac32}\left(K_{\frac32}(2\pi w \nu_H)-2\pi w\nu_H K_{\frac52}(2\pi w\nu_H) \right) \\
& = & \frac\pi3 \nu_H^3 - \sum_{w=1}^\infty \left(\frac{\nu_H^2}{w} + \frac{\nu_H}{\pi w^2} + \frac{1}{2\pi^2 w^3}\right)e^{-2\pi w \nu_H} \\
& = & \frac\pi3 \nu_H^3 -\nu_H^2\Li_1(e^{-2\pi\nu_H})-\frac{\nu_H}{\pi}\Li_2(e^{-2\pi\nu_H}) - \frac1{2\pi^2}\Li_3(e^{-2\pi\nu_H})\,,
\end{eqnarray*}
where we have used $\Gamma(-\frac32) = \frac{4\sqrt\pi}{3}$ and introduced the polylogarithm \eqref{eq:PolylogDef}.

\end{remark}

\section{Quotients of the 3-sphere: \texorpdfstring{$\Gamma\bs S^3$}{Gamma  S3}} 

\label{sec:GammaS3}

\subsection{Decomposition of \texorpdfstring{$\L^{2}(\Gamma\backslash S^{3})$}{L2(Gamma S3)}}

Let $\Gamma$ be a finite subgroup of $\mathrm{Isom}^+( S^3) \cong \SO(4)$ acting freely on $S^3$, preserving orientation. Recall that $S^3 \simeq \SU(2)$ and that
\begin{equation*}
\SO(4)\cong \left(\SU(2)\times\SU(2)\right)/\mathbb{Z}_{2}\,.
\end{equation*}
In particular, the action of $\Gamma$ on $ S^3$
induces a unitary representation of $\Gamma$ on
$\L^{2}( S^3)$. We denote by $\L^{2}( S^3)^{\Gamma}$ the subspace of $\Gamma$-invariant functions.

Since $\Gamma$ acts freely and by isometries, the quotient $\Gamma\backslash S^3$ is a compact Riemannian manifold. The quotient map induces an isomorphism of Hilbert spaces
\begin{equation}
\L^{2}(\Gamma\backslash S^3) \cong \L^{2}(S^3)^{\Gamma}\,.
\end{equation}
Indeed, every function on $\Gamma\backslash S^3$ lifts to a $\Gamma$-invariant function on $ S^3$, and conversely, every $\Gamma$-invariant function on $ S^3$ descends to the quotient. Moreover, since the action of $\Gamma$ is isometric, the Laplace-Beltrami operator commutes with the action of $\Gamma$. Therefore, under the previous identification,
\begin{equation}
\Delta_{\Gamma\backslash S^3} = \left.\Delta_{S^3}\right|_{\L^{2}( S^3)^{\Gamma}}\,.
\end{equation}
Recall that the Peter-Weyl theorem gives
\begin{equation*}
\L^{2}( S^3) \cong \widehat{\bigoplus}_{n = 0}^{\infty} \V_{n}\otimes\V_{n}^{*}\,,
\end{equation*}
where $\V_{n}$ is the irreducible representation of $\SU(2)$ of dimension $n+1$. Since the Laplace-Beltrami operator acts on
$\V_{n} \otimes \V^{*}_{n}$ by the scalar $n(n+2)$, we obtain
\begin{equation}
\L^{2}(\Gamma\backslash S^3) \cong \widehat{\bigoplus}_{n=0}^{\infty} \left(\V_{n}\otimes\V_{n}^{*}\right)^{\Gamma}\,.
\end{equation}
Therefore, the eigenvalues of $\Delta_{\Gamma\backslash S^3}$ belong to $\left\{n(n+2)\,, n\in\mathbb{N}\right\}$,
and their multiplicities are given by
\begin{equation}
m_{\Gamma}(n) = \dim\left(\V_{n}\otimes\V_{n}^{*}\right)^{\Gamma}\,.
\label{MultiplicityGeneralGamma}
\end{equation}

\begin{remark}

\begin{enumerate}
\item We now use the identification $ S^3\cong\SU(2)$
given in Equation \eqref{S3asSU2}. Under this
identification, the group $\SU(2)\times\SU(2)$
acts on $\SU(2)$ by
\begin{equation}
\left(u\,, v\right) \cdot x = uxv^{-1}\,, \qquad \left(u\,,v\,,x\in\SU(2)\right)\,.
\label{ActionSU2SU2}
\end{equation}
For each finite subgroup $\Gamma \subseteq \SO(4) \cong \SU(2) \times \SU(2) / \mathbb{Z}_{2}$ under consideration, the key will be to first describe $\Gamma$ explicitly, and then determine the action of each $\gamma \in \Gamma$ on $ S^3 \cong \SU(2)$ in terms of left and right multiplication, as in Equation \eqref{ActionSU2SU2}.
\item Suppose now that $\Gamma$ is a finite subgroup of
$\SU(2)$, acting on $ S^3 \simeq \SU(2)$ by left
multiplication. In particular, we get
\begin{equation*}
\left(\V_{n}\otimes\V^{*}_{n}\right)^{\Gamma} = \V_{n}^{\Gamma} \otimes \V_{n}^{*}\,,
\end{equation*}
i.e.
\begin{equation*}
\L^{2}(\Gamma\backslash S^3) \cong \widehat{\bigoplus}_{n=0}^{\infty} \V_{n}^{\Gamma} \otimes \V_{n}^{*}\,,
\end{equation*}
and the multiplicity formula
\eqref{MultiplicityGeneralGamma} simplifies to
\begin{equation}
m_{\Gamma}(n) = \left(n+1\right)\dim\left(\V_{n}^{\Gamma}\right)\,.
\label{EquationMGammaN}
\end{equation}
Thus, in this particular case, determining the spectral
multiplicities reduces to computing the dimensions
of the $\Gamma$-invariant subspaces of the irreducible
representations of $\SU(2)$.
\end{enumerate}
\label{RemarkGammaInSU2}

\end{remark}

\begin{remark}[Left and right quotients]

Let $\Gamma_L$ and $\Gamma_R$ be finite subgroups of $\SU(2)$ whose action on $S^3 \cong \SU(2)$ is
\begin{equation*}
x\mapsto \gamma_L x \gamma_R^{-1}\,,\qquad \left(x \in \SU(2)\,, \gamma_L \in \Gamma_L\,, \gamma_R \in \Gamma_R\right)\,.
\end{equation*}
The decomposition of square-integrable functions on the spherical space form $\Gamma_L\bs S^3/\Gamma_R$ is 
\begin{equation*}
\L^2(\Gamma_L\bs S^3 /\Gamma_R) \cong \widehat{\bigoplus}_{n=0}^{\infty} (\V_n\otimes \V_n^*)^{\Gamma_L\times \Gamma_R} = \widehat{\bigoplus}_{n=0}^\infty \V_n^{\Gamma_L} \otimes (\V_n^{\Gamma_R})^*,
\end{equation*}
whose multiplicities are therefore
\begin{equation*}
m_{\Gamma_L\times\Gamma_R}(n) = \dim(\V_n^{\Gamma_L})\dim(\V_n^{\Gamma_R})\,.
\end{equation*}
Note that $\Gamma_L\times\Gamma_R$ acts freely on $S^3$ if and only if $\tr\gamma_L\neq\tr\gamma_R$ except for $(\gamma_L,\gamma_R) = (\pm 1,\pm 1)$.
Not every freely acting finite subgroup of $\SO(4)$ can be written as $\Gamma_L\times\Gamma_R\subset \SU(2)\times\SU(2)$, for example lens spaces $L(p,q)$ with $p,q$ coprime.

\end{remark}

\noindent As we have seen, the key to determining the spectral zeta function of $\Gamma\backslash S^3$ is to compute the multiplicities $m_{\Gamma}(n)$ of the eigenvalues of the Laplace-Beltrami operator. One possible approach is to use the explicit realization of $\V_{n}$ in terms of homogeneous polynomials, with basis
\begin{equation*}
\left\{z_{1}^{n-k}z_{2}^{k}\,,\ 0\leq k\leq n\right\}\,,
\end{equation*}
and determine the subspace of $\Gamma$-invariant vectors. This method is particularly effective when $\Gamma$ consists of diagonal matrices, since the action on the monomial basis can be computed explicitly. However, for more general finite subgroups, determining the invariant subspaces directly can become considerably more complicated. An alternative approach is to use Molien series, which provide a generating function for the dimensions of the invariant subspaces and, therefore, for the spectral multiplicities. The goal of the next section is to introduce this method and explain how it can be applied to the computation of spectral zeta functions on finite quotients of $ S^3$.

\subsection{Molien Series}

\label{SectionMolienSeries}

As explained above, one of the key ingredients in the computation of the spectral zeta function of the kinetic operator on $\Gamma \backslash  S^3$ is the determination of the dimension
\begin{equation*}
\dim\left(\left(\V_{n}\otimes\V_{n}^*\right)^{\Gamma}\right),
\end{equation*}
where $\V_{n}$ is the irreducible representation of $\SU(2)$. In order to obtain explicit formulas for these dimensions, it is useful to determine the multiplicity of the trivial representation of $\Gamma$ in the spaces $\V_{n}$ or $\V_{n} \otimes \V_{n}$. 

\noindent Let $\Gamma$ be a finite group acting on a finite-dimensional complex vector space $\V$ of dimension $d$ via a representation $\rho: \Gamma \to \GL(\V)$. This action extends to the symmetric algebra $\S(\V^{*}) = \bigoplus\limits_{n=0}^{\infty} \S^{n}(\V^{*})$ by
\begin{equation}
\left(\gamma \cdot f\right)(v) = f\left(\rho(\gamma)^{-1}v\right)\,, \qquad \left(\gamma \in \Gamma\,, f \in \S(\V^{*})\,, v \in \V\right)\,,
\end{equation}
where $\V^{*}$ is the dual space of $\V$. In particular, one can see that $\rho(\Gamma)\left(\S^{n}(\V^{*})\right) \subseteq \S^{n}(\V^{*})$.

\begin{remark}

Note that once we pick a basis $v_{1}, v_{2}, \ldots, v_{d}$ of $\V$, we have $\S^{n}(\V^{*}) \cong \mathbb{C}^{n}\left[v_{1}\,, \ldots\,, v_{d}\right]$. To simplify the notations, we denote by $\V_{n}$ the set $\S^{n}(\V^{*})$. In particular,
\begin{equation*}
\V^{\Gamma}_{n} = \left\{f \in \V_{n}\,, \gamma \cdot f = f\,, \gamma \in \Gamma\right\}\,.
\end{equation*}

\end{remark}

\begin{definition}[Molien series] 

The Molien series of $\left(\Gamma\,, \rho\right)$ is the formal power series in $\mathbb{Z}_{\geq 0}\left[\left[q\right]\right]$ given by
\begin{equation}
\mathcal{M}_{\Gamma}(q) = \sum\limits_{n=0}^{\infty}\dim(\V^{\Gamma}_{n})q^{n}\,.
\label{DefinitionMGammaQ}
\end{equation}

\label{DefinitionMolienSeries}

\end{definition}

\begin{remark}

For each $n \geq 0$, we have
\begin{equation}\label{eq:dimVGamma}
\dim(\V^{\Gamma}_{n}) = \frac{1}{\left|\Gamma\right|}\sum\limits_{\gamma \in \Gamma} \chi_{n}(\gamma)\,,
\end{equation}
where $\chi_{n}(\gamma) = \tr(\rho_{n}(\gamma))$ (see Definition \ref{def:character}), where $\rho_{n} := \rho_{|_{\V_{n}}}$ is the representation of $\Gamma$ on $\V_{n}$ obtained from $\rho$. Indeed, by Schur orthogonality \cite[Theorem.~2.12]{FultonHarris}, the multiplicity of the trivial representation in a representation $\left(\pi\,, W\right)$ of $\Gamma$ is equal to 
\begin{equation*}
\frac{1}{\left|\Gamma\right|}\sum\limits_{\gamma \in \Gamma} \tr_{W}(\pi(\gamma))\,,
\end{equation*}
since the character of the trivial representation is identically $1$.

\end{remark}

\noindent The following theorem, due to Molien, gives a closed rational expression for \eqref{DefinitionMGammaQ} directly in terms of the group elements.

\begin{theo}

With the notation above, we get
\begin{equation}
\mathcal{M}_{\Gamma}(q) = \sum\limits_{n=0}^{\infty}\left(\frac{1}{\left|\Gamma\right|}\sum\limits_{\gamma \in \Gamma} \chi_{n}(\gamma)\right)q^{n} = \frac{1}{\left|\Gamma\right|}\sum\limits_{\gamma \in \Gamma} \frac{1}{\det\left(\Id_{\V} - q\rho(\gamma)\right)}\,.
\label{EquationMolien}
\end{equation}
In particular, $\mathcal{M}_{\Gamma}(q)$ is a rational function of $q$ whose power-series expansion has non-negative integer coefficients.

\label{TheoremMGammaQ}

\end{theo}

\begin{remark}

The group $\Gamma$ acts on itself by conjugation. Let $\gamma\,, \gamma' \in \Gamma$. Since
\begin{eqnarray*}
&\det\left(\Id_{\V} - q\rho(\gamma'\gamma\gamma'^{-1})\right) = \det\left(\Id_{\V} - q\rho(\gamma')\rho(\gamma)\rho(\gamma')^{-1}\right) \\
&=  \det\left(\rho(\gamma')\left(\Id_{\V} - q\rho(\gamma)\right)\rho(\gamma')^{-1}\right) = \det\left(\Id_{\V} - q\rho(\gamma)\right),
\end{eqnarray*}
it follows that Formula \eqref{EquationMolien} reduces to a finite sum over conjugacy class representatives
\begin{equation}
\mathcal{M}_{\Gamma}(q) = \frac{1}{\left|\Gamma\right|}\sum\limits_{\left[\gamma\right]} \frac{\left|\left[\gamma\right]\right|}{\det(\Id_{\V} - q\rho(\left[\gamma\right]))}\,.
\end{equation}

\label{RemarkConjugacyClasses}

\end{remark}

\subsection{The lens spaces \texorpdfstring{$L(p,q)$}{L(p,q)}}

\label{SectionLensSpace}

\subsubsection{Embedding of $\Gamma$}

We use the realization of $ S^3$ given in Equation \ref{DefinitionS3Complex}. Let $p \geq 1$ and $\omega$ be the element in $\mathbb{C}$ given by $\omega = e^{2i\pi/p}$. For all $q \in \mathbb{N}^{*}$, we define an action of the cyclic group $\Gamma \cong \mathbb{Z}/p\mathbb{Z}$ on $ S^3$ by letting a generator $\gamma$ act as
\begin{equation}
\gamma \cdot \left(z_{1}\,, z_{2}\right) = \left(\omega z_{1}\,, \omega^{q} z_{2}\right)\,.
\label{LensAction}
\end{equation}
One can show that the action \eqref{LensAction} is free if and only if $\gcd(p\,, q)=1$\,.

\begin{definition}

For $\gcd(p\,, q) = 1$, the quotient manifold
\begin{equation*}
L(p, q) := \Gamma \backslash  S^3
\end{equation*}
is called the lens space of type $\left(p, q\right)$\,.

\end{definition}

\begin{remark}

Let $u$ and $v$ be the matrices in $\SU(2)$ given by
\begin{equation*}
u = \begin{pmatrix} \alpha & 0 \\ 0 & \alpha^{-1} \end{pmatrix}\,, \qquad v = \begin{pmatrix} \beta & 0 \\ 0 & \beta^{-1} \end{pmatrix}\,,
\end{equation*}
where
\begin{equation*}
\alpha = e^{\frac{i\pi}{p}(1+q)}\,, \qquad \beta = e^{-\frac{i\pi}{p}(1-q)}\,.
\end{equation*}
In particular, the action of $u$ and $v$ on $x \in \SU(2)$ given by
\begin{equation*}
uxv^{-1}
\end{equation*}
corresponds to exactly 
\begin{equation*}
\left(z_{1}\,, z_{2}\right) \longmapsto \left(\omega z_{1}\,, \omega^{q}z_{2}\right)\,.
\end{equation*}
Thus the cyclic group $\Gamma$ is generated by the transformation
\begin{equation*}
x \longmapsto uxv^{-1}\,.
\end{equation*}

\end{remark}

\subsubsection{The coefficients $m_{\Gamma}(n)$}

Using the embedding of $\Gamma$ given above, we obtain the following proposition.

\begin{proposition}

For all $n \geq 0$, we have
\begin{equation*}
m_{\Gamma}(n) = \dim\left((\V_{n} \otimes \V_{n}^*)^{\Gamma}\right) = \#\left\{\left(a\,, b\right) \in \left\{0\,, \dots\,, n\right\}^{2}\,, a+b-n+q(a-b)\equiv 0 \pmod p \right\}\,.
\end{equation*}

\label{PropositionMultiplicityLensSpace}

\end{proposition}

\begin{proof}

Let $\left\{e_{a} := z_{1}^{n-a}z_{2}^{a}\,, 0 \leq a \leq n\right\}$ be the standard weight basis of $\V_{n}$. Then the diagonal matrix
\begin{equation*}
\begin{pmatrix} t & 0 \\ 0 & t^{-1} \end{pmatrix}
\end{equation*}
acts on $e_{a}$ by $t^{n-2a}$. Hence it follows that $u$ and $v$ acts on $e_{a}$ by $\alpha^{n-2a}$ and $\beta^{n-2a}$ respectively. 
In particular, the generator $\gamma$ acts on $e_{a} \otimes e_{b}$ by the scalar
\begin{equation*}
\alpha^{n-2a}\beta^{-(n-2b)}\,.
\end{equation*}
Using the definitions of $\alpha$ and $\beta$, we compute
\begin{equation*}
\alpha^{n-2a}\beta^{-(n-2b)} = \exp\left(\frac{i\pi}{p}(1+q)(n-2a)\right)\exp\left(\frac{i\pi}{p}(1-q)(n-2b)\right) = \omega^{n-a-b-q(a-b)}\,.
\end{equation*}
Therefore, the vector $e_{a} \otimes e_{b}$ is such that $\gamma \cdot e_{a} \otimes e_{b} = e_{a} \otimes e_{b}$ if and only if $\omega^{n-a-b-q(a-b)} = 1$, i.e. if and only if $n-a-b-q(a-b) \equiv 0 \pmod p$\,.

\end{proof}

\begin{remark}

We denote by $m_{p,q}(n)$ the multiplicity $m_{\Gamma}(n)$ corresponding to the lens space $L(p,q)$. It follows from Proposition \ref{PropositionMultiplicityLensSpace} that the spectral multiplicities depend only on the residue class of $q$ modulo $p$. In particular, for every $k\in\mathbb{Z}$, we have
\begin{equation*}
m_{p,q+kp}(n)=m_{p,q}(n)\,, \qquad \left(n \geq 0\right)\,.
\end{equation*}
Moreover, if $q^{-1}$ denotes the multiplicative inverse of $q$ modulo $p$, then
\begin{equation*}
m_{p,q^{-1}}(n) = m_{p,q}(n)\,, \qquad \left(n \geq 0\right)\,.
\end{equation*}
Indeed, multiplying the congruence
\begin{equation*}
n-a-b-q^{-1}(a-b)\equiv 0\pmod p
\end{equation*}
by $q$ and mapping $b\mapsto n-b$ we obtain the original congruence
\begin{equation*}
n-a-b-q(a-b)\equiv 0\pmod p\,.
\end{equation*}
Therefore, the Laplace-Beltrami operators on $L(p,q)$, $L(p,q+kp)$, and
$L(p,q^{-1})$ are isospectral, and their spectral zeta functions and zeta-regularized determinants coincide. This is consistent with the fact that these lens spaces are isometric.

\end{remark}

\subsubsection{Homogeneous lens spaces $L(p,1)$}

Let $n \in \mathbb{N}$. Using that $\Gamma$ is now a subgroup of $\SU(2)$, it follows from Remark \ref{RemarkGammaInSU2} and Proposition \ref{PropositionMultiplicityLensSpace} that $m_{\Gamma}(n) = (n+1)\N_{n}(p)$, with 
\begin{equation*}
\N_{n}(p) = \#\left\{b\in\left\{0\,, \dots\,, n\right\}\,, 2b \equiv n \pmod p\right\}\,.
\end{equation*}

\begin{lemma}

If $p$ is even, then
\begin{equation*}
\N_{n}(p) = \begin{cases} 0 & \text{ if } n \text{ is odd } \\ 2\left\lfloor\frac{n}{p}\right\rfloor +1 & \text{ if } n \text{ is even } \end{cases}
\end{equation*}
Hence
\begin{equation}
m_{\Gamma}(n) = \begin{cases} 0 & \text{ if } n \text{ is odd } \\ (n+1)\left(2\left\lfloor \dfrac{n}{p}\right\rfloor +1\right) & \text{ if } n \text{ is even}.
\end{cases}
\label{eq:EvenLensMult}
\end{equation}

\end{lemma}

\begin{proof}

Suppose that $p$ is even. If $n$ is odd, the congruence
\begin{equation*}
2b\equiv n\pmod p
\end{equation*}
has no solution, since both $2b$ and $p$ are even. Thus, $\N_{n}(p)=0$. Now suppose that $n$ is even. Since $p$ is even, the congruence is equivalent to
\begin{equation*}
b\equiv\frac{n}{2}\pmod{\frac{p}{2}}\,.
\end{equation*}
Hence, its solutions are given by
\begin{equation*}
b = \frac{n}{2}+k\frac{p}{2}\,, \qquad \left(k \in \mathbb{Z}\right)\,.
\end{equation*}
The condition $0\leq b\leq n$ is equivalent to
\begin{equation*}
-\frac{n}{p} \leq k \leq \frac{n}{p}\,.
\end{equation*}
Therefore,
\begin{equation*}
\N_{n}(p)=2\left\lfloor\frac{n}{p}\right\rfloor+1\,.
\end{equation*}
The multiplicity formula follows immediately from $m_{\Gamma}(n) =(n+1)\N_{n}(p)$.

\end{proof}

\begin{remark}

The case $p$ odd is similar. Indeed, write $n = kp+s$, with $0 \leq s \leq p-1$, and let $r_{s} \in \left\{0\,, \dots\,, p-1\right\}$ be the unique integer such that
\begin{equation*}
2r_{s} \equiv s \pmod p.
\end{equation*}
Then one can show that
\begin{equation*}
\N_{n}(p) = \begin{cases} k+1 & \text{ if } r_{s} \leq s \\ k & \text{ if } r_{s} > s \end{cases}
\end{equation*}
and therefore
\begin{equation*}
m_{\Gamma}(n) = (n+1)\begin{cases} k+1 & \text{ if } r_{s} \leq s \\ k & \text{ if } r_{s} > s \end{cases}
\end{equation*}

\end{remark}

\subsubsection{Even homogeneous lens spaces \texorpdfstring{$L(2p,1)$}{L(2p,1)}}

We work out the spectral zeta function, heat trace and regularized determinant of the kinetic operator on the even lens spaces $L(2p,1)$, $p\in\mathbb Z_+$.
For the proofs, refer to Section \ref{sec:EvenQuotients}, where we prove the results that appear here for general even order subgroups.
Instead, here we will highlight the key observations that allow us to obtain the general structure of the expressions for general even order quotients.

The kinetic operator on $L(2p,1)$ is
\begin{equation}
\label{eq:LensKin}
\mc K_{L(2p,1)}(\mf m) = \Delta_{L(2p,1)} + \mf m^2\ell^2,
\end{equation}
whose local behavior is identical with that of the kinetic operator on $S^3$ without the quotient since it is a local operator.
Thus, the eigenvalues remain unchanged, $c(\pi) + \mf m^2\ell^2 = n(n+2) + \mf m^2\ell^2 = (n+1)^2 - \nu^2$. 

Note that the multiplicities \eqref{eq:EvenLensMult} are expressed in terms of a floor function. 
Let us reparametrize the label $n$ to remove the dependence on the floor function as follows.
Since only even $n$ has non-zero multiplicity, let $n=2pk+2r$ with $0\leq r \leq p-1$ and $k\geq 0$ such that $\floor{\frac{n}{2p}} = k$.
Then the eigenvalues and multiplicities of the Laplace-Beltrami operator respectively read
\begin{eqnarray*}
(n+1)^2 & = & (2pk+2r+1)^2 = 4p^2\left(k+\frac{2r+1}{2p}\right)^2,\\
m_{\mathbb Z_{2p}}(n) & = & (n+1)\left(2\floor{\frac{n}{2p}}+1\right) \delta_{n\bmod 2,0}= (2pk+2r+1)(2k+1) = 4p\left(k+\frac{2r+1}{2p}\right)\left({k+\frac12}\right).
\end{eqnarray*}
To simplify notation, we introduce the following variables
\begin{equation}\label{eq:LensAlphaBeta}
\alpha_r = \frac{2r+1}{2p}\,, \qquad \beta_r = \frac12-\alpha_r\,,
\end{equation}
using which we can express the eigenvalues and multiplicities as
\begin{equation*}
(n+1)^2 = 4p^2 (k+\alpha_r)^2\,, \qquad m_{\mathbb Z_{2p}}(k,r) = 4p(k+\alpha_r)(k+\alpha_r+\beta_r)\,.
\end{equation*}
Since $0\leq r \leq p-1$, it will be important to understand the parity properties of the above variables.
In particular,
\begin{equation*}
\alpha_{p-1-r} = \frac{2(p-1-r)+1}{2p} = 1-\alpha_r\,, \qquad \beta_{p-1-r} = \frac12-\alpha_{p-1-r} = -\beta_r\,.
\end{equation*}

\begin{proposition}
    
The spectral zeta function for the kinetic operator on $L(2p,1)$ in terms of spectral data is, for $\Re(s)>\frac32$ and $\abs{\Re(\nu)}<1$, 
\begin{equation}
\zeta_{\mc K_{L(2p,1)}}(s;\nu) = \sum_{n=0}^\infty \frac{(n+1)\left(2\floor{\frac n{2p}}+1\right)\delta_{n\bmod 2,0}}{((n+1)^2-\nu^2)^s}\,,
\end{equation}
which admits the following meromorphic continuation, for a choice of $N\in\mathbb Z_{\geq 0}$ such that $2pN+1>|\nu|$,
\begin{equation}
\begin{split}
&\zeta_{\mc K_{L(2p,1)}}(s;\nu)  =\sum_{r=0}^{p-1}\sum_{k=0}^{N-1}\frac{4p\left(k+\alpha_r\right)\left(k+\frac12\right)}
{\left(4p^2\left(k+\alpha_r\right)^2-\nu^2\right)^{s}} \\
& \quad + 2(2p)^{1-2s}\sum_{r=0}^{p-1}\sum_{j=0}^\infty \frac{(s)_j}{j!}\left(\frac{\nu}{2p}\right)^{2j} \bigg[\zeta_H(2s+2j-2,\alpha_r+N) + \beta_r\zeta_H(2s+2j-1,\alpha_r+N)\bigg]\,,
\end{split}
\end{equation}
where $\alpha_r = \frac{2r+1}{2p}$, with at most simple poles at $s=\frac32-j$, $j\in\mathbb Z_{\geq 0}$. 
For the light sector, $|\nu|<1$, so $N=0$ suffices, so the finite sum is absent. 
In particular, the spectral zeta function is regular at $s=0$.

\end{proposition}

\begin{proof}

The spectral zeta function for the kinetic operator on $L(2p,1)$ can be directly built from the spectral data to get the first expression. 
Using \eqref{eq:LensAlphaBeta}, we can rewrite the zeta function to be
\begin{equation*}
\zeta_{\mc K_{L(2p,1)}}(s;\nu)  = \sum_{n=0}^\infty \frac{(n+1)\left(2\floor{\frac n{2p}}+1\right)\delta_{n\bmod 2,0}}{((n+1)^2-\nu^2)^s} = \sum_{r=0}^{p-1}\sum_{k=0}^\infty
\frac{4p\left(k+\alpha_r\right)\left(k+\frac12\right)}
{\left(4p^2\left(k+\alpha_r\right)^2-\nu^2\right)^{s}}.
\end{equation*}
Choose $N$ such that $2pN+1>|\nu|$ (in the light sector $|\nu|<1$, so $N=0$ suffices).
Perform binomial expansion for $k\geq N$, then for sufficiently large $\Re(s)$ we have
\begin{eqnarray*}
& & \zeta_{\mc K_{L(2p,1)}}(s;\nu) = \sum_{r=0}^{p-1}\sum_{k=0}^{N-1}\frac{4p\left(k+\alpha_r\right)\left(k+\frac12\right)}
{\left(4p^2\left(k+\alpha_r\right)^2-\nu^2\right)^{s}} + \sum_{r=0}^{p-1}\sum_{k=N}^\infty \frac{4p\left(k+\alpha_r\right)\left(k+\frac12\right)}
{\left(4p^2\left(k+\alpha_r\right)^2-\nu^2\right)^{s}} \\
& = & \sum_{r=0}^{p-1}\sum_{k=0}^{N-1}\frac{4p\left(k+\alpha_r\right)\left(k+\frac12\right)}
{\left(4p^2\left(k+\alpha_r\right)^2-\nu^2\right)^{s}}  + 2(2p)^{1-2s}\sum_{r=0}^{p-1}\sum_{j=0}^\infty \frac{(s)_j}{j!}\left(\frac{\nu}{2p}\right)^{2j} \sum_{k=N}^\infty (k+\tfrac12)(k+\alpha_r)^{-2s-2j+1} \\
& = & \sum_{r=0}^{p-1}\sum_{k=0}^{N-1} \frac{4p\left(k+\alpha_r\right)\left(k+\frac12\right)}
{\left(4p^2\left(k+\alpha_r\right)^2-\nu^2\right)^{s}} \\
& \quad & + 2(2p)^{1-2s}\sum_{r=0}^{p-1}\sum_{j=0}^\infty \frac{(s)_j}{j!}\left(\frac{\nu}{2p}\right)^{2j} \bigg[\zeta_H(2s+2j-2,\alpha_r+N) + \beta_r\zeta_H(2s+2j-1,\alpha_r+N)\bigg],
\end{eqnarray*}
which provides the analytic continuation of the spectral zeta function.
Observe that $\sum_{r=0}^{p-1}\beta_r = 0$, so the pole from the second Hurwitz zeta at $j=0$ and $s=1$ is removed. 
Similarly the simple poles at $s=1-j$ for $j\geq1$ are multiplied by $(s)_j$, which has a simple zero at the same location, canceling the pole. 
So the location of the possible poles are at $s=\frac32-j$, $j\in\mathbb Z_{\geq 0}$.

\end{proof}

\begin{proposition}

The regularized determinant of the massive kinetic operator on $L(2p,1)$ for $|\Re(\nu)|<1$ is
\begin{equation}
\begin{split}
-\log{\det}'(\mc K_{L(2p,1)}(\mf m)) 
&=\pd_s \zeta_{\mc K_{L(2p,1)}}(s;\nu)\big|_{s=0} \\
&= \frac{i\pi}{6p}\nu^3 + 
\frac{\nu^2}p \Li_1\left(-e^{\pi i\nu}\right) +\frac{2i\nu}{\pi p} \Li_2\left(-e^{\pi i\nu}\right)-\frac{2}{\pi^2 p} \Li_3\left(-e^{\pi i\nu}\right)\\
&\quad -2\sum_{r=0}^{p-1} \beta_r\left(\nu \Li_1(e^{\frac{\pi i\nu}{p}+2\pi i \alpha_r})+\frac{ip}{\pi} \Li_2(e^{\frac{\pi i\nu}{p}+2\pi i\alpha_r})\right),
\end{split}
\end{equation}
where $\alpha_r = \frac{2r+1}{2p}$.  The regularized determinant of the massless kinetic operator on $L(2p,1)$, i.e. $\nu=1$ or $\mf m = 0$ is
\begin{equation}
\begin{split}
-\log{\det}'(\mc K_{L(2p,1)}(0)) 
&= \pd_s \zeta_{\mc K_{L(2p,1)}}(s;1)\big|_{s=0} \\
&= \frac{\pi i (p-1)}{6}+ \left(1-\frac1p\right)\left(\log p- \frac{\pi i}2\right) +\log\frac2{\pi} - \frac{2}{\pi^2 p}\zeta_R(3) \\ &\quad 
-\sum_{r=0}^{p-2} \beta_r\left(2\Li_1(e^{\frac{\pi i}{p}+2\pi i \alpha_r})+\frac{2ip}{\pi} \Li_2(e^{\frac{\pi i}{p}+2\pi i\alpha_r})\right).
\end{split}
\end{equation}

\end{proposition}

For the proof, please see the corresponding proof in the following section, Section~\ref{sec:EvenQuotients}.

\begin{remark}

The real projective space $\mathbb{RP}^3\cong (\mathbb Z/2\mathbb Z)\bs S^3 \cong L(2,1)$ is the special case at $p=1$.
Therefore the parameter $r$ can only be $r=0$, where the twist coefficient vanishes $\beta_0 = 0$, removing the twisted sum.
Indeed, since $L(2,1)$ is the quotient of $S^3$ $\mathbb Z/2\mathbb Z = \{\pm 1\}$, which is the center of $\SO(4)$, only the central parts of the expressions for the heat trace, spectral zeta function and regularized determinant are retained. 

It will be interesting to compare the result for the regularized determinant to the result obtained from other means, such as the Wilson spool \cite{Haupfear:2025grb} or the heat kernel. 

\end{remark}

Recall the definition of the heat operator, $e^{-t \mc K_{L(2p,1)}(\mf m)}$ and the corresponding heat trace,
\begin{equation}
\label{eq:LensHeat}
H_{\mc K_{L(2p,1)}}(t;\nu) := \tr'(e^{-t\mc K_{L(2p,1)}(\mf m)})\,, \qquad \left(t > 0\right)\,.
\end{equation}

\begin{proposition} 
\label{prop:LensHeat}

The heat trace of the kinetic operator \eqref{eq:LensKin} for $t>0$ admits the following equivalent expressions, (a) in terms of the eigenvalues and multiplicities, (b) in terms of Jacobi theta functions \eqref{def:JacobiThetaDef}, and (c) in terms of a geometric parameter $w$,
\begin{subequations}    
\begin{align}
&H_{\mc K_{L(2p,1)}}(t;\nu) \nonumber\\
&=  \sum_{n=0}^\infty (n+1)\left(2\floor{\frac{n}{2p}}+1\right)\delta_{n\bmod 2,0}\, e^{-t((n+1)^2-\nu^2)} - \delta_{\nu^2,1}\\
&= -\frac{e^{t\nu^2}}{2p}\pd_t \vartheta_2\left(\frac{4it}{\pi}\right)+\frac{e^{t\nu^2}}{2\pi i}\sum_{r=0}^{p-1}(\tfrac12-\alpha_r)\pd_z \JacTh{\alpha_r}{0}{2pz}{4p^2\frac{it}{\pi}} \Bigg|_{z=0} - \delta_{\nu^2,1}\\
&= \frac{\sqrt\pi}{8p}\sum_{w\in\mathbb Z}(-1)^w \left(1-\frac{\pi^2w^2}{2t}\right)t^{-\frac32}e^{-\frac{\pi^2 w^2}{4t}+t\nu^2}+\frac{\pi^{\frac32}}{8p^2}\sum_{\substack{w\in\mathbb Z\\p\nmid w}}\frac{w}{\sin\frac{\pi w}{p}}t^{-\frac32}e^{-\frac{\pi^2 w^2}{4p^2 t}+t\nu^2} - \delta_{\nu^2,1}\,,
\end{align}
\end{subequations}
where $\alpha_r = \frac{2r+1}{2p}$, and we subtract the zero mode with a Kronecker delta if $\nu^2=1$.

In the heavy sector, $\nu_H = -i\nu > 0$, the spectral zeta function admits an expansion in Bessel functions, manifestly free of the auxiliary parameter $N$,
\begin{equation}
\begin{split}
&\zeta_{\mc K_{L(2p,1)}}(s;i\nu_H) \\
&= \frac{\sqrt{\pi}\nu_H^{3-2s} \Gamma(s-\tfrac32)}{8p\Gamma(s)} + \frac{\pi^{s-1}}{2^{s-\frac12}p\Gamma(s)} 
\sum_{w=1}^\infty (-1)^w\left(\frac{w}{\nu_H}\right)^{s-\frac32} \left(K_{\frac32-s}\left(\pi w\nu_H\right)-\pi w\nu_H K_{\frac52-s}\left(\pi w\nu_H\right)\right) \\&\quad 
+\frac{\pi^s \nu_H^{\frac32-s}}{2^{s-\frac12}p^{s+\frac12}\Gamma(s)}\sum_{\substack{w=1\\p\nmid w}}^\infty \frac{w^{s-\frac12}}{\sin\frac{\pi w}{p}}  K_{\frac32-s}\left(\frac{\pi w\nu_H}{p}\right)\,,
\end{split}
\end{equation}
where the sum over $w$ is entire in $s$, with all the poles coming from the $w=0$ term, located at $s = \frac32-j$, $j\in\mathbb Z_{\geq 0}$.
The expansion rapidly converges for $\nu_H>\abs{\Gamma}/2\pi$.

\end{proposition}

\begin{proof}

We build the heat trace using the eigenvalues and degeneracies as follows, and then use the definitions of the Jacobi theta functions with characteristics (cf.~\ref{def:JacobiThetaDef}) to rewrite the expression. 
Let $n=2pk+2r$ with $0\leq r \leq p-1$ and $k\geq 0$, note that $\floor{\frac{n}{2p}} = k$.
Defining $\alpha_r = \frac{2r+1}{2p}$, we have $n+1 = 2p(k+\alpha_r)$, and so for $t>0$ we have
\begin{eqnarray*}
H_{\mc K_{L(2p,1)}}(t;\nu) & = & \sum_{n=0}^\infty (n+1)\left(2\floor{\frac{n}{2p}}+1\right)\delta_{n\bmod 2,0}\, e^{-t((n+1)^2-\nu^2)} \\
& = & 4p e^{t\nu^2}\sum_{r=0}^{p-1}\sum_{k=0}^\infty \big((k+\alpha_r)^2+\beta_r(k+\alpha_r)\big)e^{-4p^2t(k+\alpha_r)^2}\\
& = & -\frac{e^{t\nu^2}}{2p}\pd_t \vartheta_2\left(\frac{4it}{\pi}\right)+\frac{e^{t\nu^2}}{2\pi i}\sum_{r=0}^{p-1}\beta_r \pd_z \JacTh{\alpha_r}{0}{2pz}{4p^2\frac{it}{\pi}} \Bigg|_{z=0},
\end{eqnarray*}
The first term is the ``central'' part of the quotient, i.e. generated by the $\mathbb Z/2\mathbb Z$ subgroup of $\mathbb Z/2p\mathbb Z$.
The second term is what we call the twisted sum, corresponding to the non-central parts of the quotient, where we have used $\beta_r=-\beta_{p-1-r}$ to promote the sum over positive integers to a sum over all integers with a factor of $\frac12$. 

Performing a modular $\mc S$ transform to the theta functions to perform a Poisson resum, we have
\begin{eqnarray*}
& & H_{\mc K_{L(2p,1)}}(t;\nu) \\ 
& = & -\frac{e^{t\nu^2}}{2p}\pd_t\left(\sqrt{\frac{\pi}{4t}}\vartheta_4\left(\frac{\pi i}{4t}\right)\right) + \frac{e^{t\nu^2}}{2\pi i}\sum_{r=0}^{p-1}\beta_r\pd_z\left(\sqrt{\frac{\pi}{4p^2t}} e^{-\frac{\pi^2 z^2}{t}} \JacTh{0}{\alpha_r}{\frac{\pi z}{2pit}}{\frac{\pi i}{4p^2t}}\right)\bigg|_{z=0} \\ 
& = &\sum_{w\in\mathbb Z}\left( \frac{\sqrt\pi}{8p}(-1)^w \left(1-\frac{\pi^2w^2}{2t}\right)t^{-\frac32}e^{-\frac{\pi^2 w^2}{4t}+t\nu^2}-\frac{i\pi^{\frac32}}{4p^2}wt^{-\frac32}e^{-\frac{\pi^2 w^2}{4p^2 t}+t\nu^2}\sum_{r=0}^{p-1}\beta_r e^{2\pi i w \alpha_r}\right) \\
& = & \frac{\sqrt\pi}{8p}\sum_{w\in\mathbb Z}(-1)^w \left(1-\frac{\pi^2w^2}{2t}\right)t^{-\frac32}e^{-\frac{\pi^2 w^2}{4t}+t\nu^2}+\frac{\pi^{\frac32}}{8p^2}\sum_{\substack{w\in\mathbb Z\\p\nmid w}}\frac{w}{\sin\frac{\pi w}{p}}t^{-\frac32}e^{-\frac{\pi^2 w^2}{4p^2 t}+t\nu^2}
\end{eqnarray*}
where we evaluate the sum over $r$ using
\begin{equation*}
\sum_{r=0}^{p-1}\beta_r e^{2\pi i w \alpha_r} = \begin{cases} \frac{i}{2\sin\frac{\pi w}{p}} & p\nmid w \\ 0 & p\mid w \end{cases}
\end{equation*}
which shows the twisted sum does not see the central elements of the quotient.

When $\nu^2 = 1$, the heat trace omits the $n=0$ mode, which has multiplicity 1, so we subtract 1 for the massless case by introducing a Kronecker delta.

To derive the geometric expansion of the spectral zeta function in the heavy sector, we use the Poisson resummed heat trace.
Recall that the spectral zeta function is the Mellin transform \eqref{eq:MellinDef} of the heat trace for $\Re(s)>\frac32$.
Using \eqref{eq:BesselKDef}, we find
\begin{eqnarray*}
& &\zeta_{\mc K_{L(2p,1)}}(s;i\nu_H)  = \frac1{\Gamma(s)} M[H_{\mc K_{L(2p,1)}}(t;i\nu_H)](s) \\
& = & \frac{\sqrt{\pi}\nu_H^{3-2s} \Gamma(s-\tfrac32)}{8p\Gamma(s)} + \frac{\pi^{s-1}}{2^{s-\frac12}p\Gamma(s)} 
\sum_{w=1}^\infty (-1)^w\left(\frac{w}{\nu_H}\right)^{s-\frac32} \left(K_{\frac32-s}\left(\pi w\nu_H\right)-\pi w\nu_H K_{\frac52-s}\left(\pi w\nu_H\right)\right) \\
& \qquad &  +\frac{\pi^s \nu_H^{\frac32-s}}{2^{s-\frac12}p^{s+\frac12}\Gamma(s)}\sum_{\substack{w=1\\p\nmid w}}^\infty \frac{w^{s-\frac12}}{\sin\frac{\pi w}{p}}  K_{\frac32-s}\left(\frac{\pi w\nu_H}{p}\right)\,,    
\end{eqnarray*}
which provides an $N$-independent meromorphic continuation for $s\in\mathbb C$, with simple poles at $s=\frac32-j$, $j\in\mathbb Z_{\geq 0}$ coming from the first term, and the sum over $w$ is entire in $s$.
This completes the proof. 

\end{proof}

\begin{remark}
    
The regularized determinant of the kinetic operator on $L(2p,1)$ in the heavy sector can be obtained by taking the derivative of the Bessel expansion of the spectral zeta function quoted in Proposition~\ref{prop:LensHeat}, which simply amounts to evaluating the Mellin transform of the Poisson resummed heat trace, analogous to the procedure outlined in Section~\ref{sec:3sphere}.

\end{remark}

\section{Even left quotients} \label{sec:EvenQuotients}

Let $\Gamma$ be an even subgroup of $\SU(2)$, that is $\abs{\Gamma}$ is even.
The center of $\SU(2)$ is a subgroup of $\Gamma$, i.e. $Z(\SU(2)) = \mathbb Z/2\mathbb Z \subseteq \Gamma$.
Consider the action of $\Gamma$, acting freely on $S^3$ via a left-action  $x\mapsto \gamma x, \gamma\in\Gamma,\,x\in S^3\cong\SU(2)$ (or equivalently a right-action), which defines the quotient $\Gamma\bs S^3$.
Recall that the $\Gamma$-invariant vector spaces corresponding to the irreducible representations $\pi$ of $\SU(2)$ are denoted by $\V_\pi^\Gamma$.

The kinetic operator on $\Gamma\bs S^3$ is the Laplace-Beltrami operator shifted by the mass of the scalar field,
\begin{equation}
\label{eq:KineticGamma}
\mc K_{\Gamma\bs S^3} = \Delta_{\Gamma\bs S^3} + \mf m^2\ell^2,
\end{equation}
whose eigenvalues are 
\begin{equation}
\label{eq:KineticGammaEigenVals}
n(n+2)+\mf m^2\ell^2 = (n+1)^2 - \nu^2\,, \qquad \nu=\sqrt{1-\mf m^2\ell^2}\,, \qquad \left(n\in\mathbb Z_{\geq0}\right)\,,
\end{equation}
where $\nu$ is the mass parameter defined in Def.~\ref{def:mass}.
The corresponding multiplicities are
\begin{equation}\label{eq:EvenMult1}
m_\Gamma(n) = (n+1)\dim(\V_n^\Gamma)\,.
\end{equation}

The following properties hold for every even $\Gamma\subset\SU(2)$, namely the cyclic, binary dihedral, tetrahedral, octahedral and icosahedral groups, which we explicitly compute and verify in each case.

\begin{remark}

\label{rmk:EvenVGammaDim}

\begin{enumerate}
\item  We initially carried out the computations of the spectral multiplicities, spectral zeta functions, and zeta-regularized determinants separately for each family of finite subgroups of $\SU(2)$. After completing these computations, we observed several common patterns in the resulting formulas, suggesting that the different cases could be treated within a unified framework. The purpose of this section is to exploit these observations to obtain general formulas for the meromorphic continuation of the spectral zeta function, the regularized determinant, and the heat trace for any even-order subgroup $\Gamma\subseteq\SU(2)$.

\medskip 

More precisely, we introduce the coefficients $\alpha$ and $\beta$ defined in Equation~\eqref{eq:AlphaBetaDef}, which encode the dependence on the subgroup $\Gamma$ through its spectral multiplicities. Once these coefficients are determined, the general formulas can be applied directly, without repeating the analytic computations for each individual subgroup. This approach is not necessary to obtain the results, since each case can be treated independently, but it considerably simplifies the calculations and provides a convenient and systematic way to recover the corresponding formulas for any even-order subgroup. The underlying method is essentially the one developed for the homogeneous lens spaces $L(2p,1)$ in Section~\ref{SectionLensSpace}, adapted to incorporate the coefficients $\alpha$ and $\beta$.

\medskip

\item For an even-order subgroup $\Gamma$, we have 
\begin{equation}
\dim(\V_n^\Gamma) = \begin{cases} 2\floor{\frac{n}{\abs{\Gamma}}} + \dim(\V^\Gamma_{n\bmod \abs{\Gamma}}) & \text{ if $n$ is even} \\ 0 & \text{ if $n$ is odd} \end{cases}
\end{equation}
This relationship will be verified for each family of even-order subgroups in the following sections.
\end{enumerate}

\end{remark}

We can therefore reparametrize $n$ as follows,
\begin{equation}
\label{eq:reparam}
n= k\abs{\Gamma} + 2r\,, \qquad \left(0\leq r\leq \frac12\abs{\Gamma}-1\,, k\in\mathbb Z_{\geq 0}\right)\,,
\end{equation}
which effectively relabels each even $n$ representation by two integers $k$ and $r$, where $k$ is the quotient $\floor{\frac{n}{\abs{\Gamma}}}$ and $r$ is the residue, $r =\frac12(n\bmod \abs{\Gamma})$, and therefore $\dim(\V_n^\Gamma) = 2k+\dim(\V_{2r}^\Gamma)$. 

\begin{definition}

It will be convenient to define the characteristic $\alpha_r$ and the twist coefficient $\beta_r$,
\begin{equation}
\label{eq:AlphaBetaDef}
\alpha_r = \frac{2r+1}{\abs{\Gamma}}\,, \qquad \beta_r = \frac12 \dim(\V_{2r}^\Gamma) - \alpha_r\,. 
\end{equation}

\end{definition}

\begin{remark}
\label{claim:AlphaBetaParity}

The parity properties of the characteristic and twist coefficient are as follows
\begin{equation}
\label{eq:AlphaBetaParity1}
\alpha_{\frac12\abs{\Gamma}-1-r} = 1-\alpha_r\,, \qquad \beta_{\frac12\abs{\Gamma}-1-r} = -\beta_r\,,
\end{equation}
using which, any sum of the form $\sum_{r=0}^{\frac12\abs{\Gamma}-1}f(\alpha_r)$ or $\sum_{r=0}^{\frac12\abs{\Gamma}-1} \beta_r f(\alpha_r)$ can be re-expressed as follows
\begin{equation}
\label{eq:AlphaBetaParity2}
\begin{gathered}
\sum_{r=0}^{\frac12\abs{\Gamma}-1} f(\alpha_r) = \sum_{r=0}^{\frac12\abs{\Gamma}-1}f(1-\alpha_r) = \frac12\sum_{r=0}^{\frac12\abs{\Gamma}-1}\big(f(\alpha_r)+f(1-\alpha_r)\big),\\
\sum_{r=0}^{\frac12\abs{\Gamma}-1} \beta_r f(\alpha_r) = -\sum_{r=0}^{\frac12\abs{\Gamma}-1} \beta_r f(1-\alpha_r) = \frac12\sum_{r=0}^{\frac12\abs{\Gamma}-1} \beta_r \big(f(\alpha_r)-f(1-\alpha_r)\big).
\end{gathered}
\end{equation}
We will demonstrate this property holds in each of the examples we work out below.
The data for all even quotients are listed in Table \ref{tab:EvenGammaDims}.

\end{remark}

\begin{table}
\centering
\begin{tabular*}{0.7\textwidth}{@{\extracolsep{\fill}}cccc@{}}
\toprule
$\Gamma$  & $\abs{\Gamma}$ & $\dim(\V_{2r}^\Gamma)$ & $\mathscr R_{\Gamma}\vphantom{\Big|}$ \\
\midrule
$\mathbb Z/2p\mathbb Z$ & $2p$ & 1 & --$\vphantom{\Big|}$ \\[5mm]
$D_m$                   & $4m$ & $\floor{\dfrac rm}+\delta_{r\bmod 2,0}$ & -- \\[5mm]
$\widetilde{\mc T}$     & 24   &  $\floor{\dfrac r6}+\varepsilon_{2r\bmod 12}$ & $\{0,6,8\}$ \\[5mm]
$\widetilde{\mc O}$     & 48   & $\floor{\dfrac r{12}}+\varepsilon_{2r\bmod 24}$ & $\{0,8,12,16,18,20\}$ \\[5mm]
$\widetilde{\mc I}$     & 120  & $\floor{\dfrac r{30}}+\varepsilon_{2r\bmod 60}$ & Equation~\eqref{eq:Ricosahedral}\\ \bottomrule
\end{tabular*}\medskip
\caption{The order and dimensions of the invariant subspaces of the even subgroups of $\SU(2)$.
This table is adapted from \cite{MR1317006,Lehoucq:2002wy}, which we reproduce using the methods discussed in this paper.
The parameter $r$ is defined in Equation~\eqref{eq:reparam}.
Note that $\varepsilon_n = 1$ if and only if $n\in\mathscr R_\Gamma$, and 0 otherwise.
In particular, $\mathscr R_\Gamma$ is the complement of the excluded $\bar k$ values in Table 1 of \cite{Lehoucq:2002wy}, where $n=2\bar k <\frac12 \abs{\Gamma}$.}
\label{tab:EvenGammaDims}
\end{table}

The multiplicities \eqref{eq:EvenMult1} can be re-expressed using the characteristic and twist coefficient,
\begin{equation}
\label{eq:EvenMult2}
\begin{split}
n+1 &= k\abs{\Gamma}+2r +1 = \abs{\Gamma}(k+\alpha_r) \\
\dim(\V_n^\Gamma) &= 2k+\dim(\V^\Gamma_{2r}) = 2(k+\alpha_r+\beta_r) \\
\implies m_\Gamma(n) &= 2\abs{\Gamma}(k+\alpha_r)(k+\alpha_r+\beta_r) = 2\abs{\Gamma}\big( (k+\alpha_r)^2 + \beta_r (k+\alpha_r) \big)
\end{split}
\end{equation}

\begin{proposition} 
\label{prop:ZetaGamma}
    
The spectral zeta function for the kinetic operator on $\Gamma\bs S^3$ \eqref{eq:KineticGamma} in terms of the spectral data and characteristic and twist coefficients \eqref{eq:AlphaBetaDef} is, for $|\Re(\nu)|<1$ and $\Re(s)>\frac32$,
\begin{equation}
\label{eq:ZetaGamma1}
\zeta_{\mc K_{\Gamma\bs S^3}}(s;\nu) = \sum_{n=0}^\infty \frac{m_\Gamma(n)}{((n+1)^2-\nu^2)^s}
= \sum_{r=0}^{\frac12\abs{\Gamma}-1}\sum_{k=0}^{\infty} \frac{2\abs{\Gamma}(k+\alpha_r)(k+\alpha_r +\beta_r)}{\left(\abs{\Gamma}^2(k+\alpha_r)^2-\nu^2\right)^s},
\end{equation}
which admits the following meromorphic continuation, where $N$ is chosen such that $N\abs{\Gamma} + 1 > \abs{\nu}$,
\begin{equation}
\label{eq:ZetaGamma2}
\begin{split}
&\zeta_{\mc K_{\Gamma\bs S^3}}(s;\nu) \\
&= \sum_{r=0}^{\frac12\abs{\Gamma}-1}\sum_{k=0}^{N-1} \frac{2\abs{\Gamma}(k+\alpha_r)(k+\alpha_r +\beta_r)}{\left(\abs{\Gamma}^2(k+\alpha_r)^2-\nu^2\right)^s} \\&\quad
+ 2\abs{\Gamma}^{1-2s}\sum_{r=0}^{\frac12\abs{\Gamma}-1}\sum_{j=0}^{\infty}\frac{(s)_j}{j!}\left(\frac{\nu}{\abs{\Gamma}}\right)^{2j}\Big[\zeta_H(2s+2j-2,\alpha_r+N) + \beta_r \zeta_H(2s+2j-1,\alpha_r+N)\Big],
\end{split}
\end{equation}
with at most simple poles at $s=\frac32-j$, $j\in\mathbb Z_{\geq 0}$. 
In particular, the extension is regular at $s=0$.

\end{proposition}

\begin{proof}

To obtain the meromorphic continuation of the spectral zeta function, we employ binomial expansion when possible, analogous to the 3-sphere proof.    
Note that since $0\leq r\leq \frac12 \abs{\Gamma}-1$, the characteristic $\alpha_r = \frac{2r+1}{\abs{\Gamma}}$ \eqref{eq:AlphaBetaDef} has the range $\frac1{\abs{\Gamma}}\leq \alpha_r \leq 1-\frac1{\abs{\Gamma}}$. 
The binomial expansion we want to perform is the denominator of each summand,
\begin{eqnarray*}
\left(\abs{\Gamma}^2(k+\alpha_r)^2 - \nu^2 \right)^{-s} & = & \left(\abs{\Gamma}(k+\alpha_r)\right)^{-2s}\left(1-\left(\frac{\nu}{\abs{\Gamma}(k+\alpha_r)}\right)^2 \right)^{-s} \\
& = & \left(\abs{\Gamma}(k+\alpha_r)\right)^{-2s} \sum_{j=0}^{\infty}\frac{(s)_j}{j!}\left(\frac{\nu}{\abs{\Gamma}(k+\alpha_r)}\right)^{2j},
\end{eqnarray*}
which is valid only for $\abs{\nu} < \abs{\Gamma}(k+\alpha_r)$. 
Since $\alpha_r \geq \frac1{\abs{\Gamma}}$, it is sufficient to have $\abs\nu < \abs{\Gamma}\left(k+\frac1{\abs{\Gamma}}\right) = \abs{\Gamma} k + 1$. Therefore, for $N\in\mathbb Z_{\geq 0}$ such that $\abs{\nu} < \abs{\Gamma}N+1$, we can perform the binomial expansion for any $k\geq N$, so we have for $\Re(s)>\frac32$,
\[\begin{split}
\zeta_{\mc K_{\Gamma\bs S^3}}(s;\nu)  &= \sum_{r=0}^{\frac12\abs{\Gamma}-1}\sum_{k=0}^{N-1} \frac{2\abs{\Gamma}(k+\alpha_r)(k+\alpha_r +\beta_r)}{\left(\abs{\Gamma}^2(k+\alpha_r)^2-\nu^2\right)^s} + \sum_{r=0}^{\frac12\abs{\Gamma}-1}\sum_{k=N}^{\infty} \frac{2\abs{\Gamma}(k+\alpha_r)(k+\alpha_r +\beta_r)}{\left(\abs{\Gamma}^2(k+\alpha_r)^2-\nu^2\right)^s} \\
&= \sum_{r=0}^{\frac12\abs{\Gamma}-1}\sum_{k=0}^{N-1} \frac{2\abs{\Gamma}(k+\alpha_r)(k+\alpha_r +\beta_r)}{\left(\abs{\Gamma}^2(k+\alpha_r)^2-\nu^2\right)^s} \\&\quad
+2\abs{\Gamma}^{1-2s}\sum_{r=0}^{\frac12\abs{\Gamma}-1}\sum_{k=N}^{\infty}\sum_{j=0}^\infty \frac{(s)_j}{j!}\left(\frac{\nu}{\abs{\Gamma}}\right)^{2j} \bigg((k+\alpha_r)^{2-2s-2j}+\beta_r(k+\alpha_r)^{1-2s-2j}\bigg)\\
&= \sum_{r=0}^{\frac12\abs{\Gamma}-1}\sum_{k=0}^{N-1} \frac{2\abs{\Gamma}(k+\alpha_r)(k+\alpha_r +\beta_r)}{\left(\abs{\Gamma}^2(k+\alpha_r)^2-\nu^2\right)^s} \\&\quad
+2\abs{\Gamma}^{1-2s}\sum_{r=0}^{\frac12\abs{\Gamma}-1}\sum_{j=0}^\infty \frac{(s)_j}{j!}\left(\frac{\nu}{\abs{\Gamma}}\right)^{2j} \bigg[\zeta_H(2s+2j-2,\alpha_r+N)+\beta_r\zeta_H(2s+2j-1,\alpha_r+N)\bigg],
\end{split}\]
where we swap the $k$ and $j$ sums since the sums absolutely converge for $\Re(s)>3/2$ due to our choice of $N$. 
This recovers the meromorphic extension.

The finite sum is entire in $s$, so the poles of the meromorphic continuation of the spectral zeta function come from the infinite sum over Hurwitz zetas. The ``untwisted'' contribution, 
\begin{equation*}
2\abs{\Gamma}^{1-2s}\sum_{r=0}^{\frac12\abs{\Gamma}-1}\sum_{j=0}^\infty \frac{(s)_j}{j!}\left(\frac{\nu}{\abs{\Gamma}}\right)^{2j} \zeta_H(2s+2j-2,\alpha_r+N)
\end{equation*}
provides possible simple poles at $s=\frac32-j$, $j\in\mathbb Z_{\geq0}$. The ``twisted'' contribution,
\begin{equation*}
\sum_{r=0}^{\frac12\abs{\Gamma}-1}\sum_{j=0}^\infty \frac{(s)_j}{j!}\left(\frac{\nu}{\abs{\Gamma}}\right)^{2j} \beta_r \zeta_H(2s+2j-1,\alpha_r+N) \sim \sum_{j=0}^\infty \frac{(s)_j}{j!}\left(\frac{\nu}{\abs{\Gamma}}\right)^{2j} \left(\frac{1}{2(s+j-1)}\right)\sum_{r=0}^{\frac12\abs{\Gamma}-1} \beta_r = 0,
\end{equation*}
does not provide any poles due to the parity properties of $\beta_r$    \eqref{eq:AlphaBetaParity2}, i.e. $\sum_{r=0}^{\frac12\abs{\Gamma}-1}\beta_r = 0$. 

\end{proof}
    
\begin{corollary} 
\label{cor:ZetaGamma}

For the light sector, $\abs{\nu}<1$ so $N=0$ suffices and the finite sum in \eqref{eq:ZetaGamma2} is absent.  The untwisted sum simplifies as well, to obtain
\begin{equation}
\label{eq:ZetaGammaLight}
\zeta_{\mc K_{\Gamma\bs S^3}}(s;\nu) = 2\sum_{j=0}^{\infty}\frac{(s)_j}{j!}\nu^{2j}\left[\frac{1-2^{2-2s-2j}}{\abs{\Gamma}}\zeta_R(2s+2j-2) + \abs{\Gamma}^{1-2s-2j}\sum_{r=0}^{\frac12\abs{\Gamma}-1}\beta_r \zeta_H(2s+2j-1,\alpha_r)\right].
\end{equation}
    
At the critical point $\nu=0$, so with $N=0$ only the $j=0$ term survives, 
\begin{equation}
\label{eq:ZetaGammaCrit}
\zeta_{\mc K_{\Gamma\bs S^3}}(s;0) = 2\frac{1-2^{2-2s}}{\abs{\Gamma}} \zeta_R(2s-2) +2\abs{\Gamma}^{1-2s} \sum_{r=0}^{\frac12\abs{\Gamma}-1}\beta_r \zeta_H(2s-1,\alpha_r).
\end{equation}
    
For the massless case, $\nu=1$ the spectral zeta function is defined by omitting the zero mode ($n=0$ i.e. $k=r=0$) from \eqref{eq:ZetaGamma1},
\begin{equation}
\begin{split}
\zeta_{\mc K_{\Gamma\bs S^3}}(s;1) 
&= \sum_{n=1}^\infty \frac{m_\Gamma(n)}{((n+1)^2-1)^s}\\
&= 2^{1-2s}\sum_{r=1}^{\frac12\abs{\Gamma}-1} \frac{(2r+1)(\alpha_r +\beta_r)}{(r(r+1))^s}+\sum_{r=0}^{\frac12\abs{\Gamma}-1}\sum_{k=1}^{\infty} \frac{2\abs{\Gamma}(k+\alpha_r)(k+\alpha_r +\beta_r)}{\left(\abs{\Gamma}^2(k+\alpha_r)^2-1\right)^s},
\end{split}
\end{equation}
and admits the meromorphic continuation
\begin{equation}
\label{eq:ZetaGammaMassless}
\begin{split}
\zeta_{\mc K_{\Gamma\bs S^3}}(s;1) 
&= 2^{1-2s}\sum_{r=1}^{\frac12\abs{\Gamma}-1}\frac{(2r+1)(\alpha_r +\beta_r)}{(r(r+1))^s} \\&\quad
+2\abs{\Gamma}^{1-2s}\sum_{r=0}^{\frac12\abs{\Gamma}-1}\sum_{j=0}^{\infty}\frac{(s)_j}{j!} \abs{\Gamma}^{-2j}\Big[\zeta_H(2s+2j-2,1+\alpha_r) + \beta_r \zeta_H(2s+2j-1,1+\alpha_r)\Big].
\end{split}
\end{equation}

\end{corollary}

\begin{proof}

Using $\sum\limits_{r=0}^{p-1} \sum\limits_{n=0}^\infty f(pn+r) = \sum\limits_{m=0}^\infty f(m), \, m=pn+r$, one can show
\begin{equation*}
\sum_{r=0}^{p-1} \zeta_{H}\left(s,\frac{r+q}p\right) = p^{s}\zeta_H(s,q)\,.
\end{equation*}
The light sector expression can be obtained by direct substitution of $N=0$ and the above identity, similarly the critical point can be obtained by substituting $\nu=0$ in the light sector expression. 

The massless case requires removal of the zero mode ($k=r=0$) from \eqref{eq:ZetaGamma1} and then substituting $N=1$ and the definition of the characteristic \eqref{eq:AlphaBetaDef} in the $k=0$ sum.
To obtain the meromorphic continuation the rest of the calculation is identical to the general case.

\end{proof}

\begin{theo} 
\label{thm:DetGamma}

The regularized determinant of the kinetic operator \eqref{eq:KineticGamma} on $\Gamma\bs S^3$ for $\abs{\Re(\nu)}<1$ is 
\begin{equation}
\begin{split}
&-\log\det{}'(\mc K_{\Gamma\bs S^3}(\mf m)) = \pd_s \zeta_{\mc K_{\Gamma\bs S^3}}(s;\nu)\big|_{s=0} \\
&= \frac{\pi i}{3\abs{\Gamma}}\nu^3 -\frac{4}{\pi^2\abs{\Gamma}}\sum_{q=0}^2 \frac{(-\pi i\nu)^q}{q!} \Li_{3-q}\left(-e^{\pi i\nu}\right) - \sum_{r=0}^{\frac12\abs{\Gamma}-1}\beta_r\left[
2 \nu \Li_1\left(e^{\frac{2\pi i\nu}{\abs{\Gamma}}+2\pi i\alpha_r}\right)+\frac{i\abs{\Gamma}}{\pi} \Li_2\left(e^{\frac{2\pi i\nu}{\abs{\Gamma}}+2\pi i\alpha_r}\right)\right].
\end{split}
\end{equation}
The regularized determinant of the kinetic operator \eqref{eq:KineticGamma} on $\Gamma\bs S^3$ for $\nu=1$ or $\mf m = 0$ is 
\begin{equation}
\begin{split}
-\log\det{}'(\mc K_{\Gamma\bs S^3}(0)) &= \pd_s \zeta_{\mc K_{\Gamma\bs S^3}}(s;1)\big|_{s=0} \\
&= \frac{\pi i}{12}\abs{\Gamma} - \log\pi - \frac{2\pi i}3 + \frac1{\abs{\Gamma}} \left(\pi i + \log4 - \frac{4}{\pi^2}\zeta_R(3)\right) + \left(1- \frac{2}{\abs{\Gamma}}\right)\log\abs{\Gamma}  \\&\quad
- \sum_{r=0}^{\frac12\abs{\Gamma}-2}\beta_r\left[
2\Li_1\left(e^{\frac{2\pi i}{\abs{\Gamma}}+2\pi i\alpha_r}\right)+\frac{i\abs{\Gamma}}{\pi}\Li_2\left(e^{\frac{2\pi i}{\abs{\Gamma}}+2\pi i\alpha_r}\right)\right].
\end{split}
\end{equation}

\end{theo}

\begin{proof}

We differentiate \eqref{eq:ZetaGamma2} at $s=0$ where it is regular and show that it is independent of the auxiliary integer $N$.
First we isolate the $j=0$ and $j=1$ terms,
\begin{eqnarray*}
& & \pd_s \zeta_{\mc K_{\Gamma\bs S^3}}(s;\nu)\big|_{s=0}\\
& = & -\sum_{r=0}^{\frac12\abs{\Gamma}-1}\sum_{k=0}^{N-1} 2\abs{\Gamma}\big((k+\alpha_r)^2 + \beta_r(k+\alpha_r)\big)\log\left(\abs{\Gamma}^2(k+\alpha_r)^2-\nu^2\right) \\
& \quad & - 4 \abs{\Gamma}\log\abs{\Gamma}\sum_{r=0}^{\frac12\abs{\Gamma}-1}\Big[\zeta_H(-2,\alpha_r+N) + \beta_r \zeta_H(-1,\alpha_r+N)\Big]\\
& \quad & + 4 \abs{\Gamma}\sum_{r=0}^{\frac12\abs{\Gamma}-1}\Big[\zeta_H'(-2,\alpha_r+N) + \beta_r \zeta_H'(-1,\alpha_r+N)\Big]\\
& \quad & +2\abs{\Gamma}\sum_{r=0}^{\frac12\abs{\Gamma}-1} \frac{\nu^2}{\abs{\Gamma}^2}\Big(\zeta_H(0,\alpha_r+N)+\frac12\beta_r\lim_{s\to 0}\big(\zeta_H(2s+1,\alpha_r+N)-\zeta_H(2s+1,1-\alpha_r+N)\big)\Big) \\
& \quad & + 2\abs{\Gamma}\sum_{r=0}^{\frac12\abs{\Gamma}-1}\sum_{j=2}^{\infty}\frac1j \left(\frac{\nu}{\abs{\Gamma}}\right)^{2j}\Big[\zeta_H(2j-2,\alpha_r+N) + \beta_r \zeta_H(2j-1,\alpha_r+N)\Big],
\end{eqnarray*}
where in the $j=1$ term we used the parity in $\beta_r$ to justify taking the derivative at $s=0$ since the poles of the two Hurwitz zetas evaluated at $1$ cancel with each other.
    
To treat the $j=0$ terms, we use
\begin{equation*}
\zeta_H(s,\alpha_r + N) = \zeta_H(s,\alpha_r) - \sum_{k=0}^{N-1} (k+\alpha_r)^{-s}\, \quad  \zeta_H'(s,\alpha_r + N) = \zeta_H'(s,\alpha_r) + \sum_{k=0}^{N-1} (k+\alpha_r)^{-s}\log(k+\alpha_r), 
\end{equation*}
to obtain
\[\begin{split}
&\zeta_H'(-2,\alpha_r + N) + \beta_r \zeta_H'(-1,\alpha_r + N)-\log\abs{\Gamma}\left(\zeta_H(-2,\alpha_r + N) + \beta_r \zeta_H(-1,\alpha_r + N)\right) \\
&= \zeta_H'(-2,\alpha_r ) + \beta_r \zeta_H'(-1,\alpha_r)-\log\abs{\Gamma}\left(\zeta_H(-2,\alpha_r) + \beta_r \zeta_H(-1,\alpha_r)\right) \\&\qquad 
+\sum_{k=0}^{N-1}\big((k+\alpha_r)^2 +\beta_r(k+\alpha_r)\big)\log\big(\abs{\Gamma}(k+\alpha_r)\big).
\end{split}
\]
Summing over $r$, we can use parity \eqref{eq:AlphaBetaParity2} and Hurwitz's formula \eqref{eq:HurwitzFourier} to simplify the Hurwitz zeta terms further.
From \eqref{eq:HurwitzFourier}, the even and odd combinations simplify to 
\[\begin{split}
\zeta_H(s,a) + \zeta_H(s,1-a) &= 4(2\pi)^{s-1}\Gamma(1-s)\sin\left(\frac{\pi s}2\right)\sum_{w=1}^\infty w^{s-1}\cos(2\pi w a), \qquad \left(\Re(s)<1,\; 0<a<1\right), \\
\zeta_H(s,a) - \zeta_H(s,1-a) &= 4(2\pi)^{s-1}\Gamma(1-s)\cos\left(\frac{\pi s}2\right)\sum_{w=1}^\infty w^{s-1}\sin(2\pi w a), \qquad \left(\Re(s)<1,\; 0<a<1\right), \\            
\end{split}
\]
In particular, the even combination vanishes when $s=-2$, and the odd combination vanishes for $s=-1$. 
Their $s$ derivatives at $-2$ and $-1$ are, respectively,
\[
\begin{split}
\zeta_H'(-2,a) + \zeta_H'(-2,1-a) &= -\sum_{w=1}^\infty \frac{\cos(2\pi w a)}{2\pi^2 w^3} = -\frac1{4\pi^2}\big(\Li_3(e^{2\pi i a}) + \Li_3(e^{-2\pi i a})\big),  \\
\zeta_H'(-1,a) - \zeta_H'(-1,1-a) &= \sum_{w=1}^\infty \frac{\sin(2\pi w a)}{2\pi w^2} = -\frac{i}{4\pi}\big(\Li_2(e^{2\pi i a})-\Li_2(e^{-2\pi i a})\big).
\end{split}
\]
Using the above identities and parity \eqref{eq:AlphaBetaParity2}, the following terms simplify 
\[\begin{split}
&\sum_{r=0}^{\frac12\abs{\Gamma}-1}\Big[\zeta_H'(-2,\alpha_r ) + \beta_r \zeta_H'(-1,\alpha_r)-\log\abs{\Gamma}\left(\zeta_H(-2,\alpha_r) + \beta_r \zeta_H(-1,\alpha_r)\right)\Big]\\
&= \sum_{r=0}^{\frac12\abs{\Gamma}-1}\Big[-\frac1{4\pi^2}\Li_3(e^{2\pi i \alpha_r})- \frac{i}{4\pi}\beta_r \Li_2(e^{2\pi i \alpha_r})\Big].
\end{split}
\] 
Finally, we can further perform the sum over $r$ for the $\Li_3$ term by using the definition of $\alpha_r$ \eqref{eq:AlphaBetaDef},
\[
\begin{split}
-\frac1{4\pi^2}\sum_{r=0}^{\frac12\abs{\Gamma}-1} \Li_3(e^{2\pi i\alpha_r})
&= -\frac1{4\pi^2}\sum_{w=1}^\infty \frac{e^{\frac{2\pi i w}{\abs{\Gamma}}}}{w^3} \sum_{r=0}^{\frac12\abs{\Gamma}-1} \left(e^{4\pi i w/\abs{\Gamma}}\right)^r \\ 
&= -\frac{\abs\Gamma}{8\pi^2}\sum_{w=1}^\infty \frac{e^{\frac{2\pi i w}{\abs{\Gamma}}}}{w^3} \delta_{w\bmod\frac{\abs{\Gamma}}2,0} 
= -\frac{\abs\Gamma}{8\pi^2}\sum_{m=1}^\infty \frac{(-1)^m}{\left(\frac{\abs{\Gamma}}{2}m\right)^3} = \frac{3}{4\pi^2\abs{\Gamma}^2}\zeta_R(3).
\end{split}
\]
        
The $j=1$ term simplifies using the analytic continuation of the Hurwitz zeta function to perform the derivative at $s=0$,
\[\begin{split}
&\zeta_H(0,\alpha_r+N)+\frac12\beta_r\lim_{s\to 0}\big(\zeta_H(2s+1,\alpha_r+N)-\zeta_H(2s+1,1-\alpha_r+N)\big) \\
&= \frac12 -\alpha_r-N - \frac12\beta_r\big(\psi(\alpha_r+N) -\psi(1-\alpha_r+N)\big),
\end{split}\]
where $\psi$ is the digamma function (cf. Equation~\eqref{eq:DigammaDef}).
Using the parity in $\alpha_r$ and $\beta_r$ \eqref{eq:AlphaBetaParity2} and the following identity which follows from the definition of the digamma function
\[\psi(\alpha_r+N) = \psi(\alpha_r) + \sum_{k=0}^{N-1}\frac1{k+\alpha_r},\quad N\in\mathbb Z_{\geq0},\]
we can simplify the $j=1$ terms further.
We find
\[\begin{split}
\sum_{r=0}^{\frac12\abs{\Gamma}-1}&\left(\frac12 -\alpha_r-N - \frac12\beta_r\big(\psi(\alpha_r+N) -\psi(1-\alpha_r+N)\big)\right) \\
&= -\sum_{r=0}^{\frac12\abs{\Gamma}-1}\left(N+\beta_r\psi(\alpha_r+N)\right)\\
&= \sum_{r=0}^{\frac12\abs{\Gamma}-1}\left(\frac\pi2\beta_r\cot(\pi\alpha_r)-\sum_{k=0}^{N-1}\left((k+\alpha_r)^2+\beta_r(k+\alpha_r)\right)\frac1{(k+\alpha_r)^2}\right).
\end{split}\]
        
The sum over $j\geq 2$ and $k\geq N$ converges absolutely since $\abs{\nu}<\abs{\Gamma}(k+\alpha_r)$ for $k\geq N$, so we can use $\log(1-x) = -\sum_{j=1}^\infty \frac1j x^j,\; |x|<1$, to obtain
\[\begin{split}
&\sum_{j=2}^{\infty}\frac1j \left(\frac{\nu}{\abs{\Gamma}}\right)^{2j}\Big[\zeta_H(2j-2,\alpha_r+N) + \beta_r \zeta_H(2j-1,\alpha_r+N)\Big] \\
&= \sum_{k=N}^\infty \big((k+\alpha_r)^2 + \beta_r(k+\alpha_r)\big)\sum_{j=2}^\infty \frac1j\left(\frac{\nu}{\abs{\Gamma}(k+\alpha_r)}\right)^{2j} \\
&= -\sum_{k=N}^\infty \big((k+\alpha_r)^2 + \beta_r(k+\alpha_r)\big)\left[\frac{\nu^2}{\abs{\Gamma}^2(k+\alpha_r)^2}+\log\left(1-\frac{\nu^2}{\abs{\Gamma}^2(k+\alpha_r)^2}\right)\right].
\end{split}\]
    
Putting it all together, we have
\begin{eqnarray*}
\pd_s \zeta_{\mc K_{\Gamma\bs S^3}}(s;\nu)\big|_{s=0}
& = & 2\abs{\Gamma} \sum_{r=0}^{\frac12\abs{\Gamma}-1}\Bigg[
-\sum_{k=0}^{N-1} \big((k+\alpha_r)^2+\beta_r(k+\alpha_r)\big)\log\Big(\abs{\Gamma}^2(k+\alpha_r)^2 - \nu^2\Big) \\
& + & \sum_{k=0}^{N-1}\big((k+\alpha_r)^2 +\beta_r(k+\alpha_r)\big)\log\big(\abs{\Gamma}^2(k+\alpha_r)^2\big) \\
& - & \frac1{2\pi^2}\Li_3(e^{2\pi i \alpha_r})- \frac{i}{2\pi}\beta_r \Li_2(e^{2\pi i \alpha_r}) \\
& + & \frac{\nu^2}{\abs{\Gamma}^2}\left(\frac\pi2 \beta_r \cot(\pi\alpha_r) - \sum_{k=0}^{N-1}\big((k+\alpha_r)^2+\beta_r(k+\alpha_r)\big)\frac1{(k+\alpha_r)^2}\right)\\
& - & \sum_{k=N}^\infty \big((k+\alpha_r)^2 + \beta_r(k+\alpha_r)\big)\left[\frac{\nu^2}{\abs{\Gamma}^2(k+\alpha_r)^2}+\log\left(1-\frac{\nu^2}{\abs{\Gamma}^2(k+\alpha_r)^2}\right)\right]\Bigg],
\end{eqnarray*}
which simplifies to
\[\begin{split}
&\pd_s \zeta_{\mc K_{\Gamma\bs S^3}}(s;\nu)\big|_{s=0}\\
&= 2\abs{\Gamma} \sum_{r=0}^{\frac12\abs{\Gamma}-1}\Bigg[
-\sum_{k=0}^\infty \big((k+\alpha_r)^2 + \beta_r(k+\alpha_r)\big)\left[\frac{\nu^2}{\abs{\Gamma}^2(k+\alpha_r)^2}+\log\left(1-\frac{\nu^2}{\abs{\Gamma}^2(k+\alpha_r)^2}\right)\right]\\&\quad
-\frac1{2\pi^2}\Li_3(e^{2\pi i \alpha_r})- \frac{i}{2\pi}\beta_r \Li_2(e^{2\pi i \alpha_r}) 
+\frac{\pi\nu^2}{2\abs{\Gamma}^2}\beta_r \cot(\pi\alpha_r)\Bigg], 
\end{split}        
\]
which is independent of our choice of the auxiliary integer $N$. 

Now we evaluate the sum over $k$. Let $F(\nu)$ be the sum
\begin{equation*}
F(\nu) = -\sum_{r=0}^{\frac12\abs{\Gamma}-1}\sum_{k=0}^\infty \big((k+\alpha_r)^2 + \beta_r(k+\alpha_r)\big)\left[\frac{\nu^2}{\abs{\Gamma}^2(k+\alpha_r)^2}+\log\left(1-\frac{\nu^2}{\abs{\Gamma}^2(k+\alpha_r)^2}\right)\right].
\end{equation*}
Note that $F(0) = 0$.  The derivative of $F$ with respect to $\nu$ reads
\begin{eqnarray*}    
F'(\nu) & = & -\sum_{r=0}^{\frac12\abs{\Gamma}-1}\sum_{k=0}^\infty \big((k+\alpha_r)^2 + \beta_r (k+\alpha_r)\big)\left[\frac{2\nu}{\abs{\Gamma}^2(k+\alpha_r)^2}+\frac{2\nu}{\nu^2 - \abs{\Gamma}^2(k+\alpha_r)^2}\right]\\
& = & \frac{2\nu^3}{\abs{\Gamma}^2}\sum_{r=0}^{\frac12\abs{\Gamma}-1} \sum_{k=0}^\infty \frac{1+\beta_r/(k+\alpha_r)}{\abs{\Gamma}^2(k+\alpha_r)^2-\nu^2}.
\end{eqnarray*}
Now we evaluate the sum over $r$ for the first term, using $n+1=\abs{\Gamma}(k+\alpha_r)$ \eqref{eq:EvenMult2} where $n$ is even, and using $\sum_{u=0}^\infty\frac1{(2u+1)^2-z^2}= \frac{\pi}{4z}\tan\frac{\pi z}{2}$. 
For the second term, we use a partial fraction decomposition, the parity properties \eqref{eq:AlphaBetaParity2}, and the digamma function \eqref{eq:DigammaDef}, along with the identity $\psi(z)-\psi(1-z) =-\pi\cot(\pi z)$.
We find
\begin{eqnarray*}
F'(\nu) & = & \frac{2\nu^3}{\abs{\Gamma}^2} \sum_{n=0,\text{ even}}^\infty \frac1{(n+1)^2-\nu^2} + \frac{2\nu^3}{\abs{\Gamma}^2}\sum_{r=0}^{\frac12\abs{\Gamma}-1}\sum_{k=0}^\infty \frac{\beta_r}{(k+\alpha_r)}\left(\frac1{\abs{\Gamma}^2(k+\alpha_r)^2-\nu^2}\right) \\
& = & \frac{\pi\nu^2}{2\abs{\Gamma}^2}\tan\frac{\pi\nu}2+\frac{\nu}{\abs{\Gamma}} \sum_{r=0}^{\frac12\abs{\Gamma}-1}\sum_{k=0}^\infty \beta_r \left(\frac1{\abs{\Gamma}(k+\alpha_r)+\nu}+\frac1{\abs{\Gamma}(k+\alpha_r)-\nu}-\frac2{\abs{\Gamma}(k+\alpha_r)}\right) \\
& = & \frac{\pi\nu^2}{2\abs{\Gamma}^2}\tan\frac{\pi\nu}2+\frac{\nu}{\abs{\Gamma}^2} \sum_{r=0}^{\frac12\abs{\Gamma}-1} \beta_r \left(2\psi(\alpha_r)-\psi\left(\alpha_r+\frac\nu{\abs{\Gamma}}\right)-\psi\left(\alpha_r-\frac\nu{\abs{\Gamma}}\right)\right) \\
& = & \frac{\pi\nu^2}{2\abs{\Gamma}^2}\tan\frac{\pi\nu}2+\frac{\pi\nu}{2\abs{\Gamma}^2} \sum_{r=0}^{\frac12\abs{\Gamma}-1} \beta_r \left(\cot\left(\frac{\pi\nu}{\abs{\Gamma}}+\pi\alpha_r\right)-\cot\left(\frac{\pi\nu}{\abs{\Gamma}}-\pi\alpha_r\right)-2\cot(\pi\alpha_r)\right) \\
& = & \frac{\pi\nu^2}{2\abs{\Gamma}^2}\tan\frac{\pi\nu}2+\frac{\pi\nu}{\abs{\Gamma}^2} \sum_{r=0}^{\frac12\abs{\Gamma}-1} \beta_r \left(\cot\left(\frac{\pi\nu}{\abs{\Gamma}}+\pi\alpha_r\right)-\cot(\pi\alpha_r)\right).
\end{eqnarray*}
Since $F(0) = 0$, $F(\nu) = \int_0^\nu F'(t) dt$. Note the following Fourier expansions
\begin{equation*}
\tan(\pi z) = i\left(1+2\sum_{w=1}^\infty (-1)^w e^{2\pi i w z}\right), \qquad \cot(\pi z) = -i\left(1+2\sum_{w=1}^\infty e^{2\pi i w z}\right), \qquad \left(\Im(z) > 0\right). 
\end{equation*}
Since $F'(\nu)$ contains no poles in the neighborhood of $[0,\nu]$, we shift the integration contour by $i\epsilon$ to justify the usage of the Fourier expansions and swaps between the sum and integral. We also use the parity property of $\beta_r$ \eqref{eq:AlphaBetaParity2} to eliminate any terms which are odd under the parity of $r$.
We find,
\begin{eqnarray*}
F(\nu) & = & \lim_{\epsilon\to0}\int_{0+i\epsilon}^{\nu+i\epsilon} \left[\frac{\pi t^2}{2\abs{\Gamma}^2}\tan\frac{\pi t}2+\frac{\pi t}{\abs{\Gamma}^2} \sum_{r=0}^{\frac12\abs{\Gamma}-1} \beta_r \left(\cot\left(\frac{\pi t}{\abs{\Gamma}}+\pi\alpha_r\right)-\cot(\pi\alpha_r)\right)\right]dt \\
& = & \lim_{\epsilon\to0}\int_{0+i\epsilon}^{\nu+i\epsilon} \Bigg[\frac{\pi i}{2\abs{\Gamma}^2} \left(t^2+2\sum_{w=1}^\infty (-1)^w t^2 e^{\pi i w t}\right) - \frac{2\pi i}{\abs{\Gamma}^2}\sum_{r=0}^{\frac12\abs{\Gamma}-1}\beta_r\sum_{w=1}^\infty t e^{2\pi i w \left(\frac{t}{\abs{\Gamma}} + \alpha_r\right)} \Bigg]dt \\ 
& \quad &  - \frac{\pi\nu^2}{2\abs{\Gamma}^2}\sum_{r=0}^{\frac12\abs{\Gamma}-1}\beta_r \cot(\pi\alpha_r) \\
& = & \frac{\pi i}{6\abs{\Gamma}^2} \nu^3 - \frac{\pi\nu^2}{2\abs{\Gamma}^2}\sum_{r=0}^{\frac12\abs{\Gamma}-1}\beta_r \cot(\pi\alpha_r) - \frac{2}{\pi^2\abs{\Gamma}^2}\left(\frac34\zeta_R(3)+\sum_{q=0}^2\frac{(-\pi i\nu)^q}{q!}\Li_{3-q}(-e^{\pi i\nu}) \right) \\
& \quad & + \frac{i}{2\pi}\sum_{r=0}^{\frac12\abs{\Gamma}-1}\beta_r \left(\Li_2(e^{2\pi i\alpha_r})-\Li_2(e^{\frac{2\pi i\nu}{\abs{\Gamma}}+2\pi i\alpha_r})+\frac{2\pi i\nu}{\abs{\Gamma}}\Li_1(e^{\frac{2\pi i\nu}{\abs{\Gamma}}+2\pi i\alpha_r})\right),
\end{eqnarray*}
where we use the definition of the polylogarithm \eqref{eq:PolylogDef} in the final step.
Inserting this into the derivative of the spectral zeta function, all of the $\nu$ independent terms cancel and we have
\[\begin{split}
\pd_s \zeta_{\mc K_{\Gamma\bs S^3}}(s;\nu)\big|_{s=0} 
&= 2\abs{\Gamma}F(\nu) + \frac{3}{\pi^2\abs{\Gamma}}\zeta_R(3)+ 2\abs{\Gamma}\sum_{r=0}^{\frac12\abs{\Gamma}-1}\Bigg[- \frac{i}{2\pi}\beta_r \Li_2(e^{2\pi i \alpha_r})  
+\frac{\pi\nu^2}{2\abs{\Gamma}^2}\beta_r \cot(\pi\alpha_r)\Bigg] \\
&= \frac{\pi i}{3\abs{\Gamma}} \nu^3  - \frac{4}{\pi^2\abs{\Gamma}} \sum_{q=0}^2\frac{(-\pi i\nu)^q}{q!}\Li_{3-q}(-e^{\pi i\nu})\\&\quad
- \sum_{r=0}^{\frac12\abs{\Gamma}-1} \beta_r\left(\frac{i\abs{\Gamma}}{\pi}\Li_2(e^{\frac{2\pi i\nu}{\abs{\Gamma}}+2\pi i\alpha_r})+2\nu\Li_1(e^{\frac{2\pi i\nu}{\abs{\Gamma}}+2\pi i\alpha_r})\right),
\end{split}\]
which recovers the desired expression for the massive sectors, both light and heavy.        

To recover the expression in the massless limit $\nu=1$, we have to subtract the zero mode ($k=r=0$) from the massive expression before taking the $\nu\to 1^-$ limit. 
The zero mode contributes $-\log(1-\nu^2)$ to the derivative of the spectral zeta function, which when subtracted from the derivative of the massive spectral zeta function removes the logarithmic divergence coming from the $\Li_1$ term in the central term, and the $\Li_1$ term in the twisted sum for $r=\frac12\abs{\Gamma}-1$ whose argument limits to 1.
\begin{eqnarray*}
& & \lim_{\nu\to 1^-} \Big(\pd_s\zeta_{\mc K_{\Gamma\bs S^3}}(s;\nu)\big|_{s=0}+\log(1-\nu^2)\Big)\\ 
& = & \frac{\pi i}{3\abs{\Gamma}} 
-\frac{4}{\pi^2 \abs{\Gamma}}\zeta_R(3)+\frac{4i}{\pi  \abs{\Gamma}}\zeta_R(2)-\frac{i\abs{\Gamma}}{\pi}\beta_{\frac12\abs{\Gamma}-1}\zeta_R(2)\\
& \quad & -\sum_{r=0}^{\frac12\abs{\Gamma}-2} \beta_r\left(\frac{i\abs{\Gamma}}{\pi}\Li_2\left(e^{\frac{2\pi i}{\abs{\Gamma}}+2\pi i\alpha_r}\right)+2\Li_1\left(e^{\frac{2\pi i}{\abs{\Gamma}}+2\pi i\alpha_r}\right)\right) \\ 
& \quad & +\lim_{\nu\to 1^-}\left(-\frac{2\nu^2}{\abs{\Gamma}}\log(1+e^{\pi i\nu}) +2\nu\beta_{\frac12\abs{\Gamma}-1}\log(1-e^{2\pi i \left(\frac{\nu}{\abs{\Gamma}} + \alpha_{\frac12\abs{\Gamma}-1}\right)})+\log(1-\nu^2)\right),
\end{eqnarray*}
where we have used $\Li_s(1) = \zeta_R(s)$ for $\Re(s)>1$.
Using the definitions of the twist coefficient and characteristic Equation~\eqref{eq:AlphaBetaDef} and their parity properties Equation~\eqref{eq:AlphaBetaParity1}, we find 
\begin{equation*}
\alpha_{\frac12\abs{\Gamma}-1} = 1-\frac1{\abs{\Gamma}}, \qquad \beta_{\frac12\abs{\Gamma}-1} = -\beta_0 = -\frac12 \dim(\V_0^\Gamma) + \alpha_0 = -\frac12+\frac1{\abs{\Gamma}},
\end{equation*}
which we can now insert into the limit, along with $\zeta_R(2) = \frac{\pi^2}6$, to find
\[
\begin{split}
&= \frac{\pi i}{\abs{\Gamma}} 
-\frac{4}{\pi^2 \abs{\Gamma}}\zeta_R(3)+\frac{\pi i}{12}(\abs{\Gamma}-2)
-\sum_{r=0}^{\frac12\abs{\Gamma}-2} \beta_r\left(\frac{i\abs{\Gamma}}{\pi}\Li_2\left(e^{\frac{2\pi i}{\abs{\Gamma}}+2\pi i\alpha_r}\right)+2\Li_1\left(e^{\frac{2\pi i}{\abs{\Gamma}}+2\pi i\alpha_r}\right)\right) \\ &\quad
+\lim_{\nu\to 1^-}\left(-\frac{2\nu^2}{\abs{\Gamma}}\log(1+e^{\pi i\nu}) +\nu\left(\frac2{\abs{\Gamma}}-1\right)\log(1-e^{2\pi i\frac{\nu-1}{\abs{\Gamma}}})+\log(1-\nu^2)\right)\\
&= \frac{\pi i}{\abs{\Gamma}} 
-\frac{4}{\pi^2 \abs{\Gamma}}\zeta_R(3)+\frac{\pi i}{12}(\abs{\Gamma}-2)
-\sum_{r=0}^{\frac12\abs{\Gamma}-2} \beta_r\left(\frac{i\abs{\Gamma}}{\pi}\Li_2\left(e^{\frac{2\pi i}{\abs{\Gamma}}+2\pi i\alpha_r}\right)+2\Li_1\left(e^{\frac{2\pi i}{\abs{\Gamma}}+2\pi i\alpha_r}\right)\right) \\ &\quad
+\frac{\log 4}{\abs{\Gamma}}+\log\frac{\abs\Gamma}\pi-\frac{2}{\abs{\Gamma}}\log\abs{\Gamma}-\frac{\pi i}{2},
\end{split}
\]
where we use $1+e^{\pi i \nu} \sim \pi i (1-\nu) + O((1-\nu)^2)$ and $1-e^{2\pi i \frac{\nu-1}{\abs{\Gamma}}} \sim \frac{2\pi i}{\abs{\Gamma}} (1-\nu) + O((1-\nu)^2)$, which allow all $\log(1-\nu)$ terms to cancel. 
Straightforward algebra recovers the stated expression.
This completes the proof. 

\end{proof}
    
\subsection{The heat trace and a geometric expansion}

Recall the definition of the heat operator, $e^{-t\mc K_{\Gamma\bs S^3}(\mf m)}$ and the corresponding heat trace 
\begin{equation}
\label{eq:HeatTraceDefGamma}
H_{\mc K_{\Gamma\bs S^3}}(t;\nu):=\tr'(e^{-t\mc K_{\Gamma\bs S^3}(\mf m)}), \quad t>0. 
\end{equation}

\begin{proposition}
\label{prop:HeatGamma}

The heat trace of the kinetic operator \eqref{eq:KineticGamma} for $t>0$ admits the following equivalent expressions, (a) in terms of the eigenvalues and multiplicities, (b) in terms of Jacobi theta functions \eqref{def:JacobiThetaDef}, and (c) in terms of a geometric parameter $w$,
\begin{subequations}
\begin{align}
H_{\mc K_{\Gamma\bs S^3}}(t;\nu) 
&= \sum_{n=0}^{\infty} m_\Gamma(n) e^{-t((n+1)^2-\nu^2)} - \delta_{\nu^2,1} \label{eq:HeatGammav1}\\
&= -\frac{e^{t\nu^2}}{\abs{\Gamma}}\pd_t \vartheta_2\left(\frac{4it}{\pi}\right) +\frac{e^{t\nu^2}}{2\pi i}\sum_{r=0}^{\frac12\abs{\Gamma}-1}\beta_r \pd_z \JacTh{\alpha_r}{0}{\abs{\Gamma}z}{\abs{\Gamma}^2\frac{it}\pi}\bigg|_{z=0} - \delta_{\nu^2,1}, \label{eq:HeatGammav2} \\
&= \frac{\sqrt{\pi}\,e^{t\nu^2}}{4\abs{\Gamma}}\sum_{w\in\mathbb Z}\left((-1)^w \left(1-\frac{\pi^2 w^2}{2t}\right)t^{-\frac32} e^{-\frac{\pi^2w^2}{4t}} \right.\label{eq:HeatGammav3}\\\nonumber&\qquad\left.
+ \sum_{r=0}^{\frac12\abs{\Gamma}-1}\beta_r \sin(2\pi w \alpha_r) \frac{4\pi w}{\abs{\Gamma}} t^{-\frac32} e^{-\frac{\pi^2 w^2}{\abs{\Gamma}^2 t}} \right) - \delta_{\nu^2,1}. 
\end{align}
\end{subequations}
The spectral zeta function can be resummed into a rapidly convergent expansion in the heavy sector ($\nu_H = -i\nu >0$),
\begin{equation}\label{eq:ZetaGammaHeavyBessel}
\begin{split}
&\zeta_{\mc K_{\Gamma\bs S^3}}(s;i\nu_H) \\
&= \frac{\sqrt{\pi}}{4\abs{\Gamma}}\frac{\Gamma(s-\frac32)}{\Gamma(s)}\nu_H^{3-2s} 
+ \frac{2\pi^{s-1}}{\Gamma(s)} \sum_{w=1}^\infty \left(\frac{w}{\nu_H}\right)^{s-\frac32} \left[\frac{2^{\frac12-s}(-1)^w}{\abs{\Gamma}}\left(K_{\frac32-s}(\pi w \nu_H) - \pi w \nu_H K_{\frac52-s}(\pi w \nu_H)\right) \right.\\&\qquad
\left.+ \frac{2\pi w}{\abs{\Gamma}^{s+\frac12}}\left(\sum_{r=0}^{\frac12\abs{\Gamma}-1}\beta_r\sin(2\pi w \alpha_r)\right)K_{\frac32-s}\left(\frac{2\pi w \nu_H}{\abs{\Gamma}}\right)\right].
\end{split}
\end{equation}

\end{proposition}

\begin{proof}

We build the heat trace using the eigenvalues and degeneracies as follows, and then use the definitions of the Jacobi theta functions with characteristics (cf. \ref{def:JacobiThetaDef}) to rewrite the expression. 
By the definition of the heat operator, the trace sums over $\L^2(\Gamma\bs S^3)$ omitting any modes with zero eigenvalues. 
Using the eigenvalues \eqref{eq:KineticGammaEigenVals} and their corresponding multiplicities $m_\Gamma(n)$, we obtain the first expression \eqref{eq:HeatGammav1}, where the only zero mode occurs for $n=0$ when $\nu^2=1$, whose multiplicity $m_\Gamma(0) = 1$. 
    
To obtain the theta function expression, we start by expressing the multiplicities in terms of the characteristic and twist coefficients in Equation~\eqref{eq:EvenMult2},
\begin{eqnarray*}
H_{\mc K_{\Gamma\bs S^3}}(t;\nu) & = & 2\abs{\Gamma}e^{t\nu^2}\sum_{r=0}^{\frac12\abs{\Gamma}-1}\sum_{k=0}^\infty \big( (k+\alpha_r)^2 + \beta_r (k+\alpha_r) \big) e^{-t\abs{\Gamma}^2(k+\alpha_r)^2} -\delta_{\nu^2,1}  \\
& = & \frac{2e^{t\nu^2}}{\abs{\Gamma}}\sum_{\substack{n=0\\n\text{ even}}}^\infty (n+1)^2 e^{-t(n+1)^2} + 2\abs{\Gamma}e^{t\nu^2}\sum_{r=0}^{\frac12\abs{\Gamma}-1}\beta_r \sum_{k=0}^\infty (k+\alpha_r) e^{-t\abs{\Gamma}^2(k+\alpha_r)^2} -\delta_{\nu^2,1}.
\end{eqnarray*}
Using the definitions of the Jacobi theta functions (see \ref{def:JacobiThetaDef}) and the parity property \eqref{eq:AlphaBetaParity2}, we have
\[\begin{split}
\vartheta_2(\tau) = \sum_{m\in\mathbb Z} e^{\pi i(m+\frac12)^2\tau} &= 2\sum_{m=0}^\infty e^{\frac{\pi i}4 (2m+1)^2\tau} = 2\sum_{\substack{n=0\\n\text{ even}}}^\infty e^{\frac{\pi i}4 (n+1)^2\tau},\\
\sum_{r=0}^{\frac12\abs{\Gamma}-1}\beta_r\sum_{k=0}^\infty (k+\alpha_r)e^{-t\abs{\Gamma}^2(k+\alpha_r)^2}
&= \frac12\sum_{r=0}^{\frac12\abs{\Gamma}-1}\beta_r\sum_{k\in\mathbb Z} (k+\alpha_r)e^{-t\abs{\Gamma}^2(k+\alpha_r)^2}\\
&= \frac1{4\pi i\abs{\Gamma}}\sum_{r=0}^{\frac12\abs{\Gamma}-1}\beta_r \pd_z\JacTh{\alpha_r}{0}{\abs{\Gamma}z}{\abs{\Gamma}^2 \frac{it}{\pi}} \Big|_{z=0}.
\end{split}\]
Using the above identities we find that the heat trace using the Jacobi theta functions in terms of the characteristic and twist coefficient reads
\begin{equation*}
H_{\mc K_{\Gamma\bs S^3}}(t;\nu) = -\frac{e^{t\nu^2}}{\abs{\Gamma}}\pd_t \vartheta_2\left(\frac{4it}{\pi}\right) +\frac{e^{t\nu
^2}}{2\pi i}\sum_{r=0}^{\frac12\abs{\Gamma}-1}\beta_r \pd_z \JacTh{\alpha_r}{0}{\abs{\Gamma}z}{\abs{\Gamma}^2\frac{it}\pi}\bigg|_{z=0} - \delta_{\nu^2,1},
\end{equation*}
which recovers Equation~\eqref{eq:HeatGammav2}.
    
To recover the third form of the heat trace we use the modular $\mc S$-transformation property of the Jacobi theta functions (cf.~\ref{def:JacobiThetaDef}). In particular, we use
\begin{equation*}
\vartheta_2(\tau) = \sqrt{\frac i\tau}\vartheta_4\left(-\frac1\tau\right), \qquad \JacTh{b}{-a}{z}{\tau} = \sqrt{\frac i\tau}\, e^{-2\pi iab-\frac{\pi i z^2}\tau}\JacTh{a}{b}{\frac z\tau}{-\frac1\tau}.
\end{equation*}
Starting from Equation~\eqref{eq:HeatGammav2} and using the series representations of the theta functions (cf.~\ref{def:JacobiThetaDef}) and parity \eqref{eq:AlphaBetaParity2}, we have
\[\begin{split}
&H_{\mc K_{\Gamma\bs S^3}}(t;\nu) \\
&= -\frac{e^{t\nu^2}}{\abs{\Gamma}}\pd_t \left(\sqrt{\frac{\pi}{4t}}\vartheta_4\left(\frac{\pi i}{4t}\right)\right) +\frac{e^{t\nu^2}}{2\pi i\abs{\Gamma}}\sqrt{\frac{\pi}{t}}\sum_{r=0}^{\frac12\abs{\Gamma}-1}\beta_r \pd_z\left(e^{-\frac{\pi^2 z^2}{t}} \JacTh{0}{\alpha_r}{\frac{\pi z}{\abs{\Gamma}it}}{\frac{\pi i}{\abs{\Gamma}^2 t}}\right)\Bigg|_{z=0} - \delta_{\nu^2,1}\\
&= \frac{\sqrt{\pi}\,e^{t\nu^2}}{4\abs{\Gamma}}\sum_{w\in\mathbb Z}\left((-1)^w \left(1-\frac{\pi^2 w^2}{2t}\right)t^{-\frac32} e^{-\frac{\pi^2w^2}{4t}} + \sum_{r=0}^{\frac12\abs{\Gamma}-1}\beta_r \sin(2\pi w \alpha_r) \frac{4\pi w}{\abs{\Gamma}} t^{-\frac32} e^{-\frac{\pi^2 w^2}{\abs{\Gamma}^2 t}} \right) - \delta_{\nu^2,1}.
\end{split}\]
which recovers \eqref{eq:HeatGammav3}.
    
The Bessel expansion in the heavy sector can be obtained by performing a Mellin transform to the heat trace \eqref{eq:HeatGammav3}. Analogous to the 3-sphere calculation, we have
\begin{equation*}
\zeta_{\mc K_{\Gamma\bs S^3}}(s;i\nu_H) = \frac1{\Gamma(s)}M[H_{\mc K_{\Gamma\bs S^3}}(t;i\nu_H)](s)
\end{equation*}
and using the integral representation of the modified Bessel function \eqref{eq:BesselKDef}, we recover the quoted expression and completes the proof. 

\end{proof}

\begin{remark}

The theta function representation of the heat trace \eqref{eq:HeatGammav2} separates the central part of the quotient, the $\mathbb Z/2\mathbb Z$ subgroup, and the twisted part of the quotient.
The twisted sum justifies the names characteristic and twist coefficient due to their usage in this expression. 
    
The resummed expression \eqref{eq:HeatGammav3} can be written geometrically, using the central and twisted geodesic lengths, which are  $L^{\mathrm{c}}_w = \frac{2\pi\abs{w}}{\abs{\mathbb Z/2\mathbb Z}}$ and $L^{\mathrm{tw}}_w = \frac{2\pi\abs{w}}{\abs{\Gamma}}$ respectively,
\[\begin{split}
&H_{\mc K_{\Gamma\bs S^3}}(t;\nu)\\
&= \frac{\vol(\Gamma\bs S^3)}{(4\pi t)^{\frac32}} e^{t\nu^2} \sum_{w\in\mathbb Z} \left((-1)^w\left(1-\frac{(L^{\mathrm{c}}_{w})^2}{2t}\right)e^{-\frac{(L^{\mathrm{c}}_w)^2}{4t}}+2L^{\mathrm{tw}}_w\sum_{r=0}^{\frac12\abs{\Gamma}-1}\beta_r\sin((2r+1)L^{\mathrm{tw}}_w)e^{-\frac{(L^{\mathrm{tw}}_w)^2}{4t}}\right).
\end{split}\]
Note that the twisted geodesics only occur when the sum $\sum_{r=0}^{\frac12\abs{\Gamma}-1} \beta_r \sin(2\pi w \alpha_r)$ is non-zero. 

The regularized determinant (Theorem~\ref{thm:DetGamma}) can be recovered from the Bessel expansion of the spectral zeta function in the heavy sector via direct computation, by evaluating the Bessel functions at $s=0$ using Eq.~\eqref{eq:BesselKCases}.
Using \eqref{eq:GammaExp}, $\Gamma(-\frac32)=\frac{4\sqrt\pi}3$, we have
\[\begin{split}
\pd_s\zeta_{\mc K_{\Gamma\bs S^3}}(s;i\nu_H)\big|_{s=0} &= M[H_{\mc K_{\Gamma\bs S^3}}(t;i\nu_H)](0)\\
&= \frac{\pi}{3\abs{\Gamma}}\nu_H^{3} 
+ \frac2\pi\sum_{w=1}^\infty \left(\frac{w}{\nu_H}\right)^{-\frac32} \left[\frac{2^{\frac12}(-1)^w}{\abs{\Gamma}}\left(K_{\frac32}(\pi w \nu_H) - \pi w \nu_H K_{\frac52}(\pi w \nu_H)\right) \right.\\&\qquad
\left.+ \frac{2\pi w}{\abs{\Gamma}^{\frac12}}\left(\sum_{r=0}^{\frac12\abs{\Gamma}-1}\beta_r\sin(2\pi w \alpha_r)\right)K_{\frac32}\left(\frac{2\pi w \nu_H}{\abs{\Gamma}}\right)\right] \\
&= \frac{\pi}{3\abs{\Gamma}}\nu_H^3 -\frac{4}{\pi^2\abs{\Gamma}}\sum_{q=0}^2 \frac{(\pi\nu_H)^q}{q!} \Li_{3-q}\left(-e^{-\pi\nu_H}\right) \\&\qquad 
- \sum_{r=0}^{\frac12\abs{\Gamma}-1}\beta_r\left[2i\nu_H \Li_1\left(e^{\frac{-2\pi\nu_H}{\abs{\Gamma}}+2\pi i\alpha_r}\right)+\frac{i\abs{\Gamma}}{\pi} \Li_2\left(e^{\frac{-2\pi\nu_H}{\abs{\Gamma}}+2\pi i\alpha_r}\right)\right],
\end{split}
\]
which recovers the regularized determinant found in Theorem \ref{thm:DetGamma} in the heavy sector, providing an alternative derivation in the heavy sector, $\Im(\nu)>0$, $\abs{\Re(\nu)}<1$.
\end{remark}

\subsection{Prism spaces - \texorpdfstring{$D_m\bs S^3$}{Dm\textbackslash S3}}

\subsubsection{Embedding of $D_{m}$ into $\SU(2)$}

Let $m \geq 2$, and let $\Gamma = D_{m}$ be the binary dihedral group. The group $D_{m}$ is the finite subgroup of $\SU(2)$ given by
\begin{equation*}
D_{m} := \langle a\,, b \rangle \subset \SU(2)\,,
\end{equation*}
where
\begin{equation*}
a = \begin{pmatrix} e^{\frac{i\pi}{m}} & 0 \\ 0 & e^{-\frac{i\pi}{m}} \end{pmatrix}\,, \qquad b = \begin{pmatrix} 0 & 1 \\ -1 & 0 \end{pmatrix}\,.
\end{equation*}
In particular, we get
\begin{equation*}
a^{2m} = \Id_{2}\,, \qquad b^{2} = -\Id_{2}\,, \qquad bab^{-1} =a^{-1}\,.
\end{equation*}
Hence
\begin{equation*}
D_{m} = \left\{a^{k}\,, \,a^{k}b\,,  k \in \left\{0\,, \ldots\,, 2m-1\right\}\right\}\,,
\end{equation*}
and therefore $\left|D_{m}\right| = 4m$\,.

\noindent The group $D_{m} \cong D_{m} \times \left\{\Id_{2}\right\} \subseteq \SU(2) \times \SU(2)$ acts freely on $ S^3 \cong \SU(2)$, where the action is given by left multiplication (see Equation \eqref{ActionSU2SU2}). In particular, it follows from Equation \eqref{EquationMGammaN} that
\begin{equation*}
m_{\Gamma}(n) = \left(n+1\right)\dim(\V_n^{D_{m}})
\end{equation*}

\subsubsection{The coefficients $m_{\Gamma}(n)$} We start this section with a few notations.

\begin{notation}

For all $n$, we denote by $\Lambda_{n}$ the subset of $\left\{0, \ldots, n\right\}$ given by
\begin{equation*}
\Lambda^{m}_{n} = \left\{r \in \left\{0\,, \ldots\,, n\right\}\,, n-2r \equiv 0 \pmod{2m}\right\}\,.
\end{equation*}
Moreover, let
\begin{equation*}
\epsilon_{n} = \begin{cases} 1 & \text{ if } n \text{ is even} \\ 0 & \text{otherwise} \end{cases} \qquad  \delta_{n} = \begin{cases} 1 & \text{ if } n \equiv 0 \pmod{4} \\ 0 & \text{otherwise}\end{cases}
\end{equation*}

\end{notation}

\begin{proposition}

For all $n \in \mathbb{N}$, we get
\begin{equation*}
\dim\left(\V^{D_{m}}_{n}\right) = \frac{\left|\Lambda^{m}_{n}\right| -\epsilon_{n}}{2} + \delta_{n}\,.
\end{equation*}

\label{PropositionDihedral}

\end{proposition}

\begin{proof}

Again, let $\V_{n} = \Span\left\{z^{n-r}_{1}z^{r}_{2}\,, r = 0\,, \ldots\,, n\right\}$. The generator $a$ acts by
\begin{equation*}
a \cdot z^{n-r}_{1}z^{r}_{2} = e^{\frac{i\pi(n-2r)}{m}}z^{n-r}_{1}z^{r}_{2}\,.
\end{equation*}
Therefore, a basis vector is fixed by $a$ if and only if $n-2r\equiv 0 \pmod{2m}$. In particular,
\begin{equation*}
\V^{\langle a\rangle}_{n} = \Span\left\{z^{n-r}_{1}z^{r}_{2}\,, r \in \Lambda^{m}_{n}\right\}\,.
\end{equation*}
We now determine the $b$-invariant vectors in $\V^{\langle a\rangle}_{n}$. Using that
\begin{equation*}
b \cdot z_{1} = -z_{2}\,, \qquad b\cdot z_{2} = z_{1}\,,
\end{equation*}
we obtain
\begin{equation*}
b\cdot z^{n-r}_{1}z^{r}_{2} = (-1)^{n-r}z^{r}_{1}z^{n-r}_{2}\,.
\end{equation*}
Thus, $b$ exchanges the basis vectors indexed by $r$ and $n-r$, up to a sign $(-1)^{n-r}$. Moreover, if $r \in \Lambda^{m}_{n}$, then $n-r \in \Lambda^{m}_{n}$, so $b$ preserves $\V^{\langle a\rangle}_{n}$.

\noindent Suppose first that $n$ is odd. Since $2m$ is even, the congruence
\begin{equation*}
n-2r\equiv 0\pmod{2m}
\end{equation*}
has no solution, i.e.
\begin{equation*}
\V^{D_{m}}_{n} = \left\{0\right\}\,.
\end{equation*}

\noindent Now suppose that $n$ is even. For every $r\in\Lambda^{m}_{n}$ with $r < \frac{n}{2}$, consider the polynomial
\begin{equation*}
\tau_{r} = z^{n-r}_{1}z^{r}_{2} + (-1)^{n-r}z^{r}_{1}z^{n-r}_{2}\,.
\end{equation*}
Since $n$ is even, we have $b\cdot\tau_{r}=\tau_{r}$. Moreover, the vectors $\tau_{r}$ are linearly independent, and each pair of distinct indices $\left\{r\,,n-r\right\}$ contributes exactly one dimension to $\V^{D_{m}}_{n}$.

\noindent Finally, the index $r=n/2$ always belongs to $\Lambda^{m}_{n}$. The corresponding monomial satisfies
\begin{equation*}
b\cdot z^{\frac{n}{2}}_{1}z^{\frac{n}{2}}_{2} = (-1)^{\frac{n}{2}}z^{\frac{n}{2}}_{1}z^{\frac{n}{2}}_{2}\,.
\end{equation*}
Hence, it is fixed by $b$ if and only if $n\equiv 0\pmod 4$. Therefore,
\begin{equation*}
\dim\left(\V^{D_{m}}_{n}\right) = \frac{\left|\Lambda^{m}_{n}\right|-1}{2}+\delta_{n}\,,
\end{equation*}
when $n$ is even. Combining the even and odd cases, we conclude that
\begin{equation*}
\dim\left(\V^{D_{m}}_{n}\right) = \frac{\left|\Lambda^{m}_{n}\right|-\epsilon_{n}}{2}+\delta_{n}\,,
\end{equation*}
which completes the proof.

\end{proof}

\begin{corollary}

For all $n\geq 0$, we have
\begin{equation}
m_{\Gamma}(n) = \begin{cases} 0 & \text{if $n$ is odd} \\ (n+1)\left(\displaystyle\left\lfloor\frac{n}{2m}\right\rfloor+1\right) & \text{if $n\equiv 0\pmod{4}$} \medskip \\ (n+1)\displaystyle\left\lfloor\frac{n}{2m}\right\rfloor & \text{if $n\equiv 2\pmod{4}$} \end{cases}
\label{eq:DihedralMult}
\end{equation}

\end{corollary}

\begin{proof}

Recall that
\begin{equation*}
m_{\Gamma}(n)=(n+1)\dim\left(\V_{n}^{D_{m}}\right)\,.
\end{equation*}
Using Proposition \ref{PropositionDihedral}, we have
\begin{equation*}
\dim\left(\V_{n}^{D_{m}}\right) = \frac{\left|\Lambda^{m}_{n}\right|-\epsilon_{n}}{2}
+\delta_{n}\,.
\end{equation*}
Moreover, since $2m$ is even, we obtain
\begin{equation*}
\left|\Lambda^{m}_{n}\right| = \begin{cases} 0 & \text{if $n$ is odd} \\
\displaystyle 2\left\lfloor\frac{n}{2m}\right\rfloor+1
& \text{if $n$ is even}\,.
\end{cases}
\end{equation*}
The result follows immediately from the definitions of $\varepsilon_{n}$ and $\delta_{n}$.

\end{proof}

\subsubsection{The heat trace and spectral zeta function}

The kinetic operator and the corresponding eigenvalues on $D_m\bs S^3$ are
\begin{equation}\label{eq:KinDihedral}
\mc K_{D_m}(\mf m) = \Delta_{D_m \bs S^3} + \mf m^2\ell^2,\qquad n(n+2) + \mf m^2 \ell^2 = (n+1)^2 - \nu^2,
\end{equation}
where $\nu^2 = 1-\mf m^2\ell^2$.
Since the order of the group $D_m$ is $\abs{D_m} = 4m$, the representation label $n$ can be reparametrized \eqref{eq:reparam}, 
\begin{equation}
    n=4km+2r,\quad 0\leq r \leq 2m-1.
\end{equation}
The characteristic and twist coefficient are \eqref{eq:AlphaBetaDef},
\begin{equation}
\label{eq:DmAlphaBeta}
\alpha_r = \frac{2r+1}{4m}, \qquad \beta_r = \frac12 \left(\floor{\frac rm}+\delta_{r\bmod 2,0}\right)-\alpha_r,
\end{equation}
so the multiplicities \eqref{eq:DihedralMult} read
\begin{equation}
\begin{split}
m_{D_m}(n) 
&= (n+1)\left(\floor{\frac{n}{2m}}+\delta_{n\bmod 4, 0}\right)\delta_{n\bmod 2,0} \\
&= 8m(k+\alpha_r)(k+\alpha_r + \beta_r).
\end{split}
\end{equation}
The parity of $\alpha_r$ and $\beta_r$ under the map $r\mapsto 2m-1-r$ is as follows
\begin{equation}
\begin{split}
\alpha_{2m-1-r} &= \frac{2(2m-1-r)+1}{4m} = 1-\alpha_r, \\
\beta_{2m-1-r}  &= \frac12\left(\floor{\frac{2m-1-r}{m}}+\delta_{2m-1-r\bmod 2,0} \right) - \alpha_{2m-1-r} \\
&= -\frac12\left(\floor{\frac rm}+\delta_{r\bmod2,0}\right) + \alpha_r  = -\beta_r,
\end{split}
\end{equation}
which allows us to use the parity properties we claimed in \eqref{eq:AlphaBetaParity1} and \eqref{eq:AlphaBetaParity2}.
Then the heat trace, spectral zeta function and the regularized determinant of the kinetic operator on $D_m\bs S^3$ is recovered by substituting $\abs{D_m}=4m$ and the characteristic and twist coefficient \eqref{eq:DmAlphaBeta} into Propositions~\ref{prop:HeatGamma} and \ref{prop:ZetaGamma}, Corollary \ref{cor:ZetaGamma}, and Theorem \ref{thm:DetGamma}. 

\subsection{The binary Icosahedral group}

\subsubsection{Embedding of $\widetilde{\mathcal{I}}$ into $\SU(2)$}

We start with a definition.

\begin{definition}

The binary icosahedral group $\widetilde{\mathcal{I}}$ is the unique subgroup of $\SU(2)$ of order $120$ that maps onto the icosahedral rotation group $\mathcal{I} \cong \mathrm{A}_{5}$ under the covering homomorphism $\SU(2) \to \SO(3)$. It is generated by two matrices $a\,, b \in \SU(2)$, where
\begin{equation}
a =  \begin{pmatrix}-\zeta^{3} & 0 \\ 0 & -\zeta^{-3} \end{pmatrix}\,, \qquad  b = \frac1{\sqrt 5}\begin{pmatrix} -\zeta+\zeta^{-1} & \zeta^2 - \zeta^{-2} \\ \zeta^2 - \zeta^{-2} & \zeta-\zeta^{-1}
\end{pmatrix}\,,
\label{EquationGenerators}
\end{equation}
with $\zeta = e^{\frac{2i\pi}{5}}$. These generators satisfy
\begin{equation}
a^{10} = \Id_{2}\,, \qquad a^5 = b^{2} = \left(ab\right)^{3} = -\Id_{2}\,.
\label{RelationGenerators}
\end{equation}

\label{DefinitionBinaryIcosahedralGroup}

\end{definition}

The element $-\Id_{2} \in \widetilde{\mathcal{I}}$ is central, and $\widetilde{\mathcal{I}}/\left\{\pm \Id_{2}\right\} \cong \mathcal{I} \cong \mathrm{A}_{5}$. The quotient $\widetilde{\mathcal{I}} \backslash  S^3$ is the Poincar\'e homology sphere, the unique closed orientable $3$-manifold with trivial first homology and finite non-trivial fundamental group.

\begin{remark}

Unlike the cyclic and binary dihedral cases, the method based on the explicit action of the generators on the monomial basis of $\V_{n}$ does not readily lead to simple formulas for the spectral multiplicities of the binary icosahedral group. Indeed, while the generator $a$ is diagonal and its invariant subspace can be determined explicitly, the generator $b$ sends each monomial to a linear combination of monomials. Therefore, determining the common invariant vectors requires solving a system of linear equations whose size increases with $n$, making it difficult to obtain a general formula for $m_{\Gamma}(n)$. Instead, we will use Molien series, which provide a generating function for the dimensions of the invariant subspaces $\V_{n}^{\widetilde{\mathcal{I}}}$. This approach will be developed in the next section, where we will obtain explicit formulas for the spectral multiplicities.

\end{remark}

\subsubsection{The multiplicities $m_{\Gamma}(n)$}

We now apply Theorem \ref{TheoremMGammaQ} for the binary icosahedral group introduced in Section \ref{SectionMolienSeries}. 

\begin{remark}

For any finite group $\Gamma \subset \SU(2)$ and any $g \in \SU(2)$ with eigenvalues $\lambda\,, \overline{\lambda}$ (satisfying $\left|\lambda\right|=1$), the action of $g$ on $\V_{n}$ (with $\dim(\V_{n}) = n+1$) has eigenvalues 
\begin{equation*}
\lambda^{n} \quad \lambda^{n-2} \quad \ldots \quad \lambda^{-n+2} \quad \lambda^{-n}\,,
\end{equation*}
so the character $\chi_{n}$ of $\rho_{n}$ is given for all $g \in \Gamma$ by
\begin{equation}
\chi_{n}(g) = \lambda^{n} + \lambda^{n-2} + \cdots + \lambda^{-n} = \frac{\lambda^{n+1}-\lambda^{-(n+1)}}{\lambda-\lambda^{-1}}\,.
\end{equation}
Therefore
\begin{eqnarray*}
\sum\limits_{n=0}^{\infty}\chi_{n}(g)q^{n} & = &  \sum\limits_{n=0}^{\infty} \left(\lambda^{n} + \lambda^{n-2} + \cdots + \lambda^{-n}\right)q^{n} = \sum\limits_{n=0}^{\infty}\sum\limits_{k=0}^{n}\lambda^{n-2k}q^{n} \\
& = & \sum\limits_{k=0}^{\infty}\sum\limits_{n=k}^{\infty}\lambda^{n-2k}q^{n} = \sum\limits_{k=0}^{\infty}\lambda^{-2k}\sum_{n=k}^{\infty}\left(\lambda q\right)^{n} = \sum\limits_{k=0}^{\infty}\lambda^{-2k}\frac{\left(\lambda q\right)^{k}}{1-\lambda q} \\
& = & \frac{1}{1-\lambda q}\sum\limits_{k=0}^{\infty}\left(\lambda^{-1}q\right)^{k} = \frac{1}{1-\lambda q}\frac{1}{1-\lambda^{-1}q} = \frac{1}{(1-\lambda q)(1-\lambda^{-1}q)} \\
& = & \frac{1}{1-(\lambda+\lambda^{-1})q+q^{2}} = \frac{1}{1-\tr(g)q+q^{2}} = \frac{1}{\det(\Id_{2}-qg)}\,.
\end{eqnarray*}
In particular, we recover Theorem \ref{TheoremMGammaQ} for $\widetilde{\mathcal{I}}$.

\noindent Now, since $\det(g)=1$ for $g\in\SU(2)$, it follows from the Cayley-Hamilton's theorem that
\begin{equation*}
\det(\Id-qg) = 1- \tr(g)q+q^{2}\,,
\end{equation*}
so the summand in the previous formula depends on $\tr(g)$ only\,.

\end{remark}

\noindent As explained in Remark \ref{RemarkConjugacyClasses}, the sum over $\widetilde{\mathcal{I}}$ only depends on the conjugacy classes of $\widetilde{\mathcal{I}}$. The group $\widetilde{\mathcal{I}}$ has exactly $9$ conjugacy classes.
Let $\varphi = \frac{1+\sqrt{5}}{2}$, so $\varphi^{-1} = \varphi-1 = \frac{\sqrt{5}-1}{2}$.
In Table \ref{tab:IcosahedralOrders}, we give the classes, their sizes, representative traces, and element orders, and the corresponding determinant.

\begin{table}
    \centering
    \renewcommand{\arraystretch}{1.2}
    \begin{tabular}{ccccc}
\toprule
Class & Size & Trace & Order & Determinant \\
\midrule
$\gamma_{1}$ & $1$  & $2$ & $1$ & $\left(1-q\right)^{2}$ \\
$\gamma_{2}$ & $1$ & $-2$ & $2$ & $\left(1+q\right)^{2}$ \\
$\gamma_{3}$ & $12$ & $\varphi$ & $10$ & $1-\varphi q+q^{2}$ \\
$\gamma_{4}$ & $12$ & $\varphi^{-1}$ & $5$ & $1-\varphi^{-1}q+q^{2}$ \\
$\gamma_{5}$ & $12$ & $-\varphi^{-1}$& $10$ & $1+\varphi^{-1}q+q^{2}$ \\
$\gamma_{6}$ & $12$ & $-\varphi$ & $5$ & $1+\varphi q+q^{2}$ \\
$\gamma_{7}$ & $20$ & $-1$ & $3$ & $1+q+q^{2}$ \\
$\gamma_{8}$ & $20$ & $1$ & $6$ & $1-q+q^{2}$ \\
$\gamma_{9}$ & $30$ & $0$ & $4$ & $1+q^{2}$ \\
\bottomrule
\end{tabular}
\medskip
\caption{Conjugacy classes for the binary icosahedral group and their corresponding determinant, where $\varphi^{\pm 1}=\frac{\sqrt5 \pm 1}2$.}
\label{tab:IcosahedralOrders}
\end{table}

\noindent Therefore, it follows from Theorem \ref{TheoremMGammaQ} and Remark \ref{RemarkConjugacyClasses} that
\begin{eqnarray*}
120\mathcal{M}_{\widetilde{\mathcal{I}}}(q) & = & \frac{1}{(1-q)^2} + \frac{1}{(1+q)^2} + \frac{12}{1-\varphi q+q^{2}} + \frac{12}{1-\varphi^{-1}q+q^2} \\
& + & \frac{12}{1+\varphi^{-1}q+q^{2}} + \frac{12}{1+\varphi q+q^{2}} + \frac{20}{1+q+q^{2}} \\ 
& + & \frac{20}{1-q+q^{2}} + \frac{30}{1+q^{2}}\,.
\end{eqnarray*}
We simplify this by pairing terms whose denominators differ only by the sign of the linear term.  Indeed, for all $\tau$, we get
\begin{equation}
\frac{1}{1-\tau q+q^{2}}+\frac{1}{1+\tau q+q^{2}} = \frac{2(1+q^{2})}{(1-\tau q+q^{2})(1+\tau q+q^{2})} = \frac{2(1+q^{2})}{1-(\tau^{2}-2)q^{2}+q^{4}}\,.
\label{KeyIdentity}
\end{equation}
Also, note that
\begin{equation}
\varphi^{2} = \varphi+1\,, \qquad \varphi^{-2} = 2-\varphi\,, \qquad \varphi^{2}-2 = \varphi-1 = \varphi^{-1}\,, \qquad \varphi^{-2}-2 = -\varphi\,.
\end{equation}
Using Equation \eqref{KeyIdentity} with $\tau=\varphi$ and $\tau=\varphi^{-1}$ respectively, we get
\begin{equation*}
\frac{1}{1-\varphi q+q^{2}}+\frac{1}{1+\varphi q+q^{2}} = \frac{2(1+q^{2})}{1-\varphi^{-1}q^{2}+q^{4}}\,,
\end{equation*}
and
\begin{equation*}
\frac{1}{1-\varphi^{-1}q+q^{2}}+\frac{1}{1+\varphi^{-1}q+q^{2}} = \frac{2(1+q^{2})}{1+\varphi q^{2}+q^{4}}\,.
\end{equation*}
After combining the nine terms over a common denominator and simplifying, we get
\begin{equation*}
\mathcal{M}_{\widetilde{\mathcal{I}}}(q) = 
\end{equation*}
{\small
\begin{equation*}
\frac{q^{16}+q^{14}-q^{10}-q^8-q^6+q^2+1}{(q-1)^2(q+1)^2(q^2+1)(q^2-q+1)(q^2+q+1)(q^4-q^3+q^2-q+1)(q^4+q^3+q^2+q+1)}\,.
\end{equation*}}
We now simplify the previous fraction to get an explicit formula for $\dim(\V^{\widetilde{\mathcal{I}}}_{n})$.

\begin{remark}

For all $n \geq 1$, we denote by $\Phi_{n}(q)$ the polynomial in $q$ given by
\begin{equation*}
\Phi_{n}(q) = \prod\limits_{\substack{1\leq k\leq n\\\gcd(k,n)=1}}\left(q-e^{\frac{2ik\pi}{n}}\right)\,.
\end{equation*}
It is known that
\begin{equation}
q^{n}-1 = \prod\limits_{d \mid n}\Phi_{d}(q)\,.
\end{equation}

\end{remark}

In particular, it follows that
\begin{equation*}
q^{12}-1 = (q-1)(q+1)(q^{2}+1)(q^{2}-q+1)(q^{2}+q+1)(q^{4}-q^{2}+1)
\end{equation*}
and
\begin{equation*}
q^{20}-1 = (q-1)(q+1)(q^{2}+1)(q^{4}-q^{3}+q^{2}-q+1)(q^{4}+q^{3}+q^{2}+q+1)(q^{8}-q^{6}+q^{4}-q^{2}+1)\,,
\end{equation*}
so their product is
\begin{equation*}
(1-q^{12})(1-q^{20}) = 
\end{equation*}
{\small \begin{equation*}
(q-1)^{2}(q+1)^{2}(q^{2}+1)^{2}(q^{2}-q+1)(q^{2}+q+1)(q^{4}-q^{2}+1)(q^{4}-q^{3}+q^{2}-q+1)(q^{4}+q^{3}+q^{2}+q+1)(q^{8}-q^{6}+q^{4}-q^{2}+1)\,.
\end{equation*}}
On the other hand,
\begin{equation}
1+q^{30} = (q^{2}+1)(q^{4}-q^{2}+1)(q^{8}-q^{6}+q^{4}-q^{2}+1)(q^{16}+q^{14}-q^{10}-q^{8}-q^{6}+q^{2}+1)\,.
\end{equation}
Dividing $1+q^{30}$ by $(1-q^{12})(1-q^{20})$, the common factors $(q^{2}+1)(q^{4}-q^{2}+1)(q^{8}-q^{6}+q^{4}-q^{2}+1)$ cancel, so
\begin{equation*}
\frac{1+q^{30}}{(1-q^{12})(1-q^{20})} = 
\end{equation*}
\begin{equation*}
\frac{q^{16}+q^{14}-q^{10}-q^8-q^6+q^2+1}{(q-1)^2(q+1)^2(q^2+1)(q^2-q+1)(q^2+q+1)(q^4-q^3+q^2-q+1)(q^4+q^3+q^2+q+1)}\,.
\end{equation*}
Finally, we get

\begin{lemma}

For all $q$ such that $\left|q\right| < 1$, we get
\begin{equation}
\mathcal{M}_{\widetilde{\mathcal{I}}}(q) = \frac{1+q^{30}}{(1-q^{12})(1-q^{20})}\,.
\label{LemmaMIQ}
\end{equation}

\end{lemma}

Using that $\left|q\right| < 1$, the geometric series expansion gives
\begin{equation*}
\frac{1}{1-q^{12}} = \sum\limits_{j \geq 0}q^{12j} \qquad \text{ and } \qquad \frac{1}{1-q^{20}} = \sum\limits_{k \geq 0}q^{20k}\,,
\end{equation*}
we obtain
\begin{eqnarray*}
\mathcal{M}_{\widetilde{\mathcal{I}}}(q) = \frac{1+q^{30}}{(1-q^{12})(1-q^{20})} & = & (1+q^{30})\left(\sum\limits_{j \geq 0}\sum\limits_{k \geq 0}q^{12j+20k}\right) \\ 
& = & \sum\limits_{j \geq 0}\sum\limits_{k \geq 0}q^{12j+20k} + \sum\limits_{j \geq 0}\sum\limits_{k \geq 0}q^{12j+20k+30}\,.
\end{eqnarray*}
Finally, we get the following proposition\,.

\begin{proposition}

For all $n \geq 0$, we get
\begin{equation}
\dim(\V^{\widetilde{\mathcal{I}}}_{n}) = \#\left\{\left(j\,, k\right) \in\mathbb{N}^{2}\,, 12j + 20k = n\right\} + \#\left\{\left(j\,, k\right) \in \mathbb{N}^{2}\,, 12j+20k=n-30\right\}\,.
\label{MultiplicitesVnI}
\end{equation}

\end{proposition}

\begin{remark}

In particular, one can easily see that $\dim(\V^{\widetilde{\mathcal{I}}}_{n}) = 0$ if $n$ is odd. The first non-zero values are

\begin{center}
\begin{tabular}{ccccccccccccccccc}
\toprule
$n$ & $0$ & $12$ & $20$ & $24$ & $30$ & $32$ & $36$ & $40$ & $42$ & $44$ & $48$ & $50$ & $52$ & $54$ & $56$ & $60$ \\
\midrule
$\dim(\V^{\widetilde{\mathcal{I}}}_{n})$ & $1$  & $1$  & $1$  & $1$  & $1$  & $1$  & $1$  & $1$ & $1$  & $1$  & $1$  & $1$  & $1$  & $1$ & $1$ & $2$ \\
\bottomrule
\end{tabular}
\end{center}

\end{remark}

\begin{notation}

We denote by $\mathscr{R}_{\widetilde{\mc I}}$ the finite set given by
\begin{equation}\label{eq:Ricosahedral}
\mathscr{R}_{\widetilde{\mc I}} = \left\{0\,, 12\,, 20\,, 24\,, 30\,, 32\,, 36\,, 40\,, 42\,, 44\,, 48\,, 50\,, 52\,, 54\,, 56\right\}\,.
\end{equation}

\end{notation}

\begin{proposition}

\label{PropositionMultiplicityI}

The dimension of the irreducible representation with eigenvalue $n+1$ is $\dim(\V^{\widetilde{\mathcal{I}}}_{n})$. We find $\dim(\V^{\widetilde{\mc I}}_n) = 0$ if $n$ is odd. If $n$ is even, write
\begin{equation*}
n = 60q + r\,, \qquad \left(0 \leq r < 60\,, q \in \mathbb{N}\right)\,.
\end{equation*}
Then
\begin{equation}
\label{EquationMultiplicityFormula}
\dim(\V^{\widetilde{\mathcal I}}_{n}) = q + \varepsilon_{r}\,, \text{ with } \qquad  \varepsilon_{r} = \begin{cases} 1 & \text{ if } r \in \mathscr{R}_{\widetilde{\mc I}} \\ 0 & \text{ otherwise} \end{cases}
\end{equation}
In other words, 
\begin{equation}
\label{eq:IcosahedralDim}
\dim(\V^{\widetilde{\mathcal I}}_{n}) = \begin{cases} 0 & n \text{ odd } \\
\floor{\frac{n}{60}}+\varepsilon_{n\bmod 60} & n \text{ even}\end{cases}
\end{equation}
i.e.
\begin{equation}
\label{eq:IcosahedralMult}
m_{\widetilde{\mc{I}}}(n) = \begin{cases} 0 & n \text{ odd} \\
(n+1)\left(\floor{\frac{n}{60}}+\varepsilon_{n\bmod 60}\right) & n \text{ even}.
\end{cases}
\end{equation}

\end{proposition}

\begin{remark}

In particular, one can see that for all even $n \geq 0$, we get
\begin{equation}
\label{EquationMMultiplicity}
\dim(\V^{\widetilde{\mc{I}}}_{n + 60}) = \dim(\V^{\widetilde{\mc I}}_n)+ 1\,.
\end{equation}

\end{remark}

\noindent The key step in the proof of Proposition \ref{PropositionMultiplicityI} is the following periodicity lemma.

\begin{lemma}

\label{LemmaFPeriodicity}

For all $q \in \mathbb{N}$, define
\begin{equation*}
\F(q) = \#\left\{\left(j, k\right) \in \mathbb{N}^{2}\,, 3j + 5k = q\right\}\,.
\end{equation*}
Then
\begin{equation}
\label{EquationFPeriodicity}
\F(q + 15) = \F(q) + 1\,.
\end{equation}
Moreover, the values of $\F$ on $\left\{0\,, 1\,, \ldots\,, 14\right\}$ are given by
\begin{equation*}
\F(q) = \begin{cases} 1 & q \in \left\{0\,, 3\,, 5\,, 6\,, 8\,, 9\,, 10\,, 11\,, 12\,, 13\,, 14\right\} \\ 0 & q \in \left\{1\,, 2\,, 4\,, 7\right\} \end{cases}
\end{equation*}
Therefore, writing $q = 15 l + s$ with $0 \leq s < 15$, we have
\begin{equation}
\label{EquationFFormula}
\F(q) = l + \eta_{s}\,,
\end{equation}
where $\eta_{s} = \F(s)$ is as above\,.

\label{TechnicalLemma120}

\end{lemma}

\begin{proof}

We first prove Equation \eqref{EquationFPeriodicity}. Define the map $\varphi$ as
\begin{equation*}
\varphi: \left\{\left(j\,, k\right) \in \mathbb{N}^{2}\,, 3j+5k = q\right\} \longrightarrow \left\{\left(j\,, k\right) \in \mathbb{N}^{2}\,, 3j+5k = q+15\right\}
\end{equation*}
by $\varphi\left(j\,, k\right) = \left(j+5\,, k\right)$. This map is clearly injective, so $\F(q+15) \geq \F(q)$. It remains to show that exactly one solution
$\left(j_{0}\,, k_{0}\right)$ of $3j+5k = q+15$ is not in the image of $\varphi$, i.e. satisfies $j_{0} < 5$. Let $j_{0} \in \left\{0\,, 1\,, 2\,, 3\,, 4\right\}$ be the unique element satisfying $3j_0 \equiv q+15 \pmod{5}$, i.e.\ $3j_0 \equiv q \pmod 5$ (which exists since $\gcd(3\,, 5)=1$). Set $k_{0} = \frac{q+15-3j_{0}}{5}$. Since $j_{0} \leq 4$, we have $3j_{0} \leq 12 < 15$, so $k_{0} = \frac{q+15-3j_{0}}{5} \geq \frac{q+3}{5} \geq 0$. This gives a unique solution with $j_{0} < 5$ not in the image of $\varphi$, proving $\F(q+15) = \F(q)+1$.

\noindent The values of $\F(q)$ for $0 \leq q \leq 14$ are verified by direct enumeration, listing all pairs $\left(j\,, k\right) \in \mathbb{N}^2$ with $3j+5k=q$:

\medskip

\begin{equation*}
\renewcommand{\arraystretch}{1.2}
\begin{array}{cc}
\toprule
q & \text{Solutions } \left(j\,, k\right) \\
\midrule
0 & \left(0\,, 0\right) \\
1\,,2\,, 4\,, 7 & \text{none} \\
3 & \left(1\,,0\right) \\
5 & \left(0\,,1\right) \\
6 & \left(2\,,0\right) \\
8 & \left(1\,,1\right) \\
9 & \left(3\,,0\right) \\
10 & \left(0\,,2\right) \\
11 & \left(2\,,1\right) \\
12 & \left(4\,,0\right) \\
13 & \left(1\,,2\right) \\
14 & \left(3\,,1\right) \\ 
\bottomrule
\end{array}
\end{equation*}

\medskip

Moreover, formula \eqref{EquationFFormula} follows by induction using Equation \eqref{EquationFPeriodicity}\,.

\end{proof}

\begin{proof}[Proof of Proposition \ref{PropositionMultiplicityI}]

By Proposition \ref{MultiplicitesVnI}, we have
\begin{equation*}
m(n) = \#\left\{\left(j, k\right) \in \mathbb{N}^{2}\,, 12j+20k=n\right\} + \#\left\{\left(j, k\right)\in\mathbb{N}^{2}\,, 12j+20k=n-30\right\}\,.
\end{equation*}
Since $12j+20k = 4(3j+5k)$ is always divisible by $4$, the first term is non-zero only when $4 \mid n$, and the second only when $4 \mid (n-30)$, i.e.\ $n \equiv 2 \pmod{4}$. In particular, $m(n) = 0$ whenever $n$ is odd\,.

\noindent Moreover, since the two conditions $4\mid n$ and $n \equiv 2\pmod{4}$ are mutually exclusive, at most one of the two terms is non-zero for
any given $n$. Using the function $\F$ of Lemma \ref{LemmaFPeriodicity}, we have
\begin{equation*}
\#\left\{\left(j, k\right) \in \mathbb{N}^{2}\,, 12j+20k=n\right\} = \F\left(\tfrac{n}{4}\right) \qquad \left(\text{when } 4 \mid n\right)\,,
\end{equation*}
and
\begin{equation*}
\#\left\{\left(j, k\right) \in \mathbb{N}^{2}\,, 12j+20k=n-30\right\} = \F\left(\tfrac{n-30}{4}\right) \qquad \left(\text{when } n \equiv 2 \pmod{4}\right)\,.
\end{equation*}
Now write $n = 60q+r$ with $0 \leq r < 60$\,.

\begin{itemize}
\item If $4 \mid r$\,: then $4\mid n$, and $\frac{n}{4} = 15q + r/4$. By \eqref{EquationFFormula} with $l = q$ and $s = \frac{r}{4}$,
\begin{equation*}
\dim(\V^{\widetilde{\mathcal I}}_{n}) = \F(15q + r/4) = q + \eta_{\frac{r}{4}}\,.
\end{equation*}

\item If $r \equiv 2 \pmod{4}$ and $30\leq r < 60$\,:  then $n \equiv 2 \pmod 4$, and $\frac{n-30}{4} = 15q + \frac{r-30}{4}$. Note that $r - 30 \equiv r + 2 \pmod{4}$, so $\frac{r-30}{4}$ is an integer. By \eqref{EquationFFormula} with $l = q$ and $s = \frac{r-30}{4}$,
\begin{equation*}
\dim(\V^{\widetilde{\mathcal I}}_{n})  = \F\left(15q + \frac{r-30}{4}\right) = q + \eta_{\frac{r-30}{4}}\,.
\end{equation*}
\item If $r \equiv 2 \pmod{4}$ and $0\leq r < 30$\,:  then $n \equiv 2 \pmod 4$, and $\frac{n-30}{4} = 15(q-1) + \frac{r+30}{4}$. By \eqref{EquationFFormula} with $l = q-1$ and $s = \frac{r+30}{4}$,
\begin{equation*}
\dim(\V^{\widetilde{\mathcal I}}_{n})  = \F\left(15(q-1) + \frac{r+30}{4}\right) = (q-1) + \eta_{\frac{r+30}{4}}\,.
\end{equation*}
\end{itemize}

\noindent In all three cases, $\dim(\V^{\widetilde{\mathcal I}}_{n}) = q + \varepsilon_r$ where $\varepsilon_{r} \in \left\{0\,, 1\right\}$ depends only on $r$. A direct computation for $0 \leq r < 60$ (by combining the three cases) gives $\varepsilon_{r} = 1$ precisely for
\begin{equation*}
r \in \left\{0\,, 12\,, 20\,, 24\,, 30\,, 32\,, 36\,, 40\,, 42\,, 44\,, 48\,, 50\,, 52\,, 54\,, 56\right\}\,,
\end{equation*}
and $\varepsilon_{r} = 0$ otherwise, completing the proof\,.

\end{proof}

\subsubsection{The heat trace and spectral zeta function}

The kinetic operator is
\begin{equation*}
\mc K_{\widetilde{\mc I}\backslash S^3}(\mf m) = \Delta_{\widetilde{\mc I}\backslash S^3}+\mf m^2\ell^2 
\end{equation*}
with eigenvalues $n(n+2)+\mf m^2 \ell^2 = (n+1)^2-\nu^2,\,\nu^2 = 1-\mf m^2\ell^2$, whose multiplicities are \eqref{eq:IcosahedralMult}.
Since $\abs{\widetilde{\mc I}}=120$, the reparametrization of the representation label $n$ using \eqref{eq:reparam} is $n=120k + 2r$. 
Then the characteristic and twist coefficient are \eqref{eq:AlphaBetaDef}
\begin{equation}
\label{eq:IAlphaBeta}
\alpha_r = \frac{2r+1}{120}, \qquad \beta_r = \frac12\left(\floor{\frac{r}{30}}+\varepsilon_{2r\bmod 60}\right) - \alpha_r, \qquad \left(0 \leq r \leq 59\right).
\end{equation}
Indeed, it is easily verified that the parity properties of the characteristic and twist coefficient \eqref{eq:AlphaBetaParity1} are satisfied here.
Then the heat trace, spectral zeta function and the regularized determinant of the kinetic operator on $\widetilde{\mc I}\bs S^3$ are recovered by substituting $\abs{\widetilde{\mc I}}=120$ and the characteristic and twist coefficient \eqref{eq:IAlphaBeta} into Propositions \ref{prop:HeatGamma} and \ref{prop:ZetaGamma}, Corollary \ref{cor:ZetaGamma}, and Theorem \ref{thm:DetGamma}. 

\subsection{The binary tetrahedral group}

\subsubsection{Embedding of $\widetilde{\mathcal{T}}$ into $\SU(2)$}

It will be convenient to use quaternionic notation. We identify the unit quaternion 
\begin{equation*}
x = x_{0} + x_{1}i + x_{2}j + x_{3}k = \left(x_{0} + x_{1}i\right) + \left(x_{2} + x_{3}i\right)j
\end{equation*}
with the matrix
\begin{equation}
x \longmapsto \begin{pmatrix} x_{0}+ix_{1} & x_{2}+ix_{3} \\ -x_{2}+ix_{3} & x_{0}-ix_{1} \end{pmatrix} \in \SU(2)\,,
\label{QuaternionsSU2}
\end{equation}
which is compatible with Equation \eqref{S3asSU2}. This is a group isomorphism from the group of unit quaternions onto $\SU(2)$, and $\tr(x) = 2x_{0}$. In particular, the quaternion group $\Q_{8} = \left\{\pm 1\,, \pm i\,, \pm j\,, \pm k\right\}$ is the binary dihedral group $D_{2}$.

\begin{definition}

The binary tetrahedral group $\widetilde{\mathcal{T}}$ is the subgroup of $\SU(2)$ of order $24$ that maps onto the tetrahedral rotation group $\mathcal{T} \cong \A_{4}$ under the covering homomorphism $\SU(2) \to \SO(3)$. It is generated by two matrices $a, b \in \SU(2)$, where
\begin{equation}
a = \frac{1}{2}\begin{pmatrix} 1+i & 1+i \\ -1+i & 1-i \end{pmatrix}\,, \qquad b = \begin{pmatrix} -i & 0 \\ 0 & i \end{pmatrix}\,,
\label{GeneratorsT}
\end{equation}
i.e. $a = \frac{1}{2}(1+i+j+k)$ and $b = -i$ in the notation \eqref{QuaternionsSU2}. These generators satisfy
\begin{equation}
a^{6} = \Id_{2}\,, \qquad a^{3} = b^{2} = \left(ab\right)^{3} = -\Id_{2}\,.
\label{RelationGeneratorsT}
\end{equation}

\label{DefinitionBinaryTetrahedralGroup}

\end{definition}

\begin{proposition}

We have
\begin{equation*}
\widetilde{\mathcal{T}} = \Q_{8} \sqcup a\Q_{8} \sqcup a^{2}\Q_{8} = \Q_{8} \,\sqcup\, \left\{\tfrac{1}{2}\left(\pm 1 \pm i \pm j \pm k\right)\right\}\,.
\end{equation*}
In particular, $\left|\widetilde{\mathcal{T}}\right| = 24$.

\end{proposition}

\noindent The proof follows from a straightforward computation and is omitted.

\noindent The element $-\Id_{2} \in \widetilde{\mathcal{T}}$ is central, and $\widetilde{\mathcal{T}}/\left\{\pm\Id_{2}\right\} \cong \mathcal{T} \cong \A_{4}$. The group $\widetilde{\mathcal{T}} \cong \widetilde{\mathcal{T}} \times \left\{\Id_{2}\right\} \subseteq \SU(2) \times \SU(2)$ acts freely on $S^{3} \cong \SU(2)$ by left multiplication (see Equation \eqref{ActionSU2SU2}). In particular, it follows from Equation \eqref{EquationMGammaN} that
\begin{equation*}
m_{\Gamma}(n) = \left(n+1\right)\dim(\V_{n}^{\widetilde{\mathcal{T}}})\,.
\end{equation*}

\subsubsection{The coefficients $m_{\Gamma}(n)$}

As for the binary icosahedral group, we use the Molien series. The group $\widetilde{\mathcal{T}}$ has $7$ conjugacy classes, given in the next table. Recall that, for $g \in \SU(2)$, $\det(\Id_{2}-qg) = 1-\tr(g)q+q^{2}$.

\bigskip

\begin{center}
\renewcommand{\arraystretch}{1.2}
\begin{tabular}{ccccc}
\toprule
Class & Size & Trace & Order & Determinant \\
\midrule
$\left\{1\right\}$ & $1$ & $2$ & $1$ & $(1-q)^{2}$ \\
$\left\{-1\right\}$ & $1$ & $-2$ & $2$ & $(1+q)^{2}$ \\
$\left\{\pm i\,, \pm j\,, \pm k\right\}$ & $6$ & $0$ & $4$ & $1+q^{2}$ \\
$\left[a\right]$ & $4$ & $1$ & $6$ & $1-q+q^{2}$ \\
$\left[a^{-1}\right]$ & $4$ & $1$ & $6$ & $1-q+q^{2}$ \\
$\left[a^{2}\right]$ & $4$ & $-1$ & $3$ & $1+q+q^{2}$ \\
$\left[a^{-2}\right]$ & $4$ & $-1$ & $3$ & $1+q+q^{2}$ \\
\bottomrule
\end{tabular}
\end{center}

\bigskip

\noindent Therefore, it follows from Theorem \ref{TheoremMGammaQ} and Remark \ref{RemarkConjugacyClasses} that
\begin{equation*}
24\mathcal{M}_{\widetilde{\mathcal{T}}}(q) = \frac{1}{(1-q)^{2}} + \frac{1}{(1+q)^{2}} + \frac{6}{1+q^{2}} + \frac{8}{1-q+q^{2}} + \frac{8}{1+q+q^{2}}\,.
\end{equation*}

\begin{lemma}

For all $q$ such that $\left|q\right| < 1$, we get
\begin{equation}
\mathcal{M}_{\widetilde{\mathcal{T}}}(q) = \frac{1+q^{12}}{(1-q^{6})(1-q^{8})}\,.
\label{LemmaMTQ}
\end{equation}

\end{lemma}

\begin{proof}

Using Equation \eqref{KeyIdentity} with $\tau = 2$ and $\tau = 1$, we get
\begin{equation*}
\frac{1}{(1-q)^{2}} + \frac{1}{(1+q)^{2}} = \frac{2(1+q^{2})}{(1-q^{2})^{2}}\,, \qquad \frac{1}{1-q+q^{2}} + \frac{1}{1+q+q^{2}} = \frac{2(1+q^{2})}{1+q^{2}+q^{4}}\,.
\end{equation*}
Combining the terms over a common denominator and simplifying, we obtain
\begin{equation*}
\mathcal{M}_{\widetilde{\mathcal{T}}}(q) = \frac{q^{8}-q^{4}+1}{(q-1)^{2}(q+1)^{2}(q^{2}+1)(q^{2}-q+1)(q^{2}+q+1)}\,.
\end{equation*}
On the other hand, $1+q^{12} = (q^{4}+1)(q^{8}-q^{4}+1)$ and
\begin{equation*}
(1-q^{6})(1-q^{8}) = (q-1)^{2}(q+1)^{2}(q^{2}+1)(q^{2}-q+1)(q^{2}+q+1)(q^{4}+1)\,.
\end{equation*}
After cancelling the common factor $q^{4}+1$, the two fractions coincide.

\end{proof}

\noindent Using the geometric series expansion, we obtain
\begin{equation*}
\mathcal{M}_{\widetilde{\mathcal{T}}}(q) = \sum\limits_{j \geq 0}\sum\limits_{k \geq 0} q^{6j+8k} + \sum\limits_{j \geq 0}\sum\limits_{k \geq 0} q^{6j+8k+12}\,,
\end{equation*}
so that, for all $n \geq 0$,
\begin{equation}
\dim(\V^{\widetilde{\mathcal{T}}}_{n}) = \#\left\{\left(j\,, k\right) \in \mathbb{N}^{2}\,, 6j+8k = n\right\} + \#\left\{\left(j\,, k\right) \in \mathbb{N}^{2}\,, 6j+8k = n-12\right\}\,.
\label{MultiplicitesVnT}
\end{equation}
In particular, $\dim(\V^{\widetilde{\mathcal{T}}}_{n}) = 0$ if $n$ is odd. The first values for $n$ even are

\medskip

\begin{center}
\begin{tabular}{ccccccccccccc}
\toprule
$n$ & $0$ & $2$ & $4$ & $6$ & $8$ & $10$ & $12$ & $14$ & $16$ & $18$ & $20$ & $22$ \\
\midrule
$\dim(\V^{\widetilde{\mathcal{T}}}_{n})$ & $1$ & $0$ & $0$ & $1$ & $1$ & $0$ & $2$ & $1$ & $1$ & $2$ & $2$ & $1$ \\
\bottomrule
\end{tabular}
\end{center}

\medskip

\begin{lemma}

\label{LemmaF34}

For all $l \in \mathbb{Z}$, define $\F_{3,4}(l) = \#\left\{\left(j\,, k\right) \in \mathbb{N}^{2}\,, 3j+4k = l\right\}$
(so that $\F_{3,4}(l) = 0$ if $l < 0$). Then, for all $l \geq 0$,
\begin{equation*}
\F_{3,4}(l+12) = \F_{3,4}(l)+1\,.
\end{equation*}
Moreover, writing $l = 12p+s$ with $0 \leq s \leq 11$, we have $\F_{3,4}(l) = p+\eta_{s}$, where
$\eta_{s} = 0$ if $s \in \left\{1\,, 2\,, 5\right\}$ and $\eta_{s} = 1$ otherwise.

\end{lemma}

\begin{proof}

The proof is similar to that of Lemma \ref{TechnicalLemma120}.

\end{proof}

\begin{proposition}

\label{PropositionMultiplicityT}

We denote by $\mathscr{R}_{\widetilde{\mathcal{T}}}$ the set $\mathscr{R}_{\widetilde{\mathcal{T}}} = \left\{0\,, 6\,, 8\right\}$,
and we let $\varepsilon_{r} = 1$ if $r \in \mathscr{R}_{\widetilde{\mathcal{T}}}$ and $\varepsilon_{r} = 0$ otherwise. Then
\begin{equation*}
\dim(\V^{\widetilde{\mathcal{T}}}_{n}) = \begin{cases} 0 & n \text{ odd} \\
\left\lfloor\frac{n}{12}\right\rfloor+\varepsilon_{n\bmod 12} & n \text{ even}\end{cases}
\end{equation*}
i.e.
\begin{equation}\label{eq:TetrahedralMult}
m_{\widetilde{\mathcal{T}}}(n) = \begin{cases} 0 & \text{ if n is odd} \\
(n+1)\left(\left\lfloor\frac{n}{12}\right\rfloor + \varepsilon_{n \bmod 12}\right) & \text{ if n is even} \end{cases}
\end{equation}

\end{proposition}

\begin{proof}

Let $n$ be even and write $n = 24p+2s$ with $0 \leq s \leq 11$. Unlike the case of $\widetilde{\mathcal{I}}$, both terms in Equation \eqref{MultiplicitesVnT} can be non-zero, and
\begin{equation*}
\dim(\V^{\widetilde{\mathcal{T}}}_{n}) = \F_{3,4}(12p+s) + \F_{3,4}(12p+s-6)\,.
\end{equation*}
By Lemma \ref{LemmaF34}, the first term equals $p+\eta_{s}$. The second term equals $p+\eta_{s-6}$ if $s \geq 6$, and $p$ if $s \leq 5$ (since $\eta_{s+6} = 1$, and the term vanishes when $p = 0$). Since $\eta_{s} = 1$ for $s \geq 6$, in both cases
\begin{equation*}
\dim(\V^{\widetilde{\mathcal{T}}}_{n}) = 2p + \left\lfloor\tfrac{s}{6}\right\rfloor + \eta_{s \bmod 6} = \left\lfloor\tfrac{n}{12}\right\rfloor + \varepsilon_{n \bmod 12}\,,
\end{equation*}
because $\eta_{t} = \varepsilon_{2t}$ for $0 \leq t \leq 5$.

\end{proof}

\subsubsection{The heat trace and spectral zeta function}

The kinetic operator is
\begin{equation*}\mc K_{\widetilde{\mc T}\backslash S^3}(\mf m) = \Delta_{\widetilde{\mc T}\backslash S^3}+\mf m^2\ell^2 
\end{equation*}
with eigenvalues $n(n+2)+\mf m^2 \ell^2 = (n+1)^2-\nu^2,\,\nu^2 = 1-\mf m^2\ell^2$, whose multiplicities are \eqref{eq:TetrahedralMult}.
Since $\abs{\widetilde{\mc T}}=24$, the reparametrization of the representation label $n$ using \eqref{eq:reparam} is $n=24k + 2r$. 
Then the characteristic and twist coefficient are \eqref{eq:AlphaBetaDef}
\begin{equation}
\label{eq:TAlphaBeta}
\alpha_r = \frac{2r+1}{24}, \qquad \beta_r = \frac12\left(\floor{\frac{r}{6}}+\varepsilon_{2r\bmod 12}\right) - \alpha_r, \qquad \left(0\leq r \leq 11\right).
\end{equation}
It is easily verified that the parity properties of the characteristic and twist coefficient \eqref{eq:AlphaBetaParity1} are satisfied here.
Then the heat trace, spectral zeta function and the regularized determinant of the kinetic operator on $\widetilde{\mc T}\bs S^3$ are recovered by substituting $\abs{\widetilde{\mc T}}=24$ and the characteristic and twist coefficient \eqref{eq:TAlphaBeta} into Propositions \ref{prop:HeatGamma} and \ref{prop:ZetaGamma}, Corollary \ref{cor:ZetaGamma}, and Theorem \ref{thm:DetGamma}. 

\subsection{The binary octahedral group}

\subsubsection{Embedding of $\widetilde{\mathcal{O}}$ into $\SU(2)$}

We keep the quaternionic notation \eqref{QuaternionsSU2}.

\begin{definition}

The binary octahedral group $\widetilde{\mathcal{O}}$ is the subgroup of $\SU(2)$ of order $48$ that maps onto the octahedral rotation group $\mathcal{O} \cong \mathscr{S}_{4}$ under the covering homomorphism $\SU(2) \to \SO(3)$. It is generated by two matrices $a, c \in \SU(2)$, where $a$ is given in Equation \eqref{GeneratorsT} and
\begin{equation}
c = \frac{1}{\sqrt{2}}\begin{pmatrix} i & 1 \\ -1 & -i \end{pmatrix}\,,
\label{GeneratorsO}
\end{equation}
i.e.\ $c = \frac{1}{\sqrt{2}}(i+j)$ in the notation \eqref{QuaternionsSU2}. These generators satisfy
\begin{equation}
a^{6} = \Id_{2}\,, \qquad a^{3} = c^{2} = \left(ac\right)^{4} = -\Id_{2}\,.
\label{RelationGeneratorsO}
\end{equation}

\label{DefinitionBinaryOctahedralGroup}

\end{definition}

\begin{proposition}

We have $\widetilde{\mathcal{T}} \subseteq \widetilde{\mathcal{O}}$, and
\begin{equation*}
\widetilde{\mathcal{O}} = \widetilde{\mathcal{T}} \sqcup c \widetilde{\mathcal{T}}\,, \qquad c\widetilde{\mathcal{T}} = \left\{\tfrac{1}{\sqrt{2}}\left(\pm u \pm v\right)\,, u \neq v \in \left\{1\,, i\,, j\,, k\right\}\right\}\,.
\end{equation*}
In particular, $\left|\widetilde{\mathcal{O}}\right| = 48$.

\end{proposition}

\noindent The proof follows from a straightforward computation and is omitted.

\noindent The element $-\Id_{2} \in \widetilde{\mathcal{O}}$ is central, and $\widetilde{\mathcal{O}}/\left\{\pm\Id_{2}\right\} \cong \mathcal{O} \cong \mathscr{S}_{4}$. As before, $\widetilde{\mathcal{O}}$ acts freely on $S^{3} \cong \SU(2)$ by left multiplication, and
\begin{equation*}
m_{\Gamma}(n) = \left(n+1\right)\dim(\V_{n}^{\widetilde{\mathcal{O}}})\,.
\end{equation*}

\subsubsection{The coefficients $m_{\Gamma}(n)$}

The group $\widetilde{\mathcal{O}}$ has $8$ conjugacy classes, given in the following table.

\bigskip

\begin{center}
\renewcommand{\arraystretch}{1.2}
\begin{tabular}{ccccc}
\toprule
Class & Size & Trace & Order & Determinant \\
\midrule
$\left\{1\right\}$ & $1$ & $2$ & $1$ & $(1-q)^{2}$ \\
$\left\{-1\right\}$ & $1$ & $-2$ & $2$ & $(1+q)^{2}$ \\
$\left\{\pm i\,, \pm j\,, \pm k\right\}$ & $6$ & $0$ & $4$ & $1+q^{2}$ \\
$\left\{\frac{1}{\sqrt{2}}(\pm u \pm v)\,, \ u \neq v \in \left\{i\,, j\,, k\right\}\right\}$ & $12$ & $0$ & $4$ & $1+q^{2}$ \\
$\left\{\frac{1}{2}(1 \pm i \pm j \pm k)\right\}$ & $8$ & $1$ & $6$ & $1-q+q^{2}$ \\
$\left\{\frac{1}{2}(-1 \pm i \pm j \pm k)\right\}$ & $8$ & $-1$ & $3$ & $1+q+q^{2}$ \\
$\left\{\frac{1}{\sqrt{2}}(1 \pm u)\,, \ u \in \left\{i\,, j\,, k\right\}\right\}$ & $6$ & $\sqrt{2}$ & $8$ & $1-\sqrt{2}q+q^{2}$ \\
$\left\{\frac{1}{\sqrt{2}}(-1 \pm u)\,, \ u \in \left\{i\,, j\,, k\right\}\right\}$ & $6$ & $-\sqrt{2}$ & $8$ & $1+\sqrt{2}q+q^{2}$ \\
\bottomrule
\end{tabular}
\end{center}

\bigskip

\noindent Therefore, it follows from Theorem \ref{TheoremMGammaQ} and Remark \ref{RemarkConjugacyClasses} that
\begin{eqnarray*}
48\,\mathcal{M}_{\widetilde{\mathcal{O}}}(q) & = & \frac{1}{(1-q)^{2}} + \frac{1}{(1+q)^{2}} + \frac{18}{1+q^{2}} + \frac{8}{1-q+q^{2}} \\
&  & + \frac{8}{1+q+q^{2}} + \frac{6}{1-\sqrt{2}q+q^{2}} + \frac{6}{1+\sqrt{2}q+q^{2}}\,.
\end{eqnarray*}

\begin{lemma}

For all $q$ such that $\left|q\right| < 1$, we get
\begin{equation}
\mathcal{M}_{\widetilde{\mathcal{O}}}(q) = \frac{1+q^{18}}{(1-q^{8})(1-q^{12})}\,.
\label{LemmaMOQ}
\end{equation}

\end{lemma}

\begin{proof}

Using Equation \eqref{KeyIdentity} with $\tau = 2$, $\tau = 1$ and $\tau = \sqrt{2}$, where in the last case
\begin{equation*}
\frac{1}{1-\sqrt{2}q+q^{2}} + \frac{1}{1+\sqrt{2}q+q^{2}} = \frac{2(1+q^{2})}{1+q^{4}}\,,
\end{equation*}
and combining the terms over a common denominator, we obtain
\begin{equation*}
\mathcal{M}_{\widetilde{\mathcal{O}}}(q) = \frac{q^{12}-q^{6}+1}{(q-1)^{2}(q+1)^{2}(q^{2}+1)(q^{4}+1)(q^{2}-q+1)(q^{2}+q+1)}\,.
\end{equation*}
On the other hand, $1+q^{18} = (q^{2}+1)(q^{4}-q^{2}+1)(q^{12}-q^{6}+1)$ and
\begin{equation*}
(1-q^{8})(1-q^{12}) = (q-1)^{2}(q+1)^{2}(q^{2}+1)^{2}(q^{4}+1)(q^{2}-q+1)(q^{2}+q+1)(q^{4}-q^{2}+1)\,.
\end{equation*}
After cancelling the common factor $(q^{2}+1)(q^{4}-q^{2}+1)$, the two fractions coincide.

\end{proof}

\noindent Using the geometric series expansion as before, we obtain, for all $n \geq 0$,

\medskip

\begin{equation}
\dim(\V^{\widetilde{\mathcal{O}}}_{n}) = \#\left\{\left(j\,, k\right) \in \mathbb{N}^{2}\,, 8j+12k = n\right\} + \#\left\{\left(j\,, k\right) \in \mathbb{N}^{2}\,, 8j+12k = n-18\right\}\,.
\label{MultiplicitesVnO}
\end{equation}

\noindent In particular, $\dim(\V^{\widetilde{\mathcal{O}}}_{n}) = 0$ if $n$ is odd. The first non-zero values are

\medskip

\begin{center}
\begin{tabular}{ccccccccccccc}
\toprule
$n$ & $0$ & $8$ & $12$ & $16$ & $18$ & $20$ & $24$ & $26$ & $28$ & $30$ & $32$ & $34$ \\
\midrule
$\dim(\V^{\widetilde{\mathcal{O}}}_{n})$ & $1$ & $1$ & $1$ & $1$ & $1$ & $1$ & $2$ & $1$ & $1$ & $1$ & $2$ & $1$ \\
\bottomrule
\end{tabular}
\end{center}

\medskip

\begin{lemma}

\label{LemmaF23}

For all $l \in \mathbb{Z}$, define $\F_{2,3}(l) = \#\left\{\left(j\,, k\right) \in \mathbb{N}^{2}\,, 2j+3k = l\right\}$
(so that $\F_{2,3}(l) = 0$ if $l < 0$). Then, for all $l \geq 0$,
\begin{equation*}
\F_{2,3}(l+6) = \F_{2,3}(l)+1\,.
\end{equation*}
Moreover, writing $l = 6p+s$ with $0 \leq s \leq 5$, we have $\F_{2,3}(l) = p+\theta_{s}$, where
$\theta_{1} = 0$ and $\theta_{s} = 1$ for $s \neq 1$.

\end{lemma}

\begin{proof}

The proof is similar to that of Lemma \ref{TechnicalLemma120}.

\end{proof}

\begin{proposition}

\label{PropositionMultiplicityO}

We denote by $\mathscr{R}_{\widetilde{\mathcal{O}}}$ the set
$\mathscr{R}_{\widetilde{\mathcal{O}}} = \left\{0\,, 8\,, 12\,, 16\,, 18\,, 20\right\}$, and we let $\varepsilon_{r} = 1$ if
$r \in \mathscr{R}_{\widetilde{\mathcal{O}}}$ and $\varepsilon_{r} = 0$ otherwise. Then
\begin{equation}
\dim(\V^{\widetilde{\mathcal{O}}}_{n}) = \begin{cases} 0 & n \text{ odd} \\
\left\lfloor\frac{n}{24}\right\rfloor+\varepsilon_{n\bmod 24} & n \text{ even},\end{cases}
\end{equation}
i.e.
\begin{equation}\label{eq:OctahedralMult}
m_{\widetilde{\mathcal{O}}}(n) = \begin{cases} 0 & n \text{ odd} \\
(n+1)\left(\left\lfloor\frac{n}{24}\right\rfloor + \varepsilon_{n\bmod 24}\right) & n \text{ even}.
\end{cases}
\end{equation}

\end{proposition}

\begin{proof}

The proof is similar to that of Proposition \ref{PropositionMultiplicityT}.

\end{proof}

\subsubsection{The heat trace and spectral zeta function}

The kinetic operator is
\begin{equation*}
\mc K_{\widetilde{\mc O}\backslash S^3}(\mf m) = \Delta_{\widetilde{\mc O}\backslash S^3}+\mf m^2\ell^2 
\end{equation*}
with eigenvalues $n(n+2)+\mf m^2 \ell^2 = (n+1)^2-\nu^2,\,\nu^2 = 1-\mf m^2\ell^2$, whose multiplicities are \eqref{eq:OctahedralMult}.
Since $\abs{\widetilde{\mc O}}=48$, the reparametrization of the representation label $n$ using \eqref{eq:reparam} is $n=48k + 2r$. 
Then the characteristic and twist coefficient are \eqref{eq:AlphaBetaDef}
\begin{equation}\label{eq:OAlphaBeta}
\alpha_r = \frac{2r+1}{48}, \qquad \beta_r = \frac12\left(\floor{\frac{r}{12}}+\varepsilon_{2r\bmod 24}\right) - \alpha_r, \qquad \left(0\leq r \leq 23\right).
\end{equation}
It is easily verified that the parity properties of the characteristic and twist coefficient \eqref{eq:AlphaBetaParity1} are satisfied here.
Then the heat trace, spectral zeta function and the regularized determinant of the kinetic operator on $\widetilde{\mc O}\bs S^3$ are recovered by substituting $\abs{\widetilde{\mc O}}=48$ and the characteristic and twist coefficient \eqref{eq:OAlphaBeta} into Propositions \ref{prop:HeatGamma} and \ref{prop:ZetaGamma}, Corollary \ref{cor:ZetaGamma}, and Theorem \ref{thm:DetGamma}.

\section{Extension to higher dimensions and future directions}
\label{SectionFutureProjects}

In this paper, we have used the identification $S^{3} \cong \SU(2)$ to study the spectral zeta function and the regularized determinant of the Laplace-Beltrami operator on $S^{3}$ and its quotients by finite groups.
However, as explained in Remark \ref{RemarkS3}, the sphere $S^{3}$ can also be realized as the homogeneous space
\begin{equation*}
S^{3}\cong\SO(4)/\SO(3)\,.
\end{equation*}
This second realization is particularly important, as it provides a natural framework for extending our methods to higher-dimensional spheres. Before discussing these extensions, we recall some results from the representation
theory of compact groups.

\subsection{Decomposition of
\texorpdfstring{$\L^{2}(\Gamma\backslash\G/\K)$}
{L2(Gamma backslash G/K)}}

\label{SectionLTwoGModK}

Let $\G$ be a compact Lie group and let $\K$ be a closed subgroup of $\G$. We consider the homogeneous space
\begin{equation*}
\G/\K=\left\{g\K\,,g\in\G\right\}\,.
\end{equation*}
We denote by $d\dot{g}$ the normalized $\G$-invariant measure on $\G/\K$. A function on $\G/\K$ can be identified with a right $\K$-invariant function on $\G$, i.e. a function satisfying
\begin{equation*}
f(gk)=f(g)\,, \qquad \left(g\in\G\,, k\in\K\right)\,.
\end{equation*}
Therefore,
\begin{equation}
\L^{2}(\G/\K) \cong \L^{2}(\G)^{\K}\,.
\label{IdentificationLTwoGModK}
\end{equation}
Using Equation \eqref{PeterWeylEquation}, we have
\begin{equation*}
\L^{2}(\G) = \mathop{\widehat{\bigoplus}}_{\pi\in\widehat{\G}} \V_{\pi} \otimes \V^{*}_{\pi}\,.
\end{equation*}
Taking the $\K$-invariant vectors for the right action, we obtain
\begin{equation*}
\L^{2}(\G/\K) \cong \mathop{\widehat{\bigoplus}}_{\pi\in\widehat{\G}} \left(\V_{\pi}\otimes\V_{\pi}^{*}\right)^{\K} \cong \mathop{\widehat{\bigoplus}}_{\pi\in\widehat{\G}} \V_{\pi}\otimes\left(\V_{\pi}^{\K}\right)^{*}\,.
\end{equation*}
Then
\begin{equation}
\L^{2}(\G/\K) \cong \mathop{\widehat{\bigoplus}}_{\substack{\pi\in\widehat{\G} \\
\V_{\pi}^{\K}\neq\{0\}}} \dim\left(\V_{\pi}^{\K}\right)\V_{\pi}\,.
\label{PeterWeylGModK}
\end{equation}
In particular, only those irreducible representations of $\G$ whose restriction to $\K$ contains the trivial representation contribute to this decomposition.

\noindent Now let $\Gamma$ be a finite subgroup of $\G$. Since
\begin{equation*}
\L^{2}(\Gamma \backslash \G / \K) \cong \L^{2}(\G/\K)^{\Gamma}\,,
\end{equation*}
taking the $\Gamma$-invariant vectors in Equation \eqref{PeterWeylGModK} gives
\begin{equation}
\L^{2}(\Gamma\backslash\G/\K) \cong \mathop{\widehat{\bigoplus}}_{\pi\in\widehat{\G}}
\V_{\pi}^{\Gamma} \otimes \left(\V_{\pi}^{\K}\right)^{*}\,.
\label{PeterWeylGammaGModK}
\end{equation}
In particular, the dimension of the corresponding summand is
\begin{equation*}
\dim\left(\V_{\pi}^{\Gamma}\right)\dim\left(\V_{\pi}^{\K}\right)\,.
\end{equation*}

\subsection{The case of \texorpdfstring{$S^{3}\cong\SO(4)/\SO(3)$}
{S3 = SO(4)/SO(3)}}

We now apply the previous decomposition to $\G=\SO(4)$ and $\K=\SO(3)$. Recall from Remark \ref{RemarkS3} that
\begin{equation*}
\Spin(4) \cong\SU(2) \times \SU(2)
\end{equation*}
and
\begin{equation*}
\SO(4)\cong \frac{\SU(2)\times\SU(2)}{\left\{(\Id_{2},\Id_{2}), (-\Id_{2},-\Id_{2})\right\}}\,.
\end{equation*}
Let $\V_{n}$ denote the irreducible representation
of $\SU(2)$ of dimension $n+1$, and set
\begin{equation*}
\V_{n,m} = \V_{n} \otimes \V_{m}\,.
\end{equation*}
The representations $\V_{n,m}$ exhaust the irreducible representations of $\Spin(4)$. Moreover, since
\begin{equation*}
\pi_{n,m}(-\Id_{2},-\Id_{2}) = (-1)^{n+m}\Id_{\V_{n,m}}\,,
\end{equation*}
the representation $\V_{n,m}$ factors through $\SO(4)$ if and only if $n+m$ is even.

\noindent In particular, Peter-Weyl gives
\begin{equation}
\L^{2}(\SO(4)) \cong \bigoplus\limits_{\substack{n, m \geq 0 \\ n+m \text{ even}}} \V_{n,m} \otimes \V_{n,m}^{*}\,.
\label{PeterWeylSO4}
\end{equation}
Under the covering map $\Spin(4)\longrightarrow\SO(4)$, the subgroup $\SO(3)$ corresponds to the diagonal subgroup
\begin{equation*}
\SU(2)_{\Delta} = \left\{\left(g\,, g\right)\,, g \in \SU(2)\right\}\,.
\end{equation*}
Thus, by Equation \eqref{PeterWeylGModK},
\begin{equation*}
\L^{2}(\SO(4)/\SO(3)) \cong \bigoplus\limits_{\substack{n,m \geq 0 \\ n+m \text{ even}}} \V_{n,m} \otimes \left(\V_{n,m}^{\SU(2)_{\Delta}}\right)^{*}\,.
\end{equation*}
Recall that the Clebsch--Gordan decomposition gives
\begin{equation*}
\V_{n} \otimes \V_{m} \cong \V_{n+m}\oplus\V_{n+m-2} \oplus \cdots \oplus\V_{\left|n-m\right|}\,.
\end{equation*}
It follows that the trivial representation $\V_{0}$ appears if and only if $n=m$, and in this case with multiplicity one.

\noindent We therefore obtain
\begin{equation}
\L^{2}( S^3) \cong \L^{2}(\SO(4)/\SO(3)) \cong \bigoplus\limits_{n=0}^{\infty} \V_{n,n}\,.
\label{PeterWeylS3}
\end{equation}
In particular,
\begin{equation*}
\dim(\V_{n,n})=(n+1)^{2}\,,
\end{equation*}
recovering the multiplicities of the eigenvalues $n(n+2)$ of the Laplace-Beltrami operator on $S^{3}$.

\subsection{The case of \texorpdfstring{$S^{2n-1}\cong\SO(2n)/\SO(2n-1)$}
{S2n-1 = SO(2n)/SO(2n-1)}}

The realization of $S^{3}$ as the homogeneous space $\SO(4)/\SO(3)$ extends naturally to higher-dimensional spheres. For simplicity, we restrict our attention to odd-dimensional spheres, since the representation theory of $\SO(n)$ depends on the parity of $n$.
The general case will be treated in \cite{VRAS}, along with the massive extension.

For every $n \geq 2$, the group $\SO(2n)$ acts transitively on the unit sphere
\begin{equation*}
S^{2n-1} = \left\{x\in\mathbb{R}^{2n}\,,\|x\|=1\right\}\,.
\end{equation*}
The stabilizer of the vector $e_{2n} = \left(0\,, \ldots\,, 0\,, 1\right)$ is isomorphic to $\SO(2n-1)$. Therefore, we obtain the identification
\begin{equation}
S^{2n-1} \cong \SO(2n)/\SO(2n-1)\,.
\label{HomogeneousSphereHigherDimension}
\end{equation}
\noindent Recall that (see \cite[Chapter~V]{Knapp}) the irreducible representations of $\SO(2n)$ are parametrized by dominant integral highest weights
\begin{equation*}
\lambda = \left(\lambda_{1}\,, \ldots\,, \lambda_{n}\right) \in \mathbb{Z}^{n}
\end{equation*}
satisfying
\begin{equation*}
\lambda_{1} \geq \lambda_{2} \geq \cdots \geq \lambda_{n-1} \geq \left|\lambda_{n}\right|\,.
\end{equation*}
We denote by $\L_{2n}(\lambda)$ the irreducible representation of $\SO(2n)$ with highest weight $\lambda$.

\noindent By Equation \eqref{PeterWeylGModK}, the decomposition of $\L^{2}(S^{2n-1})$ is determined by the irreducible representations $\L_{2n}(\lambda)$ whose restriction to $\SO(2n-1)$ contains the trivial representation. We recall the following result.

\begin{theo}[Branching rule]

We have
\begin{equation*}
\L_{2n}(\lambda)^{\SO(2n-1)} \neq\{0\} \quad \text{ if and only if } \quad  \lambda = \left(m\,, 0\,, \ldots\,, 0\right)\,, \quad \left(m \in \mathbb{Z}_{\geq 0}\right)\,.
\end{equation*}
Moreover, in this case,
\begin{equation*}
\dim\left(\L_{2n}(m\,, 0\,, \ldots\,, 0)^{\SO(2n-1)}\right)=1\,.
\end{equation*}

\label{TheoremBranchingRule}

\end{theo}

\begin{proof}

The proof can be found in \cite[Chapter~IX]{Knapp}.

\end{proof}

\noindent For every $m \geq 0$, let
\begin{equation*}
\lambda_{m}=(m\,,0\,,\ldots\,,0)\,.
\end{equation*}
It follows from Equations \eqref{PeterWeylEquation} and \eqref{PeterWeylGModK} that
\begin{equation}
\L^{2}(S^{2n-1}) \cong \mathop{\widehat{\bigoplus}}_{m=0}^{\infty} \L_{2n}(\lambda_{m})\,.
\label{PeterWeylHigherDimensionalSphere}
\end{equation}

\noindent As in the case of $S^{3}$, the eigenvalues of the Laplace-Beltrami operator can be obtained from the action of the Casimir element. For the standard round metric on $S^{2n-1}$, the corresponding Casimir eigenvalue is
\begin{equation}
c(\lambda_{m}) = \langle\lambda_{m}\,, \lambda_{m}+2\rho\rangle = m(m+2n-2)\,.
\label{CasimirHigherDimensionalSphere}
\end{equation}
Furthermore, the dimension of $\L_{2n}(\lambda_{m})$ is
\begin{equation}
d_{m} = \binom{m+2n-1}{m} - \binom{m+2n-3}{m-2} = \frac{(2m+2n-2)(m+2n-3)!}{m!(2n-2)!}\,,
\label{DimensionHigherDimensionalSphere}
\end{equation}
where the second binomial coefficient is understood to be zero when $m<2$.

\noindent In particular, we obtain the following expression for the spectral zeta function.

\begin{proposition}

Let $n\geq 2$. The spectral zeta function of the Laplace-Beltrami operator on $S^{2n-1}$ is given, for $\Re(s)>\frac{2n-1}{2}$, by
\begin{equation}
\zeta_{\Delta_{S^{2n-1}}}(s) = \sum\limits_{m=1}^{\infty} \frac{(2m+2n-2)(m+2n-3)!}{m!(2n-2)!\left(m(m+2n-2)\right)^{s}}\,.
\label{SpectralZetaHigherDimensionalSphere}
\end{equation}
Here the zero eigenvalue, corresponding to $m=0$, is omitted.

\end{proposition}

\noindent Indeed, the formula follows from Equations \eqref{PeterWeylHigherDimensionalSphere}, \eqref{CasimirHigherDimensionalSphere}, and \eqref{DimensionHigherDimensionalSphere}. Moreover, as $m \to \infty$, we have
\begin{equation*}
d_{m} \sim \frac{2}{(2n-2)!}m^{2n-2}\,, \qquad c(\lambda_{m})\sim m^{2}\,,
\end{equation*}
which gives the stated domain of absolute convergence.

\medskip

\noindent We now turn to the meromorphic continuation of the spectral zeta function. As in the case of $S^{3}$, the main idea is to express the multiplicities as a polynomial in a shifted spectral parameter, and then use the binomial expansion together with the Hurwitz zeta function. For every $n\geq2$, let $\P_{n}$ be the polynomial
defined by
\begin{equation}
\P_{n}(x) = \frac{2x}{(2n-2)!} \prod\limits_{j=1}^{2n-3}\left(x+j-(n-1)\right) = \sum\limits_{l=0}^{2n-2} c_{n,l}x^{l}\,.
\label{PolynomialMultiplicityHigherSphere}
\end{equation}
Setting $x_{m}=m+n-1$, we get
\begin{equation*}
d_{m} = \P_{n}(x_{m}) \qquad \text{ and } \qquad m(m+2n-2)=x_{m}^{2}-(n-1)^{2}\,.
\end{equation*}

\begin{proposition}

Let $n \geq 2$, and let $c_{n,l}$ the coefficients defined in Equation \eqref{PolynomialMultiplicityHigherSphere}. Then the spectral zeta function of the Laplace-Beltrami operator on $S^{2n-1}$ admits a meromorphic continuation to $\mathbb{C}$, given by
\begin{equation}
\zeta_{\Delta_{S^{2n-1}}}(s) = \sum\limits_{l=0}^{2n-2}c_{n,l}\sum\limits_{k=0}^{\infty}\frac{(s)_{k}}{k!}(n-1)^{2k}\zeta_{H}\left(2s+2k-l\,,n\right)\,.
\label{MeromorphicExtensionHigherSphere}
\end{equation}
The series on the right-hand side converges normally on compact subsets of $\mathbb{C}$ avoiding its possible poles.

\end{proposition}

\noindent The argument is analogous to the one used for $S^{3}$. Indeed, for $m \geq 1$,
\begin{equation*}
0 \leq \frac{(n-1)^{2}}{x_{m}^{2}} \leq \left(\frac{n-1}{n}\right)^{2} < 1\,,
\end{equation*}
and hence
\begin{equation*}
\left(x_{m}^{2}-(n-1)^{2}\right)^{-s} = x_{m}^{-2s} \sum\limits_{k=0}^{\infty} \frac{(s)_{k}}{k!}\frac{(n-1)^{2k}}{x_{m}^{2k}}\,.
\end{equation*}
Substituting the polynomial expression for $d_{m}$ and using
\begin{equation*}
\sum\limits_{m=1}^{\infty} (m+n-1)^{-z} = \zeta_{H}(z, n)
\end{equation*}
gives Equation \eqref{MeromorphicExtensionHigherSphere} in the half-plane of absolute convergence. The meromorphic continuation follows from the meromorphic continuation of the Hurwitz zeta function and the normal convergence of the resulting series.

\noindent Thus, the representation-theoretic approach developed for $S^{3}$ also provides explicit spectral zeta functions and their meromorphic continuations in every odd dimension.

\subsection{Higher-dimensional lens spaces and Molien series}

We now consider quotients of $S^{2n-1}$ by finite subgroups of $\SO(2n)$. By Equation \eqref{PeterWeylGammaGModK}, the decomposition of $\L^{2}(\Gamma \backslash S^{2n-1})$ is determined by the spaces of $\Gamma$-invariant vectors in the representations $\L_{2n}(\lambda_{m})$. More precisely, it follows from Section \ref{SectionLTwoGModK}
\begin{equation}
\L^{2}(\Gamma \backslash S^{2n-1}) \cong \widehat{\bigoplus}_{m=0}^{\infty} \L_{2n}(\lambda_{m})^{\Gamma}\,.
\label{PeterWeylGammaHigherSphere}
\end{equation}
Therefore, the multiplicity of the eigenvalue $m(m+2n-2)$ is
\begin{equation}
m_{\Gamma}(m) = \dim\left(\L_{2n}(\lambda_{m})^{\Gamma}\right)\,.
\label{MultiplicityGammaHigherSphere}
\end{equation}

\noindent To compute these multiplicities, we use the realization of the representations $\L_{2n}(\lambda_{m})$ in terms of harmonic homogeneous polynomials. Let $\mathcal{P}_{m}(\mathbb{R}^{2n})$ denote the space of complex-valued homogeneous polynomials of degree $m$ on $\mathbb{R}^{2n}$, and let
\begin{equation*}
\mathcal{H}_{m}(\mathbb{R}^{2n}) = \left\{\P \in \mathcal{P}_{m}(\mathbb{R}^{2n})\,, \Delta_{\mathbb{R}^{2n}}\P = 0\right\}\,.
\end{equation*}
It is well known that
\begin{equation}
\mathcal{H}_{m}(\mathbb{R}^{2n}) \cong \L_{2n}(\lambda_{m})\,.
\label{HarmonicRepresentationHigherSphere}
\end{equation}
In particular,
\begin{equation*}
m_{\Gamma}(m) = \dim\left(\mathcal{H}_{m}(\mathbb{R}^{2n})^{\Gamma}\right)\,.
\end{equation*}

\noindent Recall that the classical Molien series computes the dimensions of the spaces of $\Gamma$-invariant homogeneous polynomials, rather than harmonic homogeneous polynomials. However, the latter can be recovered using the Fischer decomposition. Indeed,
\begin{equation*}
\mathcal{P}_{m}(\mathbb{R}^{2n}) = \mathcal{H}_{m}(\mathbb{R}^{2n}) \oplus \|x\|^{2}\mathcal{P}_{m-2}(\mathbb{R}^{2n})\,,
\end{equation*}
where $\|x\|^{2} = x_{1}^{2} + \cdots + x_{2n}^{2}$ denotes the squared Euclidean norm of $x = \left(x_{1}\,,\ldots\,,x_{2n}\right)$. Since $\Gamma \subseteq \SO(2n)$ preserves the Euclidean norm, this decomposition is $\Gamma$-invariant.

\noindent For every $m \geq 0$, let
\begin{equation*}
a_{m} = \dim\left(\mathcal{P}_{m}(\mathbb{R}^{2n})^{\Gamma}\right)\,,
\end{equation*}
with $a_{m}=0$ for $m<0$. Taking $\Gamma$-invariant vectors in the Fischer decomposition gives
\begin{equation}
m_{\Gamma}(m)=a_{m}-a_{m-2}\,.
\label{HarmonicMultiplicityHigherSphere}
\end{equation}

\noindent Therefore, if
\begin{equation*}
\mathcal{M}_{\Gamma}(q) = \sum\limits_{m=0}^{\infty}a_{m}q^{m} = \frac{1}{\left|\Gamma\right|} \sum\limits_{\gamma\in\Gamma} \frac{1}{\det(\Id_{2n}-q\gamma)}
\end{equation*}
denotes the Molien series of $\Gamma$, we obtain
\begin{equation}
\sum\limits_{m=0}^{\infty} m_{\Gamma}(m)q^{m} = \left(1-q^{2}\right)\mathcal{M}_{\Gamma}(q) = \frac{1-q^{2}}{\left|\Gamma\right|} \sum\limits_{\gamma\in\Gamma} \frac{1}{\det(\Id_{2n}-q\gamma)}\,.
\label{HarmonicMolienHigherSphere}
\end{equation}
This formula provides a direct connection between invariant theory and the spectral multiplicities of the Laplace-Beltrami operator.

\begin{remark}

To study higher-dimensional lens spaces, we identify $\mathbb{R}^{2n}$ with $\mathbb{C}^{n}$ through
\begin{equation*}
\left(x_{1}\,, \ldots\,, x_{2n}\right) \longmapsto \left(x_{1}+ix_{2}\,, \ldots\,, x_{2n-1}+ix_{2n}\right)\,.
\end{equation*}
Under this identification, the sphere $S^{2n-1}$ is realized as
\begin{equation*}
S^{2n-1} = \left\{\left(z_{1}\,, \ldots\,, z_{n}\right) \in \mathbb{C}^{n}\,, \sum\limits_{j=1}^{n}\left|z_{j}\right|^{2} = 1\right\}\,.
\end{equation*}
Moreover, the standard embedding $\U(n)\hookrightarrow\SO(2n)$ allows us to regard finite subgroups of $\U(n)$ as finite subgroups of $\SO(2n)$. Therefore, the preceding results obtained from the realization
\begin{equation*}
S^{2n-1} \cong \SO(2n)/\SO(2n-1)
\end{equation*}
remain applicable.

\end{remark}

We illustrate this construction with higher-dimensional lens spaces. Let $p \geq 2$ and let $q_{1}\,,\ldots\,,q_{n}\in\mathbb{Z}_{\geq1}$ such that $\gcd(p\,,q_{j})=1, 1 \leq j \leq n$. Let $\zeta = e^{\frac{2\pi i}{p}}$, and let $\Gamma = \mathbb{Z}_{p}$ be the cyclic group generated by
\begin{equation*}
\gamma\cdot(z_{1}\,,\ldots\,,z_{n}) = \left(\zeta^{q_{1}}z_{1}\,, \ldots\,, \zeta^{q_{n}}z_{n}\right)\,.
\end{equation*}
Since each $q_{j}$ is relatively prime to $p$, the action of $\Gamma$ on $S^{2n-1}$ is free. The corresponding lens space is given by
\begin{equation}
L^{2n-1}(p;q_{1}\,,\ldots\,,q_{n}) = \Gamma \backslash S^{2n-1}\,.
\label{HigherDimensionalLensSpace}
\end{equation}
Note that $\Gamma$ is a subgroup of $\U(n)$. Under the standard embedding $\U(n)\hookrightarrow\SO(2n)$, the eigenvalues of the complexification of $\gamma^{k}$ are
\begin{equation*}
\zeta^{kq_{1}}\,, \zeta^{-kq_{1}}\,, \ldots\,, \zeta^{kq_{n}}\,, \zeta^{-kq_{n}}\,.
\end{equation*}
Therefore,
\begin{equation*}
\det(\Id_{2n}-q\gamma^{k}) = \prod\limits_{j=1}^{n} \left(1-\zeta^{kq_{j}}q\right)\left(1-\zeta^{-kq_{j}}q\right)\,.
\end{equation*}
Applying Equation \eqref{HarmonicMolienHigherSphere}, we obtain the following result.

\begin{proposition}

The generating function for the multiplicities of the eigenvalues $m(m+2n-2)$ of the Laplace-Beltrami operator on $L^{2n-1}(p;q_{1}\,,\ldots\,,q_{n})$ is given by
\begin{equation}
\sum\limits_{m=0}^{\infty}m_{\Gamma}(m)q^{m} = \frac{1-q^{2}}{p} \sum\limits_{k=0}^{p-1}\prod\limits_{j=1}^{n}\frac{1}{\left(1-\zeta^{kq_{j}}q\right)\left(1-\zeta^{-kq_{j}}q\right)}\,.
\label{MolienHigherDimensionalLensSpace}
\end{equation}

\end{proposition}

\noindent An equivalent expression can be obtained by counting invariant monomials. For $m\in\mathbb{Z}$, define
\begin{equation}
\N_{\Gamma}(m) = \#\bigg\{\left(\alpha\,, \beta\right) \in \mathbb{Z}_{\geq0}^{n} \times \mathbb{Z}_{\geq0}^{n}\,, \left|\alpha\right| + \left|\beta\right| = m\,, \sum\limits_{j=1}^{n} q_{j}\left(\alpha_{j}-\beta_{j}\right) \equiv0\pmod p\bigg\}\,,
\label{InvariantCountHigherLensSpace}
\end{equation}
with $\N_{\Gamma}(m)=0$ for $m<0$. Here $\left|\alpha\right| = \alpha_{1}+\cdots+\alpha_{n}$ and similarly for $|\beta|$. The monomial $z^{\alpha}\overline{z}^{\beta}$ is $\Gamma$-invariant precisely when
\begin{equation*}
\sum\limits_{j=1}^{n} q_{j}(\alpha_{j}-\beta_{j}) \equiv 0 \pmod p\,.
\end{equation*}
Combining this observation with the Fischer decomposition gives the following proposition.

\begin{proposition}

For every $m \geq 0$, the multiplicity of the eigenvalue $m(m+2n-2)$ of the Laplace-Beltrami operator on $L^{2n-1}(p; q_{1}\,,\ldots\,,q_{n})$ is
\begin{equation}
m_{\Gamma}(m) = \N_{\Gamma}(m)-\N_{\Gamma}(m-2)\,.
\label{MultiplicityHigherDimensionalLensSpace}
\end{equation}
In particular, its spectral zeta function is given, for $\Re(s)>\frac{2n-1}{2}$, by
\begin{equation}
\zeta_{\Delta_{\Gamma\backslash S^{2n-1}}}(s) = \sum\limits_{m=1}^{\infty} \frac{\N_{\Gamma}(m)-\N_{\Gamma}(m-2)}{\left(m(m+2n-2)\right)^{s}}\,.
\label{SpectralZetaHigherDimensionalLensSpace}
\end{equation}

\end{proposition}

\noindent In the particular case $q_{1}=\cdots=q_{n}=1$, the invariance condition reduces to
\begin{equation*}
\left|\alpha\right| - \left|\beta\right| \equiv 0 \pmod p\,.
\end{equation*}
Writing $a=|\alpha|$ and $|\beta|=m-a$, we obtain
\begin{equation}
\N_{\Gamma}(m) = \sum\limits_{\substack{0\leq a\leq m \\ 2a-m\equiv0\pmod p}} \binom{a+n-1}{n-1} \binom{m-a+n-1}{n-1}\,.
\label{InvariantCountHomogeneousHigherLens}
\end{equation}
Together with Equation \eqref{MultiplicityHigherDimensionalLensSpace}, this provides an explicit expression for the spectral multiplicities of $\L^{2n-1}(p;1\,,\ldots\,,1)$. Further computations, including the meromorphic continuation and regularized determinants of the corresponding spectral zeta functions, will be developed in \cite{VRAS}.

As a simple example, consider $p=2$. In this case, $\Gamma$ acts by the antipodal map, and
\begin{equation*}
L^{2n-1}(2;1\,,\ldots\,,1) \cong\mathbb{RP}^{2n-1}\,.
\end{equation*}
Since the antipodal map acts on homogeneous polynomials of degree $m$ by multiplication by $(-1)^{m}$, only even degrees contribute. Thus,
\begin{equation}
m_{\Gamma}(m) = \begin{cases} \displaystyle \binom{m+2n-1}{2n-1} - \binom{m+2n-3}{2n-1} & \text{ if $m$ is even } \\  0 & \text{if $m$ is odd} \end{cases}
\label{MultiplicityRealProjectiveHigherSphere}
\end{equation}
Therefore,
\begin{equation}
\zeta_{\Delta_{\mathbb{RP}^{2n-1}}}(s) = \sum\limits_{k=1}^{\infty} \frac{ \displaystyle\binom{2k+2n-1}{2n-1} - \binom{2k+2n-3}{2n-1}}{\left(4k(k+n-1)\right)^{s}}\,.
\label{SpectralZetaRealProjectiveHigherSphere}
\end{equation}

\subsection{Compact symmetric spaces of rank one}

The methods developed in this paper are not restricted to the exceptional identification $S^{3} \cong \SU(2)$, nor do they depend on the parity of the dimension of the sphere. Indeed, for every $d\geq2$, we have
\begin{equation*}
S^{d} \cong \SO(d+1)/\SO(d)\,,
\end{equation*}
and the same representation-theoretic approach applies. As illustrated in the preceding subsection for $S^{2n-1}$, the decomposition of $\L^{2}(\G/\K)$ into irreducible representations, together with the computation of Casimir eigenvalues and dimensions of invariant subspaces, provides the spectral data required to construct the corresponding spectral zeta functions.

More generally, this framework extends to compact symmetric spaces of rank one. Besides spheres, these include the complex projective spaces
\begin{equation*}
\mathbb{CP}^{n} \cong \SU(n+1)/\S(\U(n)\times\U(1))\,,
\end{equation*}
the quaternionic projective spaces
\begin{equation*}
\mathbb{HP}^{n} \cong \Sp(n+1)/(\Sp(n)\times\Sp(1))\,,
\end{equation*}
and the Cayley projective plane
\begin{equation*}
\mathbb{OP}^{2} \cong \F_{4}/\Spin(9)\,.
\end{equation*}
In each case, the Cartan-Helgason theorem determines the irreducible representations contributing to $\L^{2}(\G/\K)$, while their Casimir eigenvalues and dimensions provide the corresponding spectral multiplicities.

The representation-theoretic decomposition, the meromorphic continuation of the associated spectral zeta functions, and explicit computations of regularized determinants for these spaces and suitable finite quotients will be investigated in subsequent work (see also \cite{VRAS}).

\subsection{Towards non-compact symmetric spaces}

The representation-theoretic methods developed in this paper suggest a natural extension to non-compact Riemannian symmetric spaces. In the compact setting, the Peter-Weyl theorem provides a decomposition of $\L^{2}(\G/\K)$ into irreducible representations,
allowing us to determine the spectrum of the Laplace--Beltrami operator through the action of the Casimir element. In the non-compact setting, an analogous decomposition is provided by Harish-Chandra's Plancherel theorem.

\noindent Let $\G$ be a connected non-compact semisimple Lie group with finite center, and let $\K$ be a maximal compact subgroup. The quotient $\G/\K$ is a Riemannian symmetric space of non-compact type. In this setting, the analogue of the Peter--Weyl theorem is provided by Harish-Chandra's Plancherel theorem. More precisely, we have
\begin{equation}
\L^{2}(\G) \cong \displaystyle\int_{\widehat{\G}_{\mathrm{temp}}}^{\oplus} \V_{\pi}\otimes\V_{\pi}^{*} d\mu_{\mathrm{Pl}}(\pi)\,,
\label{PlancherelNonCompactGroup}
\end{equation}
where $\widehat{\G}_{\mathrm{temp}}$ denote the tempered unitary dual of $\G$ and $d\mu_{\mathrm{Pl}}$ is the Plancherel measure. Taking the $\K$-invariant subspace with respect to the right regular action, we obtain
\begin{equation}
\L^{2}(\G/\K) \cong \displaystyle\int_{\widehat{\G}_{\K}}^{\oplus} \V_{\pi}\otimes \left(\V_{\pi}^{\K}\right)^{*} d\mu_{\mathrm{Pl}}(\pi)\,,
\label{PlancherelNonCompactSymmetricSpace}
\end{equation}
where $\widehat{\G}_{\K}$ denotes the tempered spherical unitary dual, consisting of tempered irreducible unitary representations admitting non-zero $\K$-fixed vectors. The corresponding spherical Plancherel measure is described explicitly by Harish-Chandra in terms of his $c$-function (see, for example, \cite{Helgason1984})\,.

As in the compact case, the Laplace-Beltrami operator is related to the Casimir element of $\G$. Therefore, its spectral decomposition can be described through the infinitesimal characters of the representations appearing in Equation \eqref{PlancherelNonCompactSymmetricSpace}. The essential difference is that the spectrum is generally continuous, and the corresponding spectral quantities are expressed through integrals against the Plancherel measure rather than discrete sums.

An important example is the real hyperbolic space
\begin{equation*}
\mathbb{H}^{n} \cong \SO_{0}(n,1)/\SO(n)\,.
\end{equation*}
For this symmetric space, the spherical Plancherel formula provides an explicit description of the spectral decomposition of the Laplace-Beltrami operator. However, because $\mathbb{H}^{n}$ is non-compact and has infinite volume, the ordinary heat trace and spectral zeta function require suitable regularization or a relative formulation.

This framework also opens the possibility of studying Selberg zeta functions and their relations to regularized determinants. The representation-theoretic approach developed here, together with Harish-Chandra's Plancherel theory, provides a natural starting point for these investigations, which will be pursued in subsequent work.

We aim to obtain generalization of the Selberg zeta function for $\Gamma\bs\mathbb H^n$, where $\Gamma$ is a discrete subgroup of $\mathrm{Isom}^+(\mathbb H^n)$.
In the physics literature, the one-loop partition function on $\Gamma\bs\mathbb H^3$ has been studied in \cite{Giombi:2008vd} for $\Gamma\cong \mathbb Z$ using heat kernel techniques, which was related to the Selberg zeta function in \cite{Williams,PerryWilliams,Haupfear:2025grb}.
Obtaining closed form expressions for the one-loop partition function on $\Gamma\bs\mathbb H^3$ for more general $\Gamma$ will be a primary goal for our program. 
Considering regularized determinants on Lorentzian manifolds is another primary focus of our program, such as (quotients of) Minkowski and (anti-) de Sitter spaces, whose isometry groups are non-compact.

\bigskip

\bigskip

\noindent \textbf{Acknowledgements:} 
VM and RP thank the University of Iceland for its hospitality during part of this work. They also thank the Isaac Newton Institute for Mathematical Sciences, Cambridge, for support and hospitality during the workshop ``Physics and automorphic $L$-functions: gravity, conformal field theory and number theory'' (OOEW12), where part of this work was completed.

\bigskip

\noindent \textbf{Declaration of Generative AI:} LLM tools (ChatGPT and Claude) were used solely for language editing and polishing the final draft of the manuscript. All mathematical results and proofs were developed independently by the authors, without the use of any large language models (LLMs). The authors take full responsibility for the content.

\bigskip

\bigskip

\appendix

\section{Conventions and definitions}

\label{app:Defs}

Here we define some of the special functions we use in this paper, to establish conventions.

\begin{definition}

The Mellin transform is an integral transform defined as follows,
\begin{equation} 
\label{eq:MellinDef}
F(s) = M[f(t)](s) = \int_0^\infty f(t) t^{s-1} dt.
\end{equation}
If $F(s)$ converges in the fundamental strip $a < \Re(s) < b$, then for $c\in(a,b)$, one can define the inverse Mellin transform as a contour integral
\begin{equation}
f(t) = M^{-1}[F(s)](t) = \frac1{2\pi i}\int_{c-i\infty}^{c+i\infty} F(s) t^{-s}ds,
\end{equation}
where the contour must lie within the fundamental strip.

\end{definition}

\begin{definition}

The Gamma function extends the factorial ($\Gamma(n+1) = n!$, $n\in\mathbb Z_{\geq 0}$) from non-negative integers to the complex plane as the Mellin transform of the exponential,
\begin{equation}
\label{eq:GammaDef}
\Gamma(s) = M[e^{-t}](s) = \int_0^\infty t^{s-1}e^{-t}dt, \quad \Re(s) > 0. 
\end{equation}
To analytically continue to the rest of the complex plane one observes $\Gamma(s+1) = s \Gamma(s)$. 
One then finds the Gamma function is meromorphic with simple poles at non-positive integers $s\in-\mathbb Z_{\geq 0}$ and no zeroes.
\end{definition}
The absence of zeroes indicates that $\frac1{\Gamma(s)}$ is entire. 
Indeed, $\frac1{\Gamma(s)}$ admits an expansion in small $s$,
\begin{equation} 
\label{eq:GammaExp}
\frac{1}{\Gamma(s)} = s+\gamma_E s^2+\frac{1}{12} \left(6\gamma_E^2-\pi ^2\right) s^3+O\left(s^4\right),\quad \frac1{\Gamma(0)} = 0 ,\quad \pd_s \frac1{\Gamma(s)} \bigg|_{s=0} = 1, 
\end{equation}
where $\gamma_E$ is the Euler-Mascheroni constant.
Some useful identities which have been used in some calculations are
\begin{equation}\label{eq:GammaIdentities}
\Gamma(z+1) = z\Gamma(z),\quad \Gamma(z)\Gamma(1-z)= \frac{\pi}{\sin(\pi z)}, \quad \Gamma(z)\Gamma\left(z+\frac12\right) = 2^{1-2z}\sqrt{\pi} \Gamma(2z).
\end{equation}
The logarithmic derivative of the Gamma function is the digamma function $\psi$,
\begin{equation} 
\label{eq:DigammaDef}
\psi(z) = \frac{\Gamma'(z)}{\Gamma(z)} = -\gamma_E + \sum_{n=0}^\infty \left(\frac1{n+1}-\frac1{n+z}\right),\quad z\in\mathbb C\setminus(-\mathbb Z_{\geq 0})
\end{equation}

\begin{definition} 
\label{def:Pochhammer}

The Pochhammer symbol or the rising factorial is defined by
\begin{equation}
(x)_n = x(x+1)(x+2)\cdots(x+n-1),
\end{equation}
which can be extended by using the Gamma function
\begin{equation}
(x)_n = \frac{\Gamma(x+n)}{\Gamma(x)}. 
\end{equation}

\end{definition}

\begin{definition}

The Riemann zeta function has the following Dirichlet series and Euler product representation,
\begin{equation}
\zeta_R(s) = \sum_{n=1}^\infty \frac1{n^s} = \prod_{p\text{ prime}}\frac{1}{1-p^{-s}} \qquad \left(\Re(s) > 1\right)\,,
\end{equation}
which can be meromorphically continued to the entire complex plane with a simple pole at $s=1$ with unit residue.
The Riemann zeta function satisfies the  functional equation
\begin{equation}
\label{eq:RiemannZetaFunct}
\zeta_R(s) = \pi^{s-\frac12}\frac{\Gamma(\tfrac{1-s}2)}{\Gamma(\tfrac s2)} \zeta_R(1-s) = 2(2\pi)^{s-1}\Gamma(1-s)\sin\left(\frac{\pi s}{2}\right) \zeta_R(1-s).
\end{equation}
\end{definition}

\begin{definition}

A generalization of the Riemann zeta function is the Hurwitz zeta function, with Dirichlet series
\begin{equation}
\label{eq:HurwitzDef}
\zeta_H(s,a) = \sum_{n=0}^\infty \frac1{(n+a)^s}, \quad \Re(s)>1,\, a\in\mathbb C\bs(-\mathbb Z_{\geq 0}),
\end{equation}
whose meromorphic continuation to the complex plane also contains a simple pole at $s=1$ with unit residue.
Note that $\zeta_H(s,1) = \zeta_R(s)$.
\end{definition}
The Hurwitz zeta function admits a Laurent expansion around the $s=1$ pole,
\begin{equation} 
\label{eq:HurwitzLaurent}
\zeta_H(s,a) = \frac1{s-1} - \psi(a) + O(s-1), \qquad \psi(a) = \frac{\Gamma'(a)}{\Gamma(a)},
\end{equation}
where $\psi(a)$ is the digamma function. 

\begin{definition}

Polylogarithms are another generalization of the Riemann zeta function.  They are defined via a Dirichlet series
\begin{equation}
\label{eq:PolylogDef}
\Li_s(z) = \sum_{n=1}^\infty \frac{z^n}{n^s}, \quad \abs{z} < 1,
\end{equation}
analytically continued to $z\in\mathbb C\setminus[1,\infty)$,
with the following special cases which we commonly use,
\begin{equation}
\Li_1(z) = -\log(1-z),\; \abs{z}\leq 1; \quad \Li_s(1) = \zeta_R(s). 
\end{equation}
The first equality is taken in the principal branch, the second equality holds for $\Re(s)>1$ and then is understood via analytic continuation elsewhere in $s\in\mathbb C$.

\end{definition}
The Hurwitz zeta function admits a Fourier expansion for $0<a<1$, and meromorphically continues the Hurwitz zeta function to the entire complex plane, known as Hurwitz's formula, and can be expressed in terms of polylogarithms,
\begin{equation}\label{eq:HurwitzFourier}
\begin{split}
\zeta_H(s,a) 
&= 2(2\pi)^{s-1}\Gamma(1-s)\sum_{w=1}^\infty w^{s-1}\sin\left(2\pi w a + \frac{\pi s}{2}\right)\\
&= i(2\pi)^{s-1}\Gamma(1-s)\left(e^{-\frac{\pi i s}{2}}\Li_{1-s}(e^{-2\pi i a}) -e^{\frac{\pi i s}{2}} \Li_{1-s}(e^{2\pi i a}) \right).
\end{split}
\end{equation}

\begin{definition} 
\label{def:JacobiThetaDef}

We define Jacobi theta functions with characteristics, and list their transformations under the $\mathrm{SL}(2,\mathbb Z)$ generators as follows. 
\begin{equation}
\begin{split}    
\JacTh{a}{b}{z}{\tau} 
&= \sum_{n\in\mathbb Z} e^{\pi i(n+a)^2\tau + 2\pi i(n+a)(z+b)},\quad \Im(\tau)>0,\quad a,b\in\mathbb R, \\
\mc T\,\JacTh{a}{b}{z}{\tau} 
&= \JacTh{a}{b}{z}{\tau+1} = e^{-\pi i a(a+1)}\,\JacTh{a}{b+a+\frac12}{z}{\tau}, \\
\mc S\, \JacTh{a}{b}{z}{\tau} 
&= \JacTh{a}{b}{\frac z\tau}{-\frac1\tau} = \sqrt{-i\tau} \, e^{2\pi i a b+\frac{i\pi z^2}{\tau}}\JacTh{b}{-a}{z}{\tau}.
\end{split}
\end{equation}
The relationships with the classical Jacobi theta functions are
\begin{equation}
\begin{split}
\JacTh{\frac12}{\frac12}{z}{\tau} &= -\vartheta_1(z;\tau),\quad
\JacTh{\frac12}{0}{z}{\tau} = \vartheta_2(z;\tau), \\  
\JacTh{0}{0}{z}{\tau} &= \vartheta_3(z;\tau),\quad
\JacTh{0}{\frac12}{z}{\tau} = \vartheta_4(z;\tau).
\end{split}
\end{equation}

\end{definition}

Where possible we will use the classical Jacobi theta functions in place of the theta function with characteristics for brevity. 
Furthermore, if the elliptic parameter $z=0$, we will drop the argument, $\vartheta_i(0;\tau) \equiv \vartheta_i(\tau)$. 

\begin{definition}
The Bessel-K function is defined via a Mellin transform,
\begin{equation}
\label{eq:BesselKDef}
K_s(z) = \frac12 M[e^{-\frac z2 (t+\frac1t)}](s) = \frac12\int_0^\infty t^{s-1} e^{-\frac z2(t+\frac1t)} dt, \quad |\arg(z)| < \frac\pi2,
\end{equation}
which is a solution to the second order differential equation, $\big(z^2\pd_z^2 + z\pd_z -(z^2+s^2)\big)K_s(z) = 0$, and is entire in $s$ for fixed non-zero $z$.
It is also symmetric about $s=0$, satisfying $K_{s}(z) = K_{-s}(z)$.
\end{definition}
We use the following special cases in this paper, 
\begin{equation}\label{eq:BesselKCases}
K_{\frac32}(z) = e^{-z}\sqrt{\frac{\pi}{2z}} \left(1+\frac1z\right), \quad  K_{\frac52}(z) = e^{-z}\sqrt{\frac{\pi}{2z}} \left(1+\frac3z+\frac3{z^2}\right).
\end{equation}

\bibliographystyle{JHEP}
\bibliography{s3bib}

\medskip

\end{document}